\documentclass{ar-1col}

\usepackage[comma]{natbib}
\usepackage{url,amssymb}
\usepackage{mymacros}

\jname{Annu. Rev. Astron. Astrophys.}
\jvol{61}
\jyear{2023}
\begin{document}

% Page header
\markboth{\"Oberg et al.}{Protoplanetary Disk Chemistry}

% Title
\title{Protoplanetary Disk Chemistry}

%Authors, affiliations address.
\author{Karin I. \"Oberg,$^1$ Stefano Facchini,$^2$ and Dana E. Anderson$^3$
\affil{$^1$Center for Astrophysics | Harvard-Smithsonian, 60 Garden St, Cambridge, MA 02138, USA; email: koberg@cfa.harvard.edu}
\affil{$^2$Universit\`a degli Studi di Milano, via Giovanni Celoria 16, 20133 Milano, Italy}
\affil{$^3$Earth and Planets Laboratory, Carnegie Institution for Science, 5241 Broad Branch Road, NW, Washington, DC 20015, USA}} 

%Abstract
\begin{abstract}
Planets form in disks of gas and dust around young stars. The disk molecular reservoirs and their chemical evolution affect all aspects of planet formation, from the coagulation of dust grains into pebbles, to the elemental and molecular compositions of the mature planet. Disk chemistry also enables unique probes of disk structures and dynamics, including those directly linked to ongoing planet formation. Here we review the protoplanetary disk chemistry of the volatile elements HOCNSP, the associated observational and theoretical methods, and the links between disk and  planet chemical compositions. Three takeaways from this review are:\\
$\bullet$ The disk chemical composition, including the organic reservoirs, is set by both inheritance and {\it in situ} chemistry.\\
$\bullet$ Disk gas and solid O/C/N/H elemental ratios often deviate from stellar values due to a combination of condensation of molecular carriers, chemistry, and dynamics.\\
$\bullet$ Chemical, physical, and dynamical processes in disks are closely linked, which complicates disk chemistry modeling, but these links also present an opportunity to develop chemical probes of different aspects of disk evolution and planet formation.
\end{abstract}

%Keywords, etc.
\maketitle

%Table of Contents
\tableofcontents

% Heading 1
\section{INTRODUCTION}

\subsection{Motivation and Scope of Review}

Protoplanetary disks are planet-forming, potentially planet-containing, dust and gas-rich disks around young stars. Disk chemical compositions determine the elemental and molecular makeup of forming planets, and influence many other disk properties relevant to planet formation.
Disk chemistry also provides some of our best tools to characterize disk structures and dynamics, including the presence of protoplanets. To develop a predictive theory of planet formation therefore requires a deep understanding of the chemistry of protoplanetary disks. This is true for all kinds of planets, but is perhaps especially salient when considering how often temperate planets in our Galaxy may be chemically hospitable to life, i.e. have access to water and a suitable combination of organic and inorganic material.   

The aim of this review is to introduce the study of disk chemistry, to present our current state of knowledge emerging from the past two decades of observational, theoretical, and experimental work, and to propose some possible paths forward. We limit the scope of the review in two important aspects. First we only treat protoplanetary disks, i.e. circumstellar disks around 1-10 Myr old stars around which the natal envelope has been cleared, and therefore do not include the chemistry of younger protostellar disks, or the debris disks found around older stars. Second, we focus on the volatile elements and molecules and therefore largely ignore the physics and chemistry of refractory elements bound up in dust particles -- for the purpose of this review, a `volatile' element or molecule is any species that is present as a liquid or gas under typical terrestrial conditions, which includes most small and mid-sized HOCNSP-molecules. The scope of this review is then the distributions and chemistry of volatiles in protoplanetary disks. 

\subsection{Context: Protoplanetary Disk Formation and Evolution}
\label{sec:context}

Protoplanetary disks emerge within the context of star formation, when in-falling interstellar cloud material becomes distributed in disk-like structures to preserve angular momentum \citep{Shu87}. The chemical composition of the resulting disk likely depends on both disk chemical processes and chemical inheritance from the preceding interstellar and protostellar phases \citep[e.g.][]{Visser09}.  The links between interstellar, protostellar and disk compositions was recently reviewed by \citet{Oberg21_Review}, \citet{Bergin23} and \citet{vanDishoeck21b}, and protostellar organic chemistry by \citet{Jorgensen20}. Here we simply summarize a few aspects of interstellar and protostellar chemistry of especial relevance to protoplanetary disks and planet formation. 

In the diffuse interstellar medium, a substantial fraction of oxygen, and half of the carbon, are incorporated into refractory grains. The remaining carbon and oxygen, as well as most nitrogen and hydrogen become bound up in stable molecules such as H$_2$, CO, CO$_2$, N$_2$, NH$_3$, and H$_2$O at the interstellar cloud stage. Many of these carriers survive disk formation. Their chemical reactivities (or lack thereof) and different volatilities impact the distribution of major volatile elements in disks. Molecular clouds also produce a first generation of organic molecules, including carbon chains, CH$_4$, CH$_3$OH, and more complex organics. At the onset of star and planet formation in molecular clouds, the majority of these molecules are found in icy grain mantles. In the subsequent protostellar phase, additional complex organics form, which together with the interstellar volatiles and organics can become incorporated into disks, setting the initial conditions for disk chemistry. 

\begin{figure}[h]
\includegraphics[width=5in]{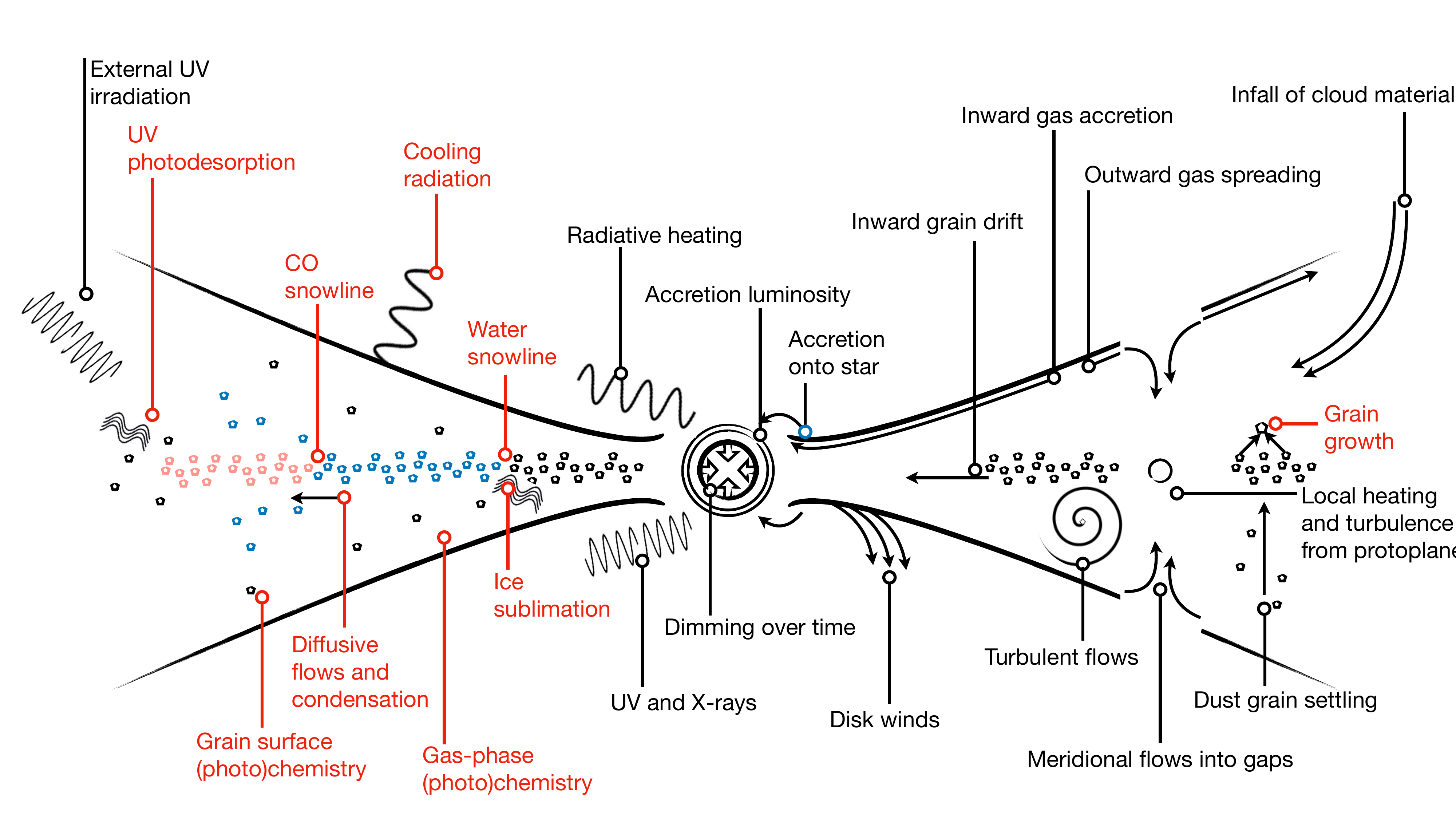}
\caption{An non-exhaustive overview of the many processes in disks that either directly impact chemistry and the distribution of molecules, depend on the disk molecular composition, are traced by molecular lines, and/or directly involve the destruction or formation of molecules or molecular phases. The processes labeled in black are stellar and disk dynamical processes that act on disk gas and dust regardless of their particular chemical makeup. Processes labeled in red are either chemical processes or processes that directly depend on the chemical composition. In the left-hand-side of the illustration bare silicate grains are grey, water-ice-dominated grains are blue, and grains with a large reservoir of hypervolatiles, including CO, are pink.}
\label{fig-cartoon}
\end{figure}

The second context for this review is the local environment within which a disk is situated -- the environment surrounding the disk sets the external radiation field, and may dynamically shape the disk structure and continue to feed the disk with gas and dust  \citep{Walsh14,Andrews15,Huang21} -- as well as the disk evolving structure and dynamical processes (Fig. \ref{fig-cartoon}). Protoplanetary disks were recently reviewed by \citet{Andrews20}, while \citet{Aikawa22} provide an introduction to many disk structures and processes relevant for disk chemistry. 

Within the disk the surface density is set by a combination of infall of material from the natal cloud and protostellar envelope, inward gas accretion, outward gas spreading, accretion onto the star, disk winds, ongoing planet formation and other gap-opening mechanisms \citep[e.g.][]{LyndenBell74,Bai13,Andrews18}. Gas and dust are transported in disks both vertically and radially. Vertical gas transport happens through large scale turbulence, diffusion, and meridonial flows into disk gaps \citep[e.g.][]{Meijerink09,Semenov11,Teague19}, while solids are preferentially moved towards the midplane due to grain growth and settling. Radially, gas and small entrained grains are transported through advection, turbulent diffusion, and disk winds, while larger dust grains partially decouple from the gas and drift towards pressure maxima, which in the absence of sub-structure entails inward transport of solids \citep[e.g.][]{dAlessio06,Ciesla06}. Taken together we should expect transport of gas and solids both between the outer low density and inner high density disk regions, and between the disk midplane and elevated layers. 

Disks are heated by stellar radiation, radiation from accretion shocks onto the star, and viscous dissipation  \citep[e.g.][]{dAlessio99}, all which decrease over time due to stellar contraction and a decreasing accretion rate. The decrease is not monotonic, however, and luminosity bursts may temporarily change the disk temperature structure \citep[e.g.][]{Armitage01}. The disk temperature further depends on the distribution of solids and molecules in the disk atmosphere, which control cooling and heating in the disk \citep{Woitke09}. At disk radii dominated by radiative heating, the disk temperature increases with disk height. Disks are also irradiated by UV and X-rays and Cosmic Rays from both the central star and from external sources \citep[e.g.][]{Kastner97,Cleeves15,Rab18}. The radiative transport of these high-energy particles and radiation is complex, especially in the presence of dust evolution and sub-structure, but generally the fluxes are high in the disk atmosphere and increasingly depleted towards the disk midplane. 

\subsection{An Introduction to Inner and Outer Disk Chemistry \label{sec:intro}}

When surveying the chemical processes that govern the distributions and evolution of volatiles in disks it is useful to consider the disk region interior to the water condensation front or snowline (the inner disk where terrestrial planets form) separately from the rest of the disk (the outer disk where gas and ice giants as well as comets originate). Dependent on the stellar luminosity, this boundary between inner and outer disk falls between a fraction of an au and $\sim$10~au. In the inner disk high densities and temperatures result in complete sublimation of volatiles and short chemical timescales in the gas phase that resets the chemistry \citep[e.g.][and references therein]{Glassgold09,Pontoppidan2014}. In contrast to most phases of star and planet formation, three-body reactions, endothermic reactions, and reactions with a substantial energy barriers may all be all possible. The inner disk may be considered to have an inner boundary at the silicate condensation line \citep{Kress10}, and in between this and the water snowline, it should  host a soot line, where large carbon molecules and small carbon grains volatilize. Despite its high densities and temperatures, the inner disk chemistry may not be at equilibrium due to strong irradiation fields impinging on the disk atmosphere \citep[e.g.][]{Adamkovics14}, and continuous transport of material from the outer disk. 

Outer disk chemistry has been reviewed by \citet{Henning13}, \citet{Dutrey14} and \citet{Aikawa22}. In the outer disk, inheritance cannot be ignored, especially in the disk midplane, where interstellar and protostellar chemical products may be preserved \citep{Visser09}. As a result, the outer disk likely begins with much of its oxygen tied up in H$_2$O and CO$_2$-rich ices and therefore taken out of circulation from the gas-phase. Outer disks are chemically stratified both radially and vertically due to two-dimensional condensation fronts, which are referred to as snowlines in the disk midplane, and the dependence of chemical reactions of density, temperature, ionization, and dissociative radiation \citep{Aikawa02}. The disk midplane may be chemically inert, simply preserving inherited reservoirs, or quite chemically active, dependent on its access to ionizing irradiation \citep{Cleeves15}. In somewhat elevated disk layers neutral-neutral and ion-molecule gas-phase reactions
operate together with grain surface chemistry to produce a so-called molecular layer, where the chemistry is neither completely reset nor completely inherited. Finally the outer disk atmosphere is characterized by relatively low densities, moderately high temperatures and high irradiation fields, which drive a rapid radical and ion based chemistry, qualitatively similar to a classical photondominated region (PDR). These different layers may all be dynamically and therefore chemically connected, however, blurring some of the described differences.  

\section{DISK CHEMISTRY MODELING \label{sec:models}}

Disk chemistry models are used to provide interpretative frameworks for observations, to expand the kinds of species that can be characterized in disks beyond what is directly observable, and connect different evolutionary phases up until and sometimes including planet formation. In this section we introduce astrochemistry modeling starting with how to set up the `physical' disk structure (\ref{sec:disk-environ}), followed by different treatments of chemistry within such a framework (\ref{sec:chem-treat}), and interactions between chemistry and dynamics (\ref{sec:models-dyn}), and finally how to compare models and observations (\ref{sec:model-obs}).

\begin{figure}[h]
\includegraphics[width=5in]{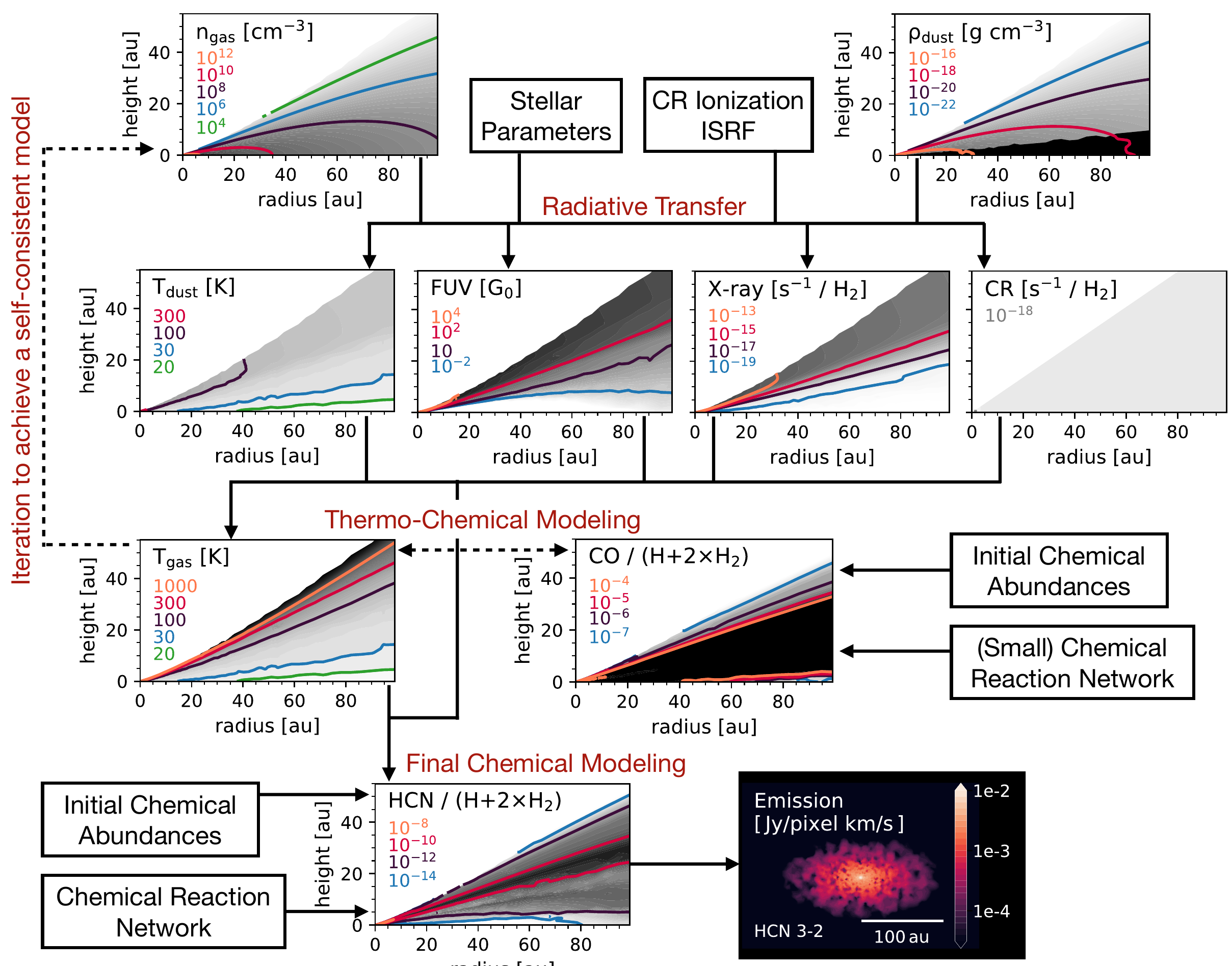}
\caption{The steps involved in a typical disk chemistry modeling framework. Starting at the top, (initial) gas and dust density structures are combined with stellar parameters and the local irradiation environment to calculated the disk temperature, radiation and ionization structures using radiative transfer. If a self-consistent model is desired these are then updated iteratively taking into account the effect of chemical abundances on cooling and heating in a thermo-chemical modeling step. The final chemical abundances are determined once the disk physical structure has converged, and the observable molecular line properties are calculated using radiative transfer.}
\label{figure:models}
\end{figure}

\subsection{Establishing the Disk Physical Structure \label{sec:disk-environ}} 

Chemical reactivity depends on the local  gas and dust grain densities, dust grain optical and material properties, gas and dust grain temperatures, energetic radiation fields, and ionization rates. Establishing these can be done relatively self-consistently taking into account the interdependence of gas and dust densities and temperatures, radiation fields and chemistry, but parametric structures are also frequently used to simplify the modeling process, or some aspect of it. The choice of approach depends ultimately on the scientific question at hand, and especially on the the acceptable computational expense of each model run. Fig. \ref{figure:models} illustrates some common approaches that are described below.

The establishment of the disk physical parameters begins with choosing an initial gas and dust density 2D structure using e.g. the steady-state solution for a protoplanetary disk undergoing viscous evolution \citep{LyndenBell74, Hartmann98}, or  parameterizations based on the minimum mass solar nebula (MMSN) or observations of disks \citep[e.g.,][]{Chiang10,Zhang21}. The vertical density distribution depends on the disk temperature, which can be accounted for by iteratively computing the gas temperature and vertical density distribution \citep[e.g.][]{Nomura07,Woitke09,Bruderer12}. The dust density structure commonly involves one or two dust populations, each described by a grain size distribution, set of opacities, and radial and vertical distribution; small dust grains (up to micron sizes) are assumed to couple to the gas, while larger grains are not \citep[e.g.,][]{dAlessio06}. The precise choice of dust properties can have a big impact on the model since dust affects ionization, heating, cooling, and chemistry \citep[e.g.][]{Gavino21}. While smoothly varying disks have been the norm in models, (sub)millimeter observations have motivated the investigation of more complex cases like transition disks, and disks with gaps opened by forming planets \citep[e.g.][]{Facchini18,Alarcon20a}.
 
 The disk gas and dust temperature, UV and X-ray radiation fields, and ionization fractions can be initially estimated through radiative transfer calculations using the aforementioned density structures  \citep[Fig.~\ref{figure:models} and e.g.][]{vanZadelhoff03,Nomura07,Bethell11a,Rab18}. The temperature calculations also need to take into account viscous heating in the dense inner disk midplane and the decoupling of gas and dust temperatures in the disk atmosphere \citep{Glassgold04}. The latter effect can be solved explicitly taking into account dust and gas heating and cooling processes \citep{Kamp04}. The gas heating and cooling processes depend on the gas chemical composition and therefore self-consistent determination of the temperature structure can only be done using thermochemical models that iteratively calculate the disk gas temperature and chemical abundances \citep{Woitke09, Bruderer12, Du14}. Finally, cosmic rays permeate through the entire disk and are typically incorporated at or below the hydrogen ionization rate in dense interstellar clouds \citep[see e.g.][]{Cleeves15}. Additional sources of ionization include radioactive decay and stellar energetic particles \citep{Cleeves14,Rab17}. 
 
 All of the above disk characteristics can also be approximated parametrically.  The dust and gas density structures can be calculated by first adopting a surface density profile and a radial temperature profile and then calculating vertical densities assuming hydrostatic equilibrium  \citep{Chiang97} or prescribing a Gaussian profile with a scale height \citep[e.g.,][]{Bruderer13}. The vertical temperature structure is frequently estimated using the prescription of \citet{Dartois03}. Gas and dust thermal decoupling can also be approximated with a parametric function \citep{Pinte06,Cleeves15}. The energetic radiation fields and ionization rates can be parameterized as a function of hydrogen column density and a chosen dust-to-gas ratio \citep[e.g.][]{Wakelam16,LeGal19}. In practice, most models combine parametric elements with self-consistent calculations that take into account some aspects of the interdependencies between chemistry, temperatures, dust and gas vertical density structures, and/or UV and X-ray radiation. 

\subsection{Treatments of chemistry \label{sec:chem-treat}}

 Early theoretical work on chemistry in protoplanetary disks largely relied on thermochemical equilibrium calculations \citep[see][]{Prinn93}. While potentially appropriate for the inner disk midplane, inner disk atmospheres and all layers in the outer disk require a non-equilibrium treatment. Building on interstellar chemistry codes \citep{vanDishoeck93, Bergin95}, \citet{Aikawa96} developed one of the first time dependent, rate equation-based disk chemistry models where abundances are calculated numerically based on formation and destruction rates, which depend on rate coefficients and abundances supplied from the previous time step. This kind of model then require initial chemical abundances and an appropriate reaction network, in addition to pre- or co-calculated disk density, temperature and radiation structures (Fig.~\ref{figure:models}).

There are two common choices of initial disk chemical abundances: inheritance of protostellar abundances, where the abundances are supplied by interstellar or protostellar observations or models  \citep[e.g.,][]{Nomura09,Walsh15} and "reset" compositions using high-temperature thermal equilibrium products \citep{Aikawa99} or assuming a fully atomic composition (except for H$_2$) \citep[e.g.,][]{Walsh10,Ballering21}.
Based on chemical timescales, the reset approach is a reasonable approximation for models focusing on the chemistry of the inner disk or disk atmosphere, while  molecular inheritance becomes increasingly important towards the cold and dense outer midplane \citep{Visser09,Drozdovskaya16,Eistrup16}. An additional aspect of the initial conditions concern the elemental abundance ratios, which are often assumed to be non-Solar \citep[e.g.,][]{Najita11,Cleeves18}, based on observational evidence  \citep[e.g.,][]{Du15,Miotello19}.  

The complexity of the model chemical reaction kinetic rate network can range from only considering phase changes of major volatile species \citep[e.g.,][]{Piso15} to incorporating large and comprehensive kinetic rate networks of chemical reactions \citep[see][and references therein]{Henning13}. The simpler chemical models have been used to establish qualitative interpretative frameworks, and to couple disk dynamics and time dependent chemistry.  ``Full'' chemical reaction networks always include various forms of two-body chemical reactions, UV, X-ray, and cosmic ray driven chemistry, adsorption and desorption of gas molecules on grain surfaces, and simple surface chemistry such as the formation of H$_2$ \citep[e.g.,][]{McElroy13}. Reaction data are available in databases like UMIST \citep{McElroy13} and KIDA \citep{Wakelam12}, from which appropriate chemical reaction networks can be developed \citep[e.g.,][]{Semenov04,Kamp17}. Additional reactions are then added dependent on the particular science question. Models developed to treat the formation of organic molecules and/or the chemical evolution of icy grain mantles add   surface chemistry reactions \citep[][]{Walsh15,Ruaud19,Furuya22}, often inspired by protostellar chemistry models \citep[][and references therein]{Jorgensen20}.  Models focused on the inner disk often expand their networks to include high temperature gas-phase chemistry, 3-body reactions, and/or warm surface chemistry \citep{Agundez08,Thi20,Anderson21}. Isotope fractionation is included in models where the precise distributions of minor isotopologues is important \citep[e.g.][]{Aikawa99,Lyons05,Willacy09,Miotello14,Visser18}.  

To compare the calculated chemical structures (densities and column densities) with observations, the   disk gas and dust density and temperature structures, the chemical abundance of the emitting species, and molecular line excitation data can be input into a radiative transfer code \citep[e.g.][]{,Brinch10} and then an observational simulator (the final step in Fig. \ref{figure:models}).

\subsection{Modeling Chemistry in a Dynamical Disk Environment}\label{sec:models-dyn}

There is a growing list of observations that are difficult to explain with static models, and a frontier in disk chemistry modeling is the co-modeling of chemistry and dynamical processes that occur on similar timescales to chemical reactions. Simulating the combined effects of all major physical, chemical, and dynamical processes throughout the entire disk is currently computationally too expensive, however, and models therefore need to make choices which physical and chemical processes to include and exclude. There are several models that incorporate some aspects of vertical mixing \citep[e.g.,][]{Semenov11,Furuya14}, radial gas diffusion \citep[e.g.,][]{Aikawa99,Ilgner04,Nomura09,Bosman18,Price20} and dust evolution \citep[e.g.,][]{Vasyunin11,Akimkin13,Krijt18, Booth_R19,Eistrup22,vanClepper22}, but many of these consider a simplified chemistry, i.e. they are not necessarily more `complete'  than the static models with larger chemical networks.  
 
In addition to local mass transport processes, some models consider how global environmental changes affect the chemistry. The effects of the stellar evolution have been explored by solving for chemical abundances while altering the disk environment at specified time-steps \citep{Price20}. The impact of short-lived accretion outbursts from the young star on disk compositions has also been investigated \citep{Rab17,Cleeves17}. The outbursts are found to generate chemical changes that persist beyond the duration of the event \citep{Molyarova18}, but the potential for long-term chemical changes or alteration of planetary compositions by such phenomena has yet to be determined.  

%\subsection{Comparison of Model Outputs and Observations \label{sec:model-obs}}

%There are multiple methods to compare calculated atomic and molecular density and column density structures with astronomical data. Full forward modeling of observables, such as radially resolved molecular emission maps, (the final step in Fig. \ref{figure:models}) feed the disk gas and dust density and temperature structures, the chemical abundance of the emitting species, and molecular line excitation data into a radiative transfer code and then an observational simulator. The radiative transfer step vary in complexity from LTE slab models \citep[e.g.,][]{Salyk11} to non-LTE radiative transfer in 3D \citep[e.g.,][]{Brinch10}. The simulated observations can then be used to prepare observations, or to constrain chemical or  structural properties of a particular disk or disk sample.  There are multiple strategies to achieve these goals, ranging from detailed modeling of individual sources using highly tuned models \citep[e.g.,][]{Cleeves18,Calahan21}, to the generation of large grids of models that can then be compared with observed disks with a range of properties \citep{Miotello16,Woitke19}. For many applications the computational cost of full forward modeling is, however, prohibitive (though see recent initiatives to reduce computational cost using machine learning \citep{Smirnov22}) and several techniques also exist to extract molecular column densities and abundances from disk observations that can then be compared to model predictions (see \ref{sec:retrievals}).

\section{DISK CHEMISTRY OBSERVATIONS \label{sec:observations}}

Disk chemistry is observationally characterized through a range of techniques, which are reviewed in \S\ref{sec:obs-tech}. These observations directly produce molecular line emission fluxes or absorption depths, from which molecular column densities, abundance structures, or higher level constraints on the disk chemistry and its environment can be retrieved (\S\ref{sec:retrievals}). In \S\ref{sec:obs-inventory} we present the molecules detected to date, and in \S\ref{sec:obs-demo} provide an overview of how molecular inventories, column densities and abundances vary between disks. Finally, \S\ref{sec:sub-structures} and \S\ref{sec:vert_gradients} review observations of radial and vertical structures in disks.

\subsection{Observational Techniques \label{sec:obs-tech}}

An observational chemical characterization of protoplanetary disks can be achieved by spectroscopic studies at a wide range of frequencies and energy scales probing electronic, vibrational and rotational transitions of atoms, molecules and ions. Each wavelength and technique probes a unique aspect of the disk chemistry, and a comprehensive disk chemical characterization requires observations across the electromagnetic spectrum.

Starting at the high-energy end of the spectrum, X-ray and UV transitions are used to probe the elemental abundances of gas and dust that is being accreted onto the stellar surface, which gives access to the composition of inner disk refractories. \citep[e.g.,][]{Drake2005,Gunther2018,Ardila2013,Kama16}. UV observations of fluorescent H$_2$ and CO transitions are also used to probe hot gas in the innermost disk region including its C/O/H ratio \citep{France2012,Arulanatham21}. Optical spectroscopy can be used to constrain the the elemental composition of the innermost disk regions \citep[e.g.,][]{Facchini16}, and to characterize the compositions and dynamics of disk winds from the upper layers of the inner disk regions \citep[][and references therein]{Pascucci22}.

The NIR-MIR regime enables observations of ro-vibrational and highly excited rotational lines of simple molecules in the upper disk layers of inner few au of disks, where temperatures and densities are high enough to collisionally excite these transitions \citep[][and references therein]{Pontoppidan2014}. From space, the Spitzer mission was instrumental in surveying a large number of disks in MIR emission lines \citep[e.g.][]{Carr08,Salyk11}, but at fairly low spectral resolution ($\lambda/\Delta \lambda<700$). These observations have been complemented with high resolution spectroscopy from the ground, which avoids line blending, and enables spectro-tomographic characterizations of the radial origin of the emission lines \citep[e.g.][]{Brittain07,Pontoppidan10,Najita18,Salyk19,Banzatti22}. Line absorption observations are possible in disks with favorable inclinations and are used to provide complementary constraints on the gas inventory \citep[e.g.][]{Gibb07,Najita21}. 
The NIR-MIR regime also enables observations of PAHs \citep[][and references therein]{Tielens08} and disk solid state features, including ice compositions through 1) absorption spectroscopy against an IR bright source, and thus usually performed on edge-on disks \citep{Thi02,Pontoppidan05,Aikawa12,terada17}, and 2) scattered light observations of water ice across the 3\,$\mu$m range \citep{Honda09}. Observations at these wavelengths are set to be transformed by the James Webb Space Telescope (JWST).

In the Far-Infrared (FIR), most of the spectroscopic surveys have been performed with the {\it Herschel} mission, due to the prohibitive atmospheric transmission at these wavelengths. FIR wavelengths provide access to ground-state transitions of small hydrides, including HD and H$_2$O \citep{Bergin13,Salinas16,vanDishoeck21}, [OI], [CI] and [CII], and highly excited CO rotational transitions \citep{Bruderer12,Meeus13,vanderWiel14,Fedele16}. The FIR can also be used to probe water ice thermal emission features \citep{vandenancker00,Min16}. 

Finally, sub-millimeter and millimeter observations enable the spectroscopic characterization of cold gas, which includes most of the disk gas reservoir. Many small and abundant molecules present rotational transitions at these wavelengths, with typical upper energy levels of 5-500\,K. Dependent on line and dust optical depths, these observations can probe all the way to the disk midplane, but more often access the outer disk upper layers. Millimeter lines were originally surveyed in disks using single dish telescopes  \citep{ Dutrey96,Thi04,Guilloteau13,Guilloteau16}. The development of (sub)millimeter interferometers enabled the first spatially resolved disk chemistry studies at high spectral resolution ($\lambda/\Delta \lambda<10^7$) \citep[e.g.][]{Sargent87,Qi03,Dutrey07}. In 2011--2014 (sub)millimeter observations of disks were transformed by the arrival of ALMA, whose collecting area and long baselines enables the detections of rarer molecules and the study of disk chemistry at higher spatial resolution studies on scales $<$10~au \citep{Oberg15Natur,Huang18}. Disk chemistry has so far not been accessible at longer wavelengths, but a future more sensitive radio array may provide access to NH$_3$ and large organic molecules in the disk midplane.

\subsection{Retrieval of Molecular Column Densities and Abundances \label{sec:retrievals}}

 Molecular line emission strengths depend on molecular column densities and excitation, and on whether both are homogeneous within the beam. Deriving molecular column densities, abundances and/or information about the disk structure from molecular line observations is therefore not trivial, and full forward modeling (Fig. \ref{figure:models}), is the only way to fully account for these complexities when retrieving disk chemical properties. Forward modeling has been used to derive disk chemical properties of individual sources using highly tuned models \citep[e.g.,][]{Cleeves18,Calahan21}, as well as the generation of large grids of models that can then be compared with observed disks with a range of properties \citep{Miotello16,Woitke19}. For many applications the computational cost of full forward modeling is, however, prohibitive (though see recent initiatives to reduce computational cost using machine learning \citep{Smirnov22}), or unpractical due to poor constraints on the disk structure. In these cases useful retrievals can still be made under some simplifying assumptions. 

The simplest retrievals assume that the molecule excitation is well described by local thermodynamic equilibrium (LTE), optically thin lines, optically thin continuum, and a simple emission geometry that results in the emission either completely filling the beam or filling some fraction of the beam in a well-defined way. In this scenario the molecular column density can be calculated either by adopting an excitation temperature, or through rotation diagram analysis using multiple lines \citep[e.g.][]{Goldsmith99,Najita03}. The calculations can be done using disk averaged fluxes to derive a disk averaged column density, or spatially resolved fluxes to derive a column density profile. A modified method can be used when the lines are marginally optically thick \citep{Najita03,Loomis18a}. Column densities can also be derived by fitting disk  spectra assuming LTE and adopting some intrinsic line width, including the fitting of hyperfine lines in high resolution spectra \citep{Kastner14,Hily-Blant19,Bergner19,Teague20}. For many lines detected in the (sub-)mm regime, this LTE retrieval approach is valid as long as molecular column densities vary slowly with radius and the lines originate below the disk atmosphere where the gas density exceeds the line critical density. 

In cases where LTE cannot be assumed, a limited forward modeling approach is needed to predict line excitation using the local disk density and temperature, and molecular collisional excitation coefficients. Cases where lines are likely in non-LTE include millimeter observations of the disk atmosphere, and ro-vibrational lines. A common  approach in these cases is to use a LVG (Large Velocity Gradient) approximation \citep{Pietu07,vanderTak07} or Monte Carlo-based radiative transfer codes \citep{Hogerheijde00,Brinch10} coupled with an observation simulator to extract molecular abundances. The disk abundance structure can be parametric \citep[e.g.][]{Qi11} or the product of a disk chemistry code \citep{Bruderer12,Du15,Cleeves18}, but in either case the retrieval is achieved by quantifying the goodness of fit of the different abundance structures to the observations. For IR lines, fast line ray-tracers can also be coupled to thermo-chemical codes to predict the rich line emission from disk regions with large velocity gradients \citep{Pontoppidan09,Bosman17,Woitke18}. In addition to molecular abundances, these kind of retrieval methods are also used to constrain disk environmental parameters such as gas mass, thermal structure, CO depletion, elemental ratios and ionization \citep{Pietu07,Kamp10,Dutrey14,Cleeves15,Anderson19,Miotello19,Woitke19}.

Finally, we note, that in all these cases, retrievals rely on detailed spectroscopic data for the observed molecular lines, which are provided in databases such as HITRAN \citep{Gordon22}, the Cologne Database for Molecular Spectroscopy \citep[CDMS;][]{Muller01,Muller05,Endres16}, JPL spectral line catalog \citep{Pickett98}, and the Leiden Atomic and Molecular Database \citep[LAMDA][]{Schoier05}. These in their turn rely on laboratory experiments and computations.

\subsection{Disk Molecular Inventories \label{sec:obs-inventory}}

\begin{figure}[h]
\includegraphics[width=5in]{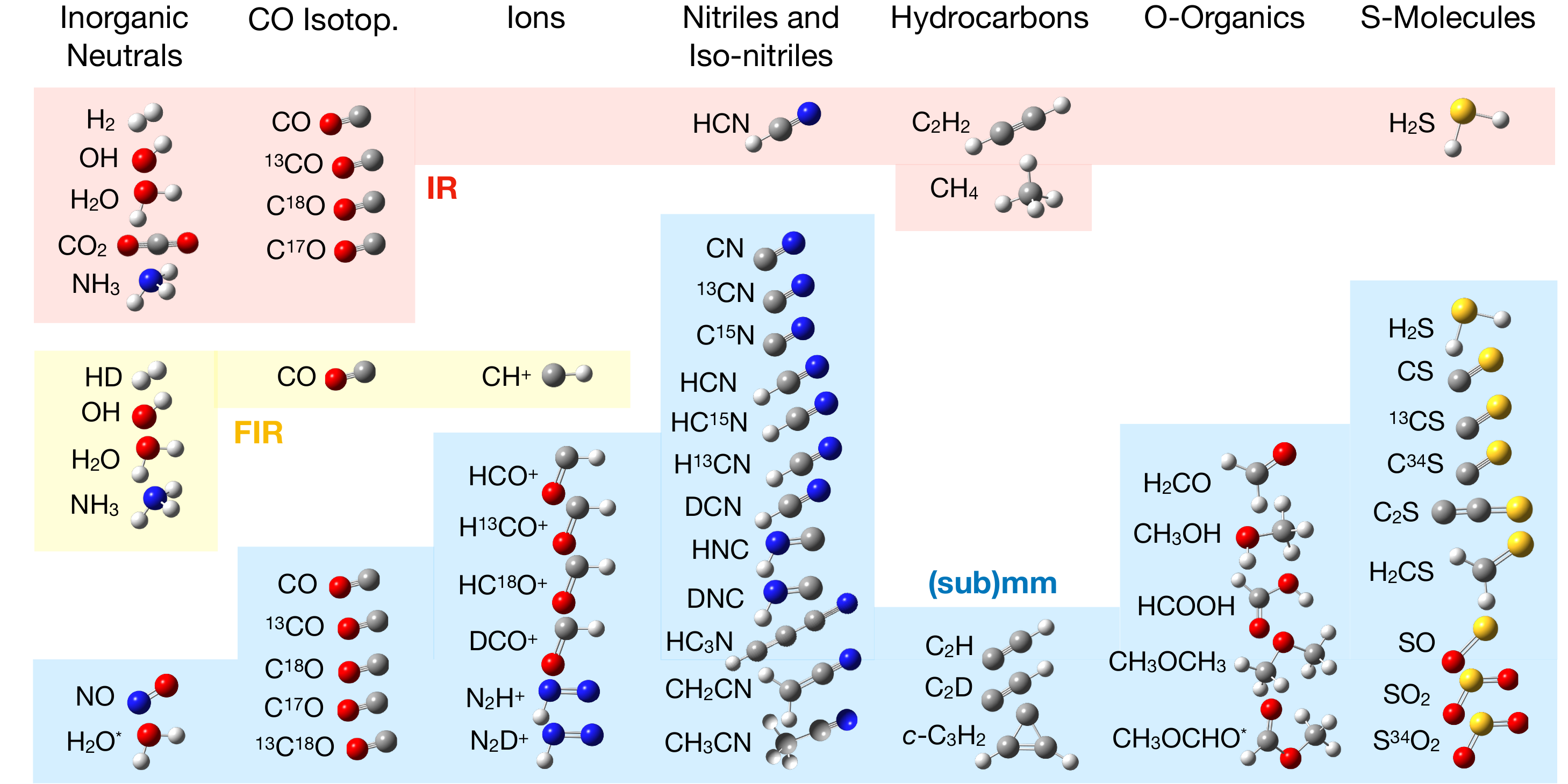}
\caption{Illustration of the molecules detected so far in disks at IR (red), FIR (yellow), and (sub)mm (blue) wavelengths. An `*' indicates a tentative detection.}
\label{fig-invent}
\end{figure}

There are 31 detected molecules in disks, not counting isotopologues, most (26) of which are presented in \citet[][and references therein]{McGuire22} with recent additions from \citet{Booth21-so}, \citet{Canta21}, \citet{Phuong21} and \citet{Brunken22}. The majority of these are detected at millimeter wavelengths (24), followed by IR (10), and far-IR (6). Figure \ref{fig-invent} shows these molecules and their observed isotopologues (also presented in \citet{McGuire22} except for $^{13}$CO$_2$ \citep{Grant22} and HC$^{18}$O$^+$ \citep{Furuya22}), 
 organized into seven chemical `families': inorganic neutrals (excluding CO and S-molecules), CO isotopologues, molecular ions, nitriles and iso-nitriles, hydrocarbons, O-containing organics, and S-molecules. %Of these species, only CO is detected with all three techniques and most molecular are only observed with one, demonstrating the complementarity of these different observation modes. %Compared to the number of molecules detected in the interstellar medium or around protostars, this is a modest number of molecules (27 not counting isotopologues). Still the number is large enough to provide many windows on the chemical evolution of disks, the abundance of gas-phase elements, and the physical processes that impact the chemistry. Figuring out how to unlock the combined potential of this collection of molecules to characterize disks must be seen as one the prime objectives going forward.

IR observations have resulted in large disk samples with detections of gas-phase inorganic neutrals (H$_2$, OH, H$_2$O, CO, $^{13}$CO, C$^{18}$O, C$^{17}$O, NH$_3$, and CO$_2$), but also of the small organics HCN and C$_2$H$_2$, and of H$_2$S \citep[e.g.][]{Lahuis07,Carr11,Mandell12,Najita21}, which together can be used to constrain the inner disk chemistry. In addition to narrow gas emission and absorption lines, IR spectra towards some disks also contain features from polycyclic aromatic hydrocarbons (PAHs) \citep{Acke04,Geers06}, and of ice-phase H$_2$O, CO$_2$ and CO \citep{Pontoppidan05,Aikawa12}.  We expect both the gas and ice IR inventory to grow rapidly in the coming years through JWST observations, and already early JWST science has presented one new inner disk molecule detection: $^{13}$CO$_2$ \citep{Grant22}.

At Far-IR wavelengths the only molecule detected towards more than a handful objects is CO \citep{Meeus13}. The few HD, OH, H$_2$O and NH$_3$ detections that do exist do provide unique constraints on disk physics and chemistry, however, and have been the subjects of a large number of studies \citep[e.g.][]{Fedele13,Bergin13, Riviere15,Salinas16, vanDishoeck21}. The importance of this group of hydrides is that they currently present the only constraints we have on cold water and ammonia in disks, as well as provide a unique handle on the disk gas mass (HD).

Millimeter and submillimeter disk observations have resulted in a relatively large sample of CO isotopologue, HCO$^+$, HCN and C$_2$H detections, and $\gtrsim$10 detections of N$_2$H$^+$, DCN, CN, H$_2$CO and CS (see \ref{sec:obs-demo}. The remaining 28 molecules and isotopologues have been either rarely targeted (e.g. HNC), or rarely detected. One aspect in Fig \ref{fig-invent} worth noting is the prominence of the nitrile and iso-nitrile family, especially compared to O-bearing organics. This can be compared to the IR observations, where the detection rate of O-bearing molecules, nitriles and hydrocarbons appear more balanced, indicative of very different chemical environments in inner and outer disk regions.

\subsection{Molecular Demographics \label{sec:obs-demo}}

Molecular line detection rates and intensities across disk samples provide a first measure of the diversity of chemical environments during planet formation, even though any interpretation is complicated by possible differences in line excitation.
 At far-IR wavelengths the detection rate has been too small to establish clear demographic patterns, though e.g. the low detection rate of water in T Tauri disks is in itself informative. This section therefore focuses on IR and (sub)mm constraints. 

The inner disk molecular demographics probed by IR observations was reviewed by \citet{Pontoppidan2014} and have not been substantially revised since. Lines from water and small organics are common (30--50\% detection rates with Spitzer) in disks around Solar-type stars, and we speculate this will increase to close to 100\% with JWST. The detection rate is much lower for more massive stars, and somewhat lower for cooler stars. In the latter case there is evidence for a different disk chemistry based on distinct HCN/C$_2$H$_2$ ratios compared to more luminous stars \citep{Pascucci09,Pascucci13}. There is also  evidence of a lower detection rate towards so-called transition disks, disks with a large central cavity, though the sample is small. These differences in line detections and intensities across different disk families may be due to  different inner disk chemistry because of different levels of stellar irradiation, differences in line excitation, or different levels of influx of icy pebbles, or a combination of all three  \citep[e.g.][]{Salyk11,Pascucci13,Antonellini16, Najita18,Banzatti20}, but some of the observed differences may also be due to  molecular excitation. Distinguishing between these different explanations is currently complicated by the low spectral resolution and limited sensitivity of Spitzer, and by small observing windows of higher resolution ground-based observations. Medium-resolution observations with JWST should directly address these difficulties and a new demographic understanding of inner disk chemistry is likely forthcoming.

\begin{figure}[h]
\includegraphics[width=4in]{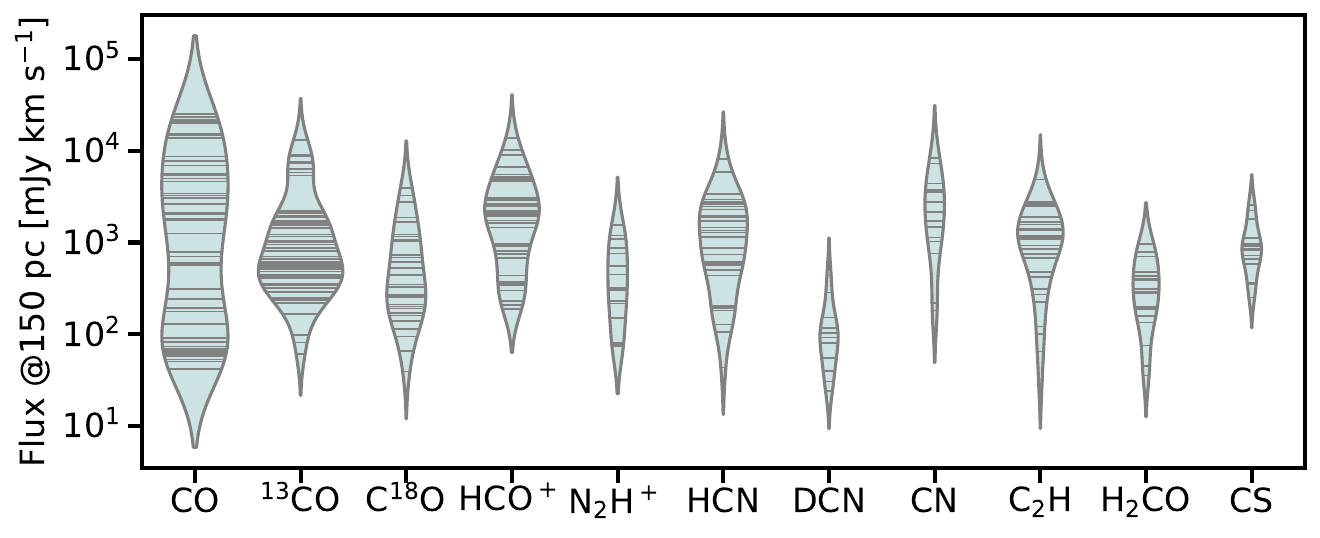}
\includegraphics[width=4in]{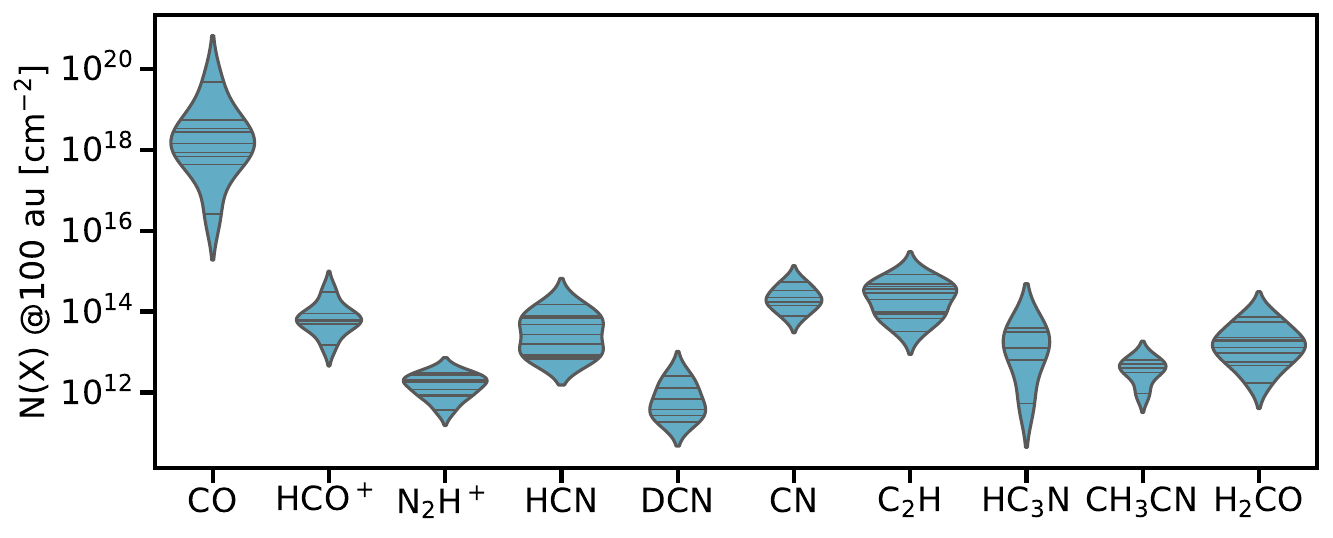}
\caption{{\it Upper panel:} Disk integrated fluxes from 1~mm transitions, corresponding to ALMA B6, normalized to a distance of 150 pc. {\it Lower panel:} Outer disk column densities extracted from spatially resolved ALMA observations. In these violin plots the width of the distribution denotes the number of disks that fall within a certain flux or column density range, and each horizontal line signifies an individual measurement.}
\label{fig-demo}
\end{figure}

Millimeter line surveys of disks are not yet mature enough to enable determinations of detection rates and how they vary across populations. Instead Fig. \ref{fig-demo} (upper panel) shows the distributions of reported disk integrated fluxes normalized to 150 pc for a sub-set of commonly observed molecules using their brightest 1~mm transition (e.g. CO 2--1)\footnote{When a different line flux was instead reported, e.g. CO 1--0 or 3--2, the 1~mm line flux was estimated assuming optically thick emission for `bright' lines and optically thin emission for `weak' lines, and an excitation temperature of 20~K. }.  These observations were taken from a large number of publications including \citet{Oberg21-maps,Bergner20,Pegues21,Miotello19,Ansdell16,Guilloteau16,Anderson22,Long17,Barenfeld16,Huang17,Oberg10c,Oberg11a}. When interpreting these distributions it is important to note that the line flux floor for each species is set by the survey sensitivity and hence not very informative.
The maximum fluxes for each species are informative, however, and show that the CO line is about an order of magnitude brighter than $^{13}$CO, HCO$^+$, HCN and CN, followed by C$^{18}$O, C$_2$H and CS. N$_2$H$^+$, H$_2$CO and DCN present the weakest emission among these commonly observed molecules. Many of the flux distributions are bimodal due to a focus in the literature on either small samples of large and bright disks \citep[e.g.][]{Dutrey07,Bergner20}, or on larger samples that are dominated by the more common smaller-sized and less bright disks \citep{Ansdell16,vanTerwisga19}. The range of observed line fluxes for a single species must therefore be interpreted with caution, but it is still interesting to note that it frequently spans two orders of magnitude, indicative of substantially different molecular reservoirs among planet-forming disks. 

A similar source-to-source variation is seen among the small number of disks that have well constrained column density radial profiles in multiple species. Fig. \ref{fig-demo} (lower panel) uses data from \citet{Zhang18,Oberg21_TWHya,Terwisscha21,Qi13c,Zhang21, Cataldi21, Qi19,Guzman21, Bergner21,Qi11,Phuong21,Zhang19,Bergner19,Facchini21} to show the range in column densities at $\sim$100~au\footnote{For two disks where 100~au data was not available we instead use reported column densities at 70 and 240 au}. Even within this small and highly biased sample there are orders of magnitude differences in column densities. Possible causes of this chemical diversity include differences in temperature structures, levels of CO freeze-out, high-energy radiation fluxes, disk ages, and initial chemical conditions across the disk sample. To obtain more detailed demographics, we need surveys that address the current sample biases and have the sensitivity to detect a range of molecular abundances.

\subsection{Molecular Sub-Structures}\label{sec:sub-structures}

\begin{figure}[h]
\includegraphics[width=4.5in]{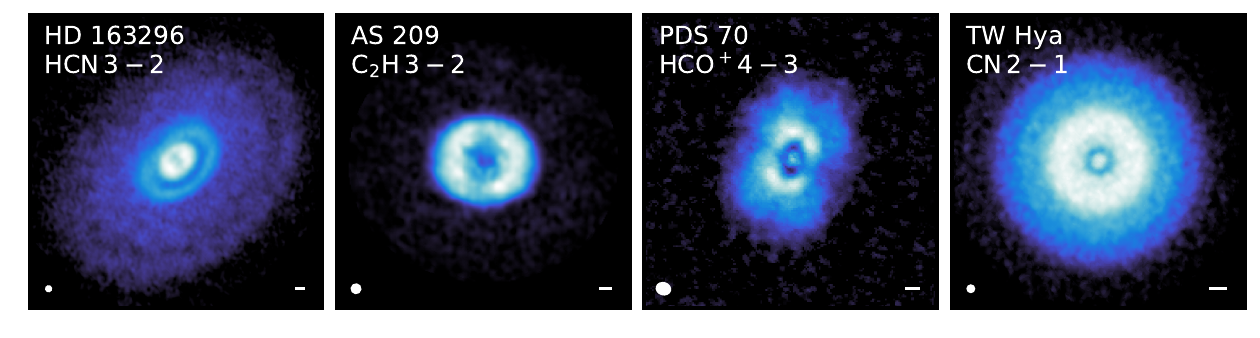}
\caption{Examples of chemical sub-structures towards disks showing the four rings in HCN emission towards HD 163296 \citep{Oberg21-maps}, a single broad ring in C$_2$H emission towards AS 209 \citep{Oberg21-maps}, a ring and central peak in HCO$^+$ emission towards PDS 70 \citep{Facchini21}, and a ring, a plateau and a central peak in CN emission towards TW Hya \citep{Nomura21}. The scale bar in the lower right corner signifies 20 au, and the oval in the lower left corner the synthesized beam.}
\label{fig-substruc}
\end{figure}

Within both the inner and outer disk regions we expect chemical sub-structure due to gradients in radiation flux, ionization and temperature, as well as dust and gas surface density gaps and rings (\S\ref{sec:context}). Analogous to dust sub-structures, which can be identified both through resolved images and through modeling of spectral energy distributions \citep{Andrews20}, molecular sub-structures can be observed through chemical imaging (Fig. \ref{fig-substruc}), and line excitation analysis. Furthermore, in a Keplerian disk, the spectral line shape can be used for spectroastrometry, and to map different spectral line regions to different disk radii, which provides additional tools to characterize chemical disk structures at smaller scales than are accessible with imaging alone. 

Spectroastrometry, line excitation modeling, and spectral profile analysis of spectrally resolved lines are been especially important to access sub-structure in the inner disk
 \citep[e.g.][]{Pontoppidan08-spec, Najita10,Thi14,Fedele13}. Three examples of how line excitation and line spectral profiles can reveal chemical sub-structure are 1) the finding of \citet{Salyk11} that  
OH, C$_2$H$_2$, HCN, CO$_2$, and H$_2$O have different excitation temperatures, suggestive of a radially progressive inner disk composition, 2) an observed trend between CO emitting radii and the stellar luminosity based on the CO spectral profiles \citep{Pontoppidan11,Salyk11b}, and 3) the discovery of a  radially varying H$_2$O/CO ratio based on a combination of excitation and spectral profile analysis \citep{Banzatti22}. Spectral line profile analysis has also been applied to millimeter line observations to map out the molecular emission structure in disks \citep[e.g.][]{Dutrey08,Rosenfeld12-kinematics,Bosman21_XV}, which has revealed gas and/or CO chemistry sub structure on scales of a few au, spanning the radial gap between typical IR and millimeter sub-structure constraints.

Most millimeter wavelength evidence for chemical sub-structure comes from spatially resolved observations, however, using millimeter interferometers.
Early examples include the discoveries of a H$_2$CO ring towards DM Tau \citep{Aikawa03}, a DCO$^+$ ring towards TW Hya \citep{Qi08} and C$_2$H rings towards several disks \citep{Henning10}. Higher resolution images emerged with the arrival of ALMA and currently $\sim$10 disks have been chemically characterized at scales of 0".1--0".2. Among these disks, chemical sub-structure is ubiquitous. The ALMA Large Program MAPS alone, which surveyed five disks, identified $\sim$250 rings, gaps, and shoulders \citep{Law21_radprof}. Figure \ref{fig-substruc} illustrates the diversity of chemical sub-structures that has been observed, ranging from single rings to four-ringed systems, and ring and gap widths from unresolved ($<$20 au), to 100s of au. 

\begin{figure}[h]
\includegraphics[width=4.2in]{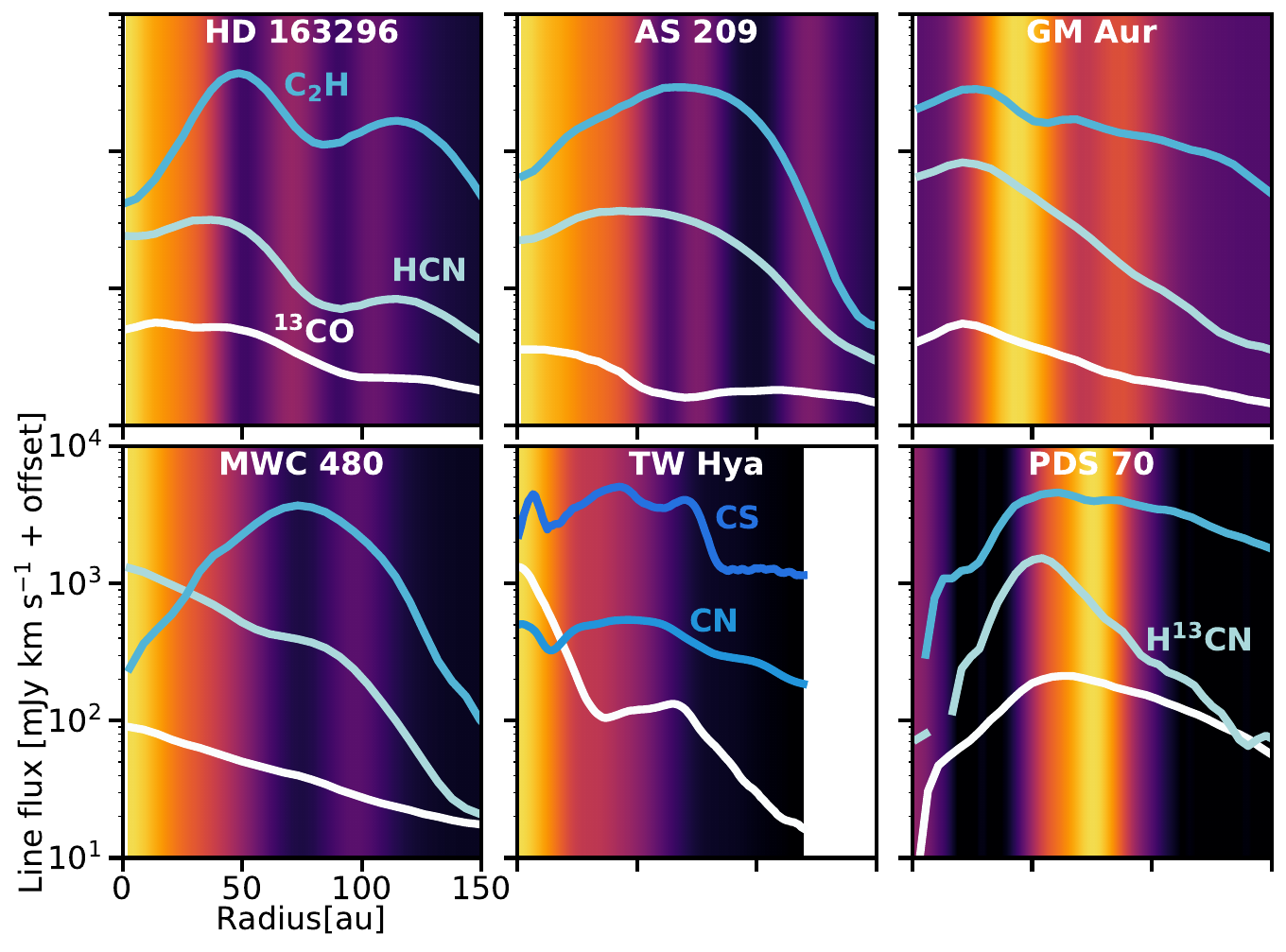}
\caption{Comparisons of dust sub-structure (background) and molecular emission radial profiles (lines) towards six disks \citep{Law21_radprof,Nomura21,Facchini21}. The beam size of the dust and lines have been homogenized and is between 10 and 25 au towards the different disks.}
\label{fig-profiles}
\end{figure}

There are multiple possible origins of the observed inner and outer disk chemical disk structures. The (gas-phase) formation and destruction of some species is temperature sensitive and this should result in peak abundances at certain disk radii. This may explain some of the inner disk chemical differentiation as well as broader chemical rings in the outer disk. Similarly, molecules that depend on photochemistry for formation or destruction should vary across the disk due a decreasing UV flux with disk radius and this may explain other broad chemical structures  \citep[e.g.][]{Cazzoletti18,Bergner21}. A third potential cause of sub-structure is snowlines (see \ref{sec:snow_lines}), though they are excluded as explanations for a majority of the currently observed chemical sub-structures \citep{Long18,Huang18b,Law21_radprof}.  Finally, gaps in dust and gas may produce chemical sub-structure due to enhanced photochemistry in dust gaps \citep{Bosman21_VII},  different thermal structures in dust gaps and rings \citep{Facchini18,Alarcon20a}, decreased grain surface chemistry in dust gaps, and lower molecular gas column densities in (H$_2$) gas gaps \citep{Teague17}.  Observational tests of these proposed relationships are somewhat ambiguous. There is no one-to-one correspondence between dust and chemical structures \citep{Jiang22}, but as shown in Fig. \ref{fig-profiles} there are some coincidences between dust and chemical sub-structures, suggestive of that specific dust gap and ring properties may be needed to shape the chemical structures of disks. Conversely, it may be possible to use molecular emission sub-structure as probes of disk gap and ring properties, including whether a gap is formed by a planet \citep{Bergner19}. In addition, these observations suggest that the local chemical environment within which a planet assembles could be quite distinct from the chemistry at nearby disk radii.

\subsection{Vertical Chemical Gradients and Sub-structure}
\label{sec:vert_gradients}

As introduced in \S\ref{sec:intro}, the disk chemistry is vertically highly structured.  In the disk atmosphere, molecular abundances are regulated by photodissociation, which results in molecule-specific and, for CO and N$_2$ isotopologue specific, photodissociation fronts \citep[e.g.][]{Visser09,Miotello14}. Towards the disk midplane, molecular abundances are bounded by vertical condensation fronts or snow surfaces,  though a combination of turbulence and non-thermal desorption may maintain a small amount of volatiles also in the coldest part of the disk midplane \citep[e.g.][]{Hersant09}. As a result most molecules should be present in distinct vertical layers,  which together constitute the disk `molecular layer' \citep{Aikawa02}. Characterizing the resulting vertical chemical stratification is key to benchmarking disk chemistry models and constraining the overall disk chemistry evolution. % The vertical dimension is, however, more difficult to probe than the radial dimension due to smaller spatial scales; many molecules emit from disk heights $z/r<$0.2. Most present constraints are therefore either indirect, or involve molecules with very bright emission, which can be observed at high spatial resolution. 

\begin{figure}[h]
\includegraphics[width=0.7\textwidth]{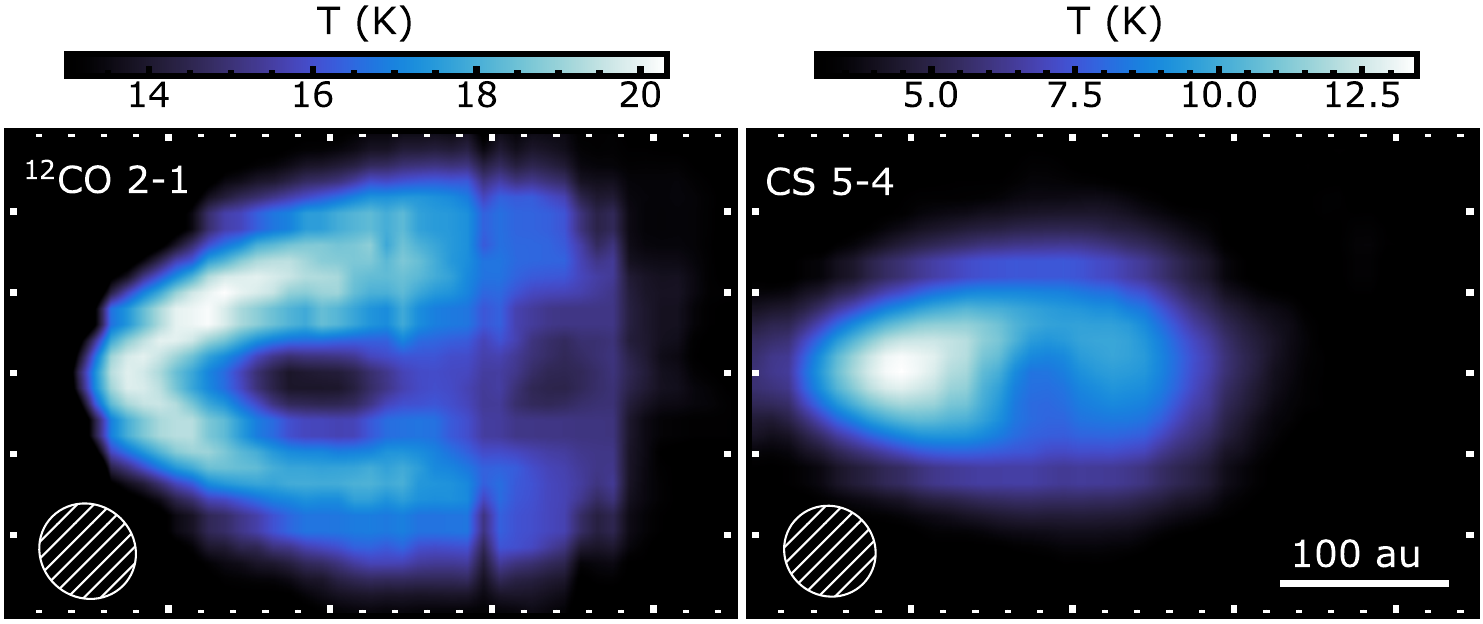}
\caption{Tomographically reconstructed distribution of $^{12}$CO and CS in the Flyng Saucer \citep[data from][]{Dutrey17}.}
\label{fig-vertical_grad}
\end{figure}

At (sub)millimeter wavelengths, interferometers provide high enough angular resolution to spatially resolve vertical gradients in bright molecular line emission, and by exploiting the position and velocity information, it is possible to reconstruct the full 3D tomography of the line intensity \citep{Dutrey17,Teague20c}. Figure \ref{fig-vertical_grad} shows a text-book case of CO and CS emitting from different layers in the edge-on disk the Flying Saucer \citep{Dutrey17}, where the lack of CO emission in the disk midplane is explained by CO freeze-out. The emission of CS 5-4 originates from a $z/r\sim0.1$ layer, close to the disk midplane, demonstrating that the molecular layer can extend deep into the disk. Emission surfaces can also be extracted for disks at other inclinations  by exploiting the Keplerian rotation pattern and disk geometry \citep{Pinte18b}. This methodology requires high angular resolution and signal-to-noise and has therefore only been applied to particularly bright, optically thick molecular lines. As a consequence, disk (frontside) emission heights exist for $^{12}$CO towards a fairly large sample, for rare CO isotopologues towards a handful of sources, and for other molecules towards even fewer disks \citep{Huang20,Law21_surf,Law22,Paneque22}. In these sources CO (isotopologue) emission becomes optically thick at a range of $z/r$, indicating that the extent of the warm molecular layer and the CO abundance vary substantially between disks. The molecular emission height derived from the disk backside should trace the snow surface, but this has only been achieved for CO towards three disks so far, though there is active development of this technique \citep{Pinte18,Casassus21,Izquierdo22}.  In the few disks for which constraints exist, CN is present  in the disk atmosphere, in agreement with a main formation route of CN being UV-pumped H$_2$ interacting with atomic N \citep{Cazzoletti18}, while other small organics, especially HCN, resides closer to the planet-forming midplane. 

For cases where direct imaging of vertical chemical structures are impractical or impossible, which includes inner disk line observations as well as weaker (sub)millimeter lines emitting in the outer disk, line excitation analysis offers an alternative approach to characterize the disk vertical chemical structure as long as the disk temperature structure is constrained. The approach has so far not been extensively used in the inner disk due to large uncertainties in the vertical temperature gradient, but this may soon change as medium-resolution JWST disk spectra analysis techniques are developed. The chemistry of the warm and highly irradiated disk atmosphere of the outer disk has been probed by  high-$J$ CO lines, and the atomic oxygen line at $63\,\mu$m in the far-IR \citep{Bruderer12,Kamp13}. The excitation temperatures of other small and mid-sized molecules have been retrieved from medium and high spatial resolution data, and at 15--40~K they are  consistent with models of the disk midplane and molecular layer  \citep{Loomis18b, Bergner19,Pegues21,Guzman21,Ilee21,Cataldi21,Facchini21}. Interestingly the excitation temperatures of several of the more complex organic molecules, including CH$_3$CN,  place them close to the midplane at $z/r\sim$0--0.2 \citep{Loomis18b,Ilee21}, which suggest they may contribute to the organic budget of forming planets. Interpretation of such data is complicated, however, by the possibility of non-LTE excitation and e.g. for CN there is a range of derived excitation temperatures and inferred emission heights \citep{Chapillon12,Hily-Blant17,Teague20c}.

\section{THE DISTRIBUTIONS AND CHEMISTRY OF DISK VOLATILES}

The distribution of volatiles in disks, including volatile organics, provides the initial conditions for planet volatile compositions and chemistry. The perhaps most important concept to predict these distributions is condensation fronts or snowlines, which is introduced in \S\ref{sec:snow_lines}.  \S\ref{sec:ocnsp} reviews the existing observational evidence for how the major volatile elements OCNSP are distributed across the disk.  \S\ref{sec:organics} focuses in on the volatile organics in disks, and the importance of inheritance and {\it in situ} UV chemistry for disk organic reservoirs. An important tool to trace the origins and evolution of volatiles in disks are stable isotope ratios, and \S\ref{sec:isotope} presents observed disk isotopologue ratios and the associated constraints on isotope fractionation chemistry.

\subsection{Snowlines}
\label{sec:snow_lines}

Volatile condensation fronts were briefly introduced in \S\ref{sec:sub-structures} and \S\ref{sec:vert_gradients}, as causes of chemical substructure and vertical chemical gradients. In this section we focus on the radial dimension i.e. on snowlines, and review the theory and observations that have been developed to predict and constrain snowline locations. These locations matter for several reasons. First, freeze-out of major volatiles changes both the gas and grain surface chemical trajectories. For example freeze-out of CO both enables the formation of O-rich organics in the ice, and may promote a C-rich gas-phase organic chemistry \citep[e.g.][]{Walsh15,Schwarz18}. Second, in planet formation models, snowlines of major carriers of O, C and N  determine the volatile compositions of primary atmospheres and planetary envelopes, and have emerged as a major explanatory framework for observed exoplanet atmospheric compositions \citep{Oberg11c,Cridland20}. Third, snowlines, especially the water snowline, are also expected to affect dust grain evolution  \citep[e.g.][]{Gundlach15,Okuzumi19}, and may regulate when and where planet cores assemble, and hence formation locations of terrestrial planets vs Neptune-like planets and Gas Giants \citep{Drazkowska22}. 

\begin{figure}[h]
\includegraphics[width=5in]{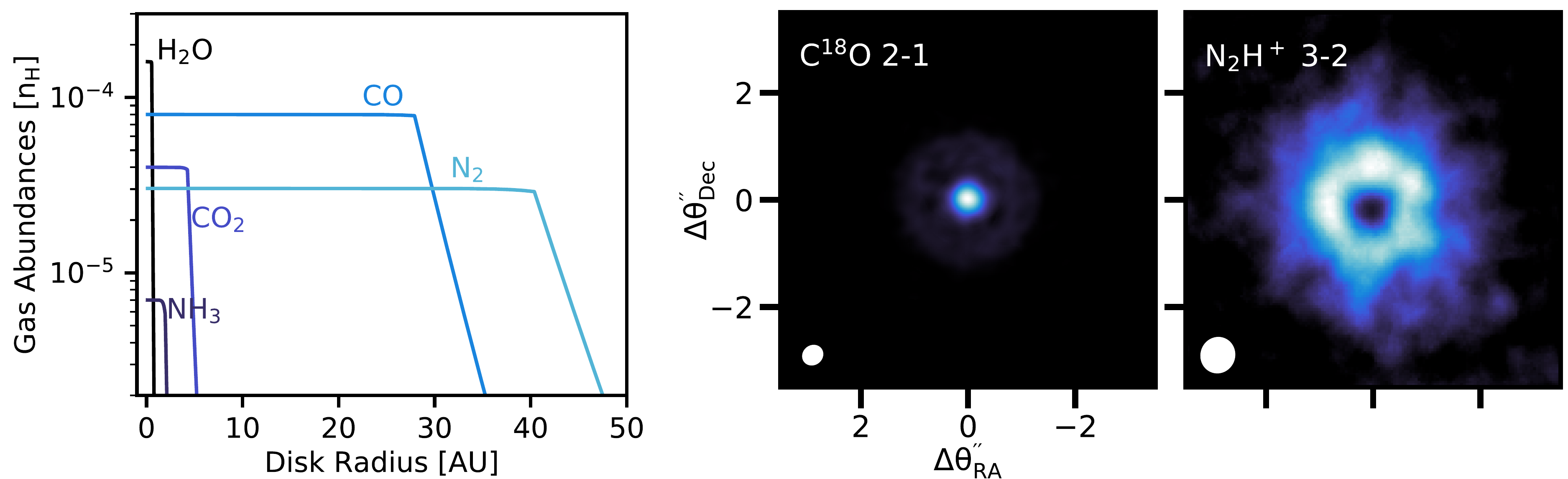}
\caption{{\it Left panel:} Expected gas-phase abundances of five major CNO carriers in the disk midplane assuming a simple T Tauri disk temperature model (see text) and inheritance of interstellar volatiles. The sharp abundance overturns mark the respective snowlines. {\it Right panels:} Observations of C$^{18}$O and N$_2$H$^+$ with ALMA towards TW Hya. The outer edge of C$^{18}$O and the inner edge of the N$_2$H$^+$ should both trace the CO snowline \citep{Qi13c,Schwarz16}}.
\label{fig-snowline}
\end{figure}

Figure \ref{fig-snowline} (left panel) illustrates the five major snowline locations in a generic disk model for a solar-type star using the disk framework, binding energies and abundances from \citet{Oberg19} and \citet{Oberg21_Review}, except that we adopt a midplane temperature profile with $T_{\rm 1au}=150$K and a power of -0.47. This simple parametric model provides a useful starting point for discussing snowline locations, but it is important to keep in mind that it presumes a specific temperature profile that is not generally applicable, and that in reality snowline models are complicated by a range of dynamical and chemical effects. To begin with, the disk temperature structure  changes over time, sometimes abruptly due an accretion burst. Such accretion bursts would move snowlines outwards, while stellar evolution results in an inward migration of snowline locations over time \citep{Ciesla06,Price21,Cieza16,Banzatti15}. Furthermore, inward drift of pebbles often occurs on timescales similar to desorption, which can readily move the effective snowline a factor of two inwards \citep{Piso16}. Snowline locations also depend on the chemical compositions of icy grains, since volatile binding energies depend on the grain surface material \citep[e.g.][]{Collings03,Fayolle16,Kamp17,Potapov18}. In addition H$_2$O and CO$_2$ ices can entrap hypervolatiles effectively producing two snowlines, one at the expected location and one at the location where the ice matrix desorbs or crystallizes \citep{Bar-Nun85,Lunine85,Collings04,Simon19}. The impact of snowlines on gas and solid abundances is also more complex than is indicated in Fig. \ref{fig-snowline}, which only considers a balance between desorption and adsorption. If the disk is somewhat turbulent,  gas diffusion should result in a depletion of vapor inside of the snowline, and build-up of ice outside of the snowlines \citep[e.g.][]{Cuzzi04,Ros13,Krijt16}. Finally, the inherited volatiles may chemically evolve over the disk lifetime \citep{Schwarz18,Eistrup16}, which will change the abundances of volatiles and therefore the relative importance of different snowlines. 

Snowline locations have been observationally constrained through at least six different techniques: 1) images of major volatiles with important snowlines, 2) images of gas-phase chemical tracers (Fig. \ref{fig-snowline}), 3) tomography using spectrally resolved lines \citep{Carr18,Salyk19}, 4) line excitation studies \citep{Zhang13,Blevins16}, 5) changes in dust spectroscopic properties at a snowline \citep{Cieza16}, and 6) association of disk dust sub-structures with snowlines \citep{Zhang15}. Each of these techniques has advantages and disadvantages. Direct imaging of snowlines is currently only possible for CO using rare isotopologues, and even for CO it is often unclear if the observed drop-off in CO vapor corresponds to the CO snowline, a decrease in C and O elemental abundances, or an overall decrease in gas surface density. There are plausible gas-phase chemical probes of H$_2$O, CO and N$_2$ snowlines that can be used to extract snowline locations  \citep{Qi13c,Bjerkeli16,Qi19,Leemker21}, but these probes are not always unique to depletion of the targeted volatile at its snowline and therefore require careful interpretation \citep{vantHoff17}. Spectral tomography (the estimation of the spacial distribution from spectral information taking advantage of Keplerian rotation) of snowlines is limited by the possible presence of non-Keplerian motion. Furthermore the water IR spectral lines that are accessible from the ground, and hence at high spectral resolution, are not well matched to snowline excitation conditions. Snowline constraints from line excitation depend on the assumed disk temperature structure, which is often uncertain. The fifth approach, observing changes in dust spectroscopic properties across snowlines depend on models of e.g. increased fragmentation at major snowlines, and are hence quite indirect. Finally, a proposed association between disk dust sub-structure and snowlines has not yet been demonstrated and this approach hence remains highly speculative \citep{Long18,Huang18b}. Given the inherent uncertainties in all these techniques, a snowline location should ideally be observed using at least two different methods to be considered secure.

\subsection{Elemental O, C, N, S and P Disk Abundances \label{sec:ocnsp}}

The elemental abundances of OCNSP across disks regulate the volatile compositions of forming planets and planetesimals. There are at least four factors that influence the OCNSP abundance patterns: the nature of the OCNSP carriers inherited from the interstellar and protostellar phases, chemical transformations in the disks, locations of condensation fronts, and transport processes. In this section we review existing constraints on elemental OCNSP disk abundances, organized by disk location (outer vs. inner disk) and phase (solid vs. gas). We proceed from the expected least processed volatiles (outer disk solids) to most (inner disk gas), but note that efficient radial and vertical transport may blur distinctions between processed and pristine volatile reservoirs in disks.

In the outer disk, OCNSP are present in both refractory and icy solids. Oxygen is a major component of silicates, which is observed to be abundant in disks, while C is present in refractory carbon.  Based on comet observations \citep{Altwegg19}, the latter is abundant and may either be inherited from the interstellar medium, or a product of disk organic chemistry, or a combination of both. There is some evidence, based on observed high gas-phase C/O ratios, that a portion of this refractory C reservoir is transformed into C gas during the disk life time \citep{Bosman21_c2h}. Nitrogen is often assumed to not have a significant refractory phase, but recent cometary measurements, as well as previous interstellar ice spectroscopy, suggests that ammonium salts may be an important carrier of N \citep{Boogert15,Altwegg20}. The outer disk ice reservoir is constrained by interstellar ice observations (setting the initial conditions), a small set of disk ice observations \citep[e.g.][]{Pontoppidan05,Aikawa12}, comet compositions \citep{Mumma11,Altwegg19}, and disk gas-phase observations of sublimated ice. The latter is possible following a stellar luminosity burst, and in sources where the outer disk is exposed to high levels of stellar radiation due to the combination of a higher-mass star and a large inner cavity \citep{Banzatti12,Banzatti15,Cieza16,Lee19, Booth21,vanderMarel21}. The disk and comet observations indicate, that similar to the interstellar medium, disk icy grains mainly consist of H$_2$O, CO$_2$, CO, NH$_3$ and O-rich organics. Based on theory we also expect  N$_2$ to be present in the coldest disk regions. The H$_2$O ice is likely largely inherited from the ISM based on models \citep[see e.g.][]{Visser09}, and Solar System H$_2$O D/H ratios \citep{Cleeves14}, while the contribution of inheritance vs in situ chemistry to the other major O, C and N carriers is less well constrained. The outer disk ices also appear to contain substantial amounts of S \citep{Altwegg19,Booth21-so}, as well as some P \citep{Altwegg19}, though most S and P are expected to be present in refractory carriers (see below).

The outer disk gas OCNSP composition can be directly constrained by observing CO, H$_2$O and NH$_3$, the three expected major carriers that are observable at far-IR and millimeter wavelengths. CO gas appears to be often depleted in disks (see \S\ref{sec:mass}), reducing the C and O gas abundances. This may be explained by a combination of CO processing, forming less volatile species, and dynamics \citep{Schwarz18,Krijt20}. H$_2$O vapor has only been detected in one T Tauri disk and a handful of Herbig Ae disks, providing independent evidence that the outer upper disk layers are generally dry and therefore O-poor \citep{Hogerheijde11,Meeus12,Du14,vanDishoeck21,Pirovano22}, though the inferred H$_2$O depletion level does depend on the assumed disk structure and water excitation \citep{Kamp13}. Finally NH$_3$ has been observed in only one disk \citep{Salinas16}, and its contribution to the outer disk N reservoir remains rather unclear. There have been no observations of P-bearing species in disks, and observations of gas-phase S molecules, such as CS and H$_2$S, indicate that less than 1\% of S is in the gas-phase \citep{Phuong18,LeGal21,Riviere22}, and hence 99\% is present in refractory grains and ices. Most constraints on the outer disk gas is less direct, however, and originates from observations of chemical probes of gas-phase elemental ratios, such as C$_2$H/CO and CS/SO abundance ratios \citep{Cleeves18,Miotello19,Fedele20,LeGal21}. These studies confirm that the outer disk gas is depleted in O, resulting in enhanced C/O ratios, which can exceed unity \citep{Bosman21_c2h}. Disks may also be somewhat depleted in Cm while there is so far no indication of N depletion, implying super-solar N/O and N/C ratios in the outer disk gas \citep{Cleeves18}.

In the inner disk, the solid composition is constrained by disk IR spectroscopy of silicates, gas-phase abundances inside of refractory dust sublimation fronts, and Solar System abundance patterns. Based on IR spectroscopic observations, silicate grains are an important O-carrier in the inner disk \citep{vanBoekel04,Natta07,Bouwman08}, while the abundance of refractory C is less clear; the Solar System record suggests that the inner Solar Nebula solids were C-poor, and observations of low C/O ratios on white dwarfs polluted by infalling planet debris indicate that C-poor solids in inner disk regions is a general phenomenon \citep{Lodders03,Wilson16}. The N content in inner Solar System solids also appears to have been low. By contrast, meteoritic measurements suggest that almost all P and about half of S are present in refractory grains in inner disks  \citep{Lodders03}. The abundance of refractory S in inner disks has also been estimated by observations of S elemental abundances in accretion flows onto Herb Ae stars, confirming that it is mainly present in refractory grains in disks \citep{Kama19}. 

If the inner disk, all O, C and N not bound up in refractory grains are expected to be present in the gas-phase. Major O and C carriers H$_2$O, CO and CO$_2$ are directly accessible in the inner disk atmosphere through IR observations, while the major N carrier, N$_2$ is not, and NH$_3$ is present at low abundances \citep{Pontoppidan19,Najita21}. The inferred gas composition around Solar-like stars is generally dominated by CO and H$_2$O \citep{Pontoppidan2014}, but the gas-phase C/O ratio is currently not well-constrained and may vary substantially between disks due to different levels of icy pebble flux and sequestration of water in the outer disk \citep{Najita11,Banzatti20}. S and P carriers have not been detected in the inner disk gas to date. It is finally important to note that the link between these disk atmospheric abundances and midplane reservoirs is not obvious. 

 Combining the above constraints, the inner disk refractory solids appear C- and N-poor, and O-, S- and P-rich. The inner disk gas composition varies between disks and likely depends on a combination of local processes and the influx (or lack thereof) of icy grains from the outer disk. As a result it may present a range of C/N/O ratios. In the outer disk the icy solids are initially O-dominated, but become more and more enriched in C and N towards the outermost disk regions where the most volatile C and N carriers (CO, small hydrocarbons, and N$_2$) freeze out. Conversely the gas is generally O-poor, but also appears somewhat depleted in C, probably due to CO depletion through a combination of freeze-out and chemical conversions. Only in the outermost disk would we expect the gas to also be depleted in N$_2$ based on the presence of N$_2$ snowlines \citep{Qi19}.

\subsection{Organic Disk Chemistry \label{sec:organics}}

Access to organic feedstock molecules constitute a key aspect of chemical habitability. Disk organic molecules can be delivered to terrestrial planets through impacts of planetesimals that originate in both the inner and outer disk regions, as well as accretion of a primary atmosphere from the local disk gas. This motivates explorations of organic chemistry across all disk radii. Protoplanetary disks serve both as conduits of the inherited interstellar and protostellar organic chemistry to planets and planetesimals, and as active producers of new organic molecules using the disk inorganic and organic carbon reservoirs. This combination of inheritance and local organic chemistry results in a distribution of organics in disks that is complicated and evolving over time.

Carbon enters the disk in the form of volatile inorganic and organic gas and ice (e.g. CO and CH$_3$OH), and more refractory  large aromatic and aliphatic hydrocarbons, PAHs and carbon grains. In the chemically active layers of the disk, some of the inherited refractory carbon may be vaporized and feed a C-rich top-down gas-phase organic chemistry \citep{Siebenmorgen12,Bosman21_c2h}. What remains provides disks with a unique organic reservoir that is characterized by aromatic groups, and relatively low levels of oxygen, especially compared to inherited organic ices. Interstellar and protostellar ices often contain high abundances of simple organics in the forms of CH$_4$, CH$_3$OH and perhaps HCOOH \citep{Oberg11c,Boogert15}. These can be energetically and non-energetically processed to form more complex organic molecules \citep{Bernstein02,MunozCaro02,Oberg15,Chuang17,Ioppolo21,Jorgensen20}. Importantly this complex organic chemistry takes place in an O-rich environment, i.e. the H$_2$O, CO and CO$_2$ dominated ice mantle, and therefore tend favor the production of O-rich organics such as alcohols, organic acids, aldehydes, ethers, and ketones.  Early evidence for the inheritance of interstellar and protostellar ices came from Solar System observations of comet organics, which have an O-rich contingent \citep{Mumma11}, and this is supported by more recent comparisons between comet and protostellar inventories  \citep[e.g.][]{Drozdovskaya19}. O-bearing organics that likely originate from interstellar or protostellar ice sublimation have also been observed in young disks \citep{vantHoff18, Lee19, Podio20}. The strongest evidence to date for substantial inheritance of interstellar and  protostellar ice chemistry comes from recent observations of gas-phase CH$_3$OH and more complex organic molecules in disks that are too warm to sustain CH$_3$OH production from CO ice \citep{vanderMarel21,Booth21-so, Brunken22}, the major CH$_3$OH production channel in the ISM \citep{Tielens82,Hidaka04,Fuchs09}. The observed COMs in these disks are hence inferred to originate from sublimation of inherited organic ices. Such ices can be further processed in disks when icy grains are lofted up from the relatively chemically inert midplane into the disk upper layers \citep{Ciesla12,Bergner21-model}. In either case, icy planet building blocks should generally contain substantial amounts of O-rich complex organics, which may soon be confirmed by JWST observations disk ice compositions.

Most current inner and outer disk observations probe the gas compositions of disk layers with relatively short chemical timescales \citep[e.g.][]{Henning10}, where detected organics are products of disk {\it in situ} organic chemistry. This chemistry is regulated by high temperatures and high energy radiation in the inner disk and by high energy radiation (UV and X-rays) alone in the outer disk \citep[e.g.,][]{Agundez08,Semenov11,Walsh15,Bergner19, Bosman21_c2h}. As reviewed in greater detail by \citet{Henning13}, the presence of the small organics C$_2$H$_2$, HCN and CH$_4$ in the inner disk, is readily explained by high-temperature gas chemistry \citep[e.g.][]{Willacy98,Ilgner04,Najita17,Walsh15-model}. There are ongoing efforts to also explore the role of irradiation fields, detailed radiative transfer, non-solar elemental ratios, transport and grain-gas interactions on this chemistry \citep[e.g.][]{Bruderer15,Najita17,Woitke18,Wei19,Price20,Anderson21,Duval22}. In either case, the presence of a hot inner disk organic chemistry implies a gas-phase organic reservoir in the terrestrial planet forming zone that is completely different from the outer disk and interstellar medium.

In the outer disk, C$_2$H, HCN, and H$_2$CO have been widely detected in sources across a range of stellar masses and ages \citep{Bergner19, Pegues20, Guzman21}. Larger  organics, such as HC$_3$N, c-C$_3$H$_2$ and CH$_3$CN, are also quite commonly detected \citep{Chapillon12,Oberg15, Bergner18, Ilee21}. High spatial-resolution, multi-line observations have located these larger nitriles and hydrocarbons in the disk molecular layer \citep{Oberg21-maps,Ilee21}. By contrast larger O-containing organics are relatively rare with only one detection of CH$_3$OH in a disk not obviously experiencing extensive ice desorption \citep{Walsh15}. Outer disk {\it in situ} gas chemistry is then characterized by an O-poor and N-rich organic chemistry, consistent with other evidence of O-poor gas in outer disk layers (see \S\ref{sec:ocnsp}). There is increasing evidence that this chemistry is qualitatively similar to that of classical photondominated regions (PDRs) \citep{Kamp04,Jonkheid04,Chapillon12,Agundez18,LeGal19b}. The organic products of this PDR-like and O-poor disk gas chemistry, including the prebiotically interesting nitriles, may become incorporated into planet atmospheres in the outer disk, and also freeze out onto comet-forming pebbles. This latter scenario would provide planetesimals with a second reservoir of organics that is distinct from the inherited protostellar one, which may at a later stage be delivered to terrestrial planets through impacts. The importance of this reservoir would depend on a combination of the disk lifetime production of nitriles and downward transport of disk atmosphere chemistry products. 

\subsection{Isotope Fractionation Chemistry \label{sec:isotope}}

\begin{figure}[h]
\includegraphics[width=5in]{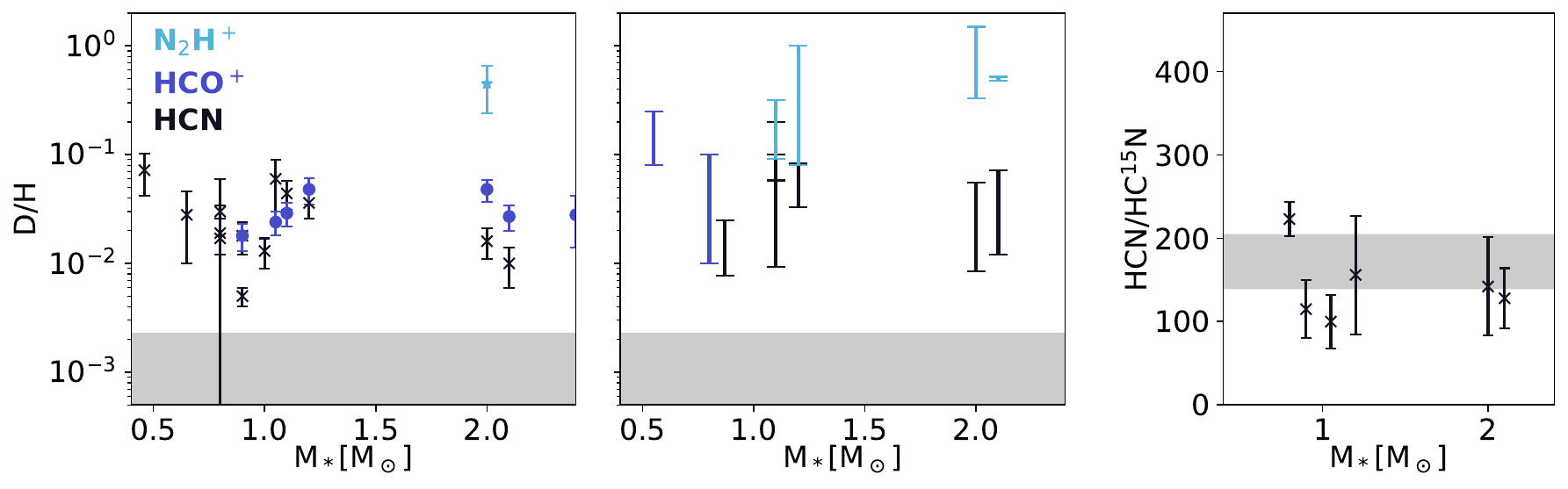}
\includegraphics[width=5in]{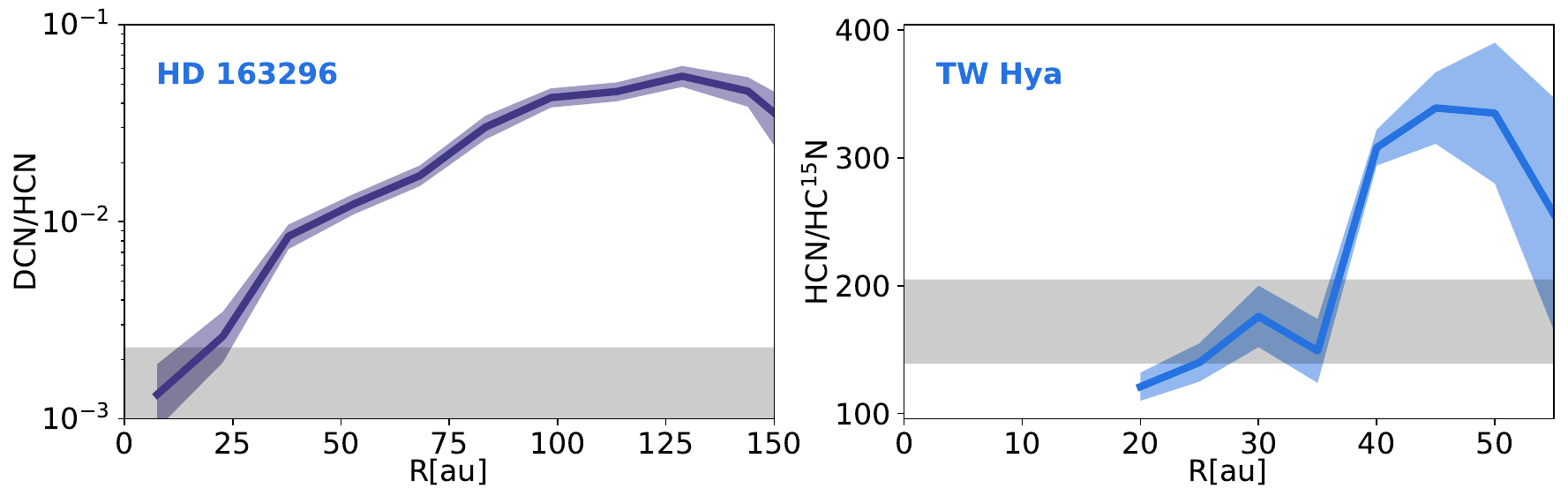}
\caption{{\it Top Panels:} Disk integrated D/H ratios for HCN, HCO$^+$ and N$_2$H$^+$ (left panel shows disk averaged values and middle panel shows the extracted ranges from spatially resolved observations), and $^{14}$N/$^{15}$N ratios for HCN towards a sample of protoplanetary disks. The shaded region depicts D/H values in comet volatiles, and the range of $^{14}$N/$^{15}$N in cometary HCN   \citep{Bockelee15,Altwegg19}. Data is from \citet{Fuente10,Teague15,Huang17,Guzman17,Bergner20,Cataldi21,Salinas17,Hily-Blant17,Hily-Blant19,Pegues21}. {\it Bottom Panels:} radially resolved DCN/HCN and and HCN/HC$^{15}$N ratios from \citet{Cataldi21,Hily-Blant17,Hily-Blant19}.}
\label{fig-fract}
\end{figure}

Isotopic fractionation occurs at low temperatures due to small differences in zero-point energies between heavier and lighter molecules, and in photon-dominated regions due to isotopologue-specific photo dissociation \citep[e.g.][]{Aikawa01,Willacy07,Visser09,Ceccarelli14,Miotello14}. Isotopic fractionation patterns in molecules in disks are generally used for two purposes: to assess the current thermal or irradiation environment, and to link together different evolutionary phases \citep{Ceccarelli14}. The former use case assumes fast chemical timescales such that observed D/H and other isotopic ratios reflect disk environmental conditions. This assumption is generally valid for gas-phase molecules in the disk atmosphere and intermediate vertical layers that are probed by mm and IR gas observations. The second use case  makes the opposite assumption such that the observed isotopologue ratio reflects the environmental conditions of an earlier stage and can be used to infer inheritance. In disks, this might hold for ices in disk midplane, dependent on the degree of vertical and radial mixing. This could be tested through isotopic gas observations in disks where ice sublimation  control the gas abundances \citep{Booth21-so}, and perhaps also through ice observations of e.g. deuterated water with JWST.  In the meantime all our isotopic measurements in disks originates from the chemically active disk atmosphere and intermediate layers. 

Five molecules have been detected in both their deuterated and non-deuterated forms: HCO$^+$, HCN, N$_2$H$^+$, HNC, and C$_2$H \citep{Qi08,Huang15,Loomis20}. The latter two have single detections, however and are not further considered here. For HCO$^+$ and HCN, the disk integrated D/H ratios are elevated above the cosmic D/H ratio by several orders of magnitude. This implies an active deuterium chemistry in the outer disk -- deuterium fractionation of these and related species can occur through the cold H$_2$D$^+$ channel or the warmer CH$_2$D$^+$ channel and both appear to be active in disks \citep[e.g.][]{Willacy07,Huang17,Salinas17,Aikawa18,Cataldi21}. The D/H ratios in HCN and HCO$^+$ are quite similar, indicative of a shared formation environment, while N$_2$H$^+$ is substantially more deuterated (Fig. \ref{fig-fract}). The latter is expected if most N$_2$H$^+$ originates close to the cold disk midplane between the CO and N$_2$ snowlines \citep{Aikawa18}. There is no clear pattern in degree of deuteration with stellar mass and luminosity, as might be expected from a cold-chemistry tracer. This lack of a pattern may be an effect of the present small and biased sample, however, which only includes hotter stars with large disks, and hence substantial cool disk regions.  Within individual disks, the D/H ratio in all three molecules generally increase with disk radius as would be expected from their status as low-temperature tracers \citep[Fig. \ref{fig-fract} and][]{Qi08,Cataldi21}, but the sample is still very small.

There are theoretical reasons to expect non-Solar C and O isotopologue ratios in disks \citep[e.g.][]{Miotello16}, and observations of $^{13}$C isotopologues in disks exist for CO, HCO$^+$, HCN, CN and CS, and of $^{18}$O and/of $^{17}$O isotopologues for CO and HCO$^+$. Fractionation in carbon and oxygen could therefore in theory be extracted. In practice this has proven difficult due to high or unknown optical depths of the main isotopologue lines, and instead minor isotopologues are often used to constrain the optical depth of the major isotopologue assuming local ISM iostopic ratios \citep[e.g.][]{Williams14,Booth19,Zhang21}. An exception is \citet{Smith15}, who used IR absorption line observations to derive a non-Solar $^{13}$C/$^{12}$C in CO in a couple of disks, but the explanations of these isotopic heterogenities is not yet clear. $^{15}$N/$^{14}$N ratios have been extracted for HCN towards a handful of disks, and HCN is always enriched in $^{15}$N. In the two disks with spatially resolved HCN fractionation observations, the HCN/HC$^{15}$N ratio increases with radius, i.e. the disk gas is the most fractionated in $^{15}$N close to the star \citep{Guzman18,Hily-Blant17,Hily-Blant19}. This is the opposite behavior of deuterium enrichment in HCN, and strongly indicates that HCN enrichment in $^{15}$N is due to isotopologue selective photodissociation \citep{Heays14}, rather than cold isotope fractionation. In one disk, TW Hya, both CN and HCN fractionation in $^{15}$N has been measured, and CN is much less enriched than HCN \citep{Hily-Blant17}, which indicates the presence of two different N reservoirs in disks.

Isotopologue ratios could be used to assess disk chemistry contributions to comet inventories. In the case of HCN the disk averaged $^{15}$N/$^{14}$N is consistent with comet values, and for disks with radially resolved emission, disk values agree with comets in the inner 10s of au of the disk (Fig. \ref{fig-fract}), which is where most comets formed in the Solar System \citep{Mumma11}. If disk fractionation chemistry is important for cometary isotopic ratios, the lower fractionation observed at larger disk radii should imprint on planetesimals assembling in the outskirts of disks, which could be tested by measuring $^{15}$N/$^{14}$N in Solar System bodies that assembled beyond 40~au. In the case of HCN D/H ratios, disk averaged values are orders of magnitude higher than comet ratios, but radially resolved data has shown cometary D/H values in the inner 30 au in some disks \citep{Cataldi21}. 
In summary, isotopologue ratios in disks and comets appear consistent, but this is not enough on its own to ascribe a causal link between the two. More detailed data on the fate of disk gas-phase fractionation products, as well as disk chemistry models that take into account both dynamics and isotopic fractionation in ices are needed to address this question \citep{Faure15b}. In addition, if disk H$_2$O and CH$_3$OH D/H values become available (see above), comparison between those and comet values would provide constraints on the relative contributions of inheritance and {\it in situ} disk chemistry for volatiles and small O-rich organics.

\section{CHEMICAL PROBES OF DISKS AND PLANET FORMATION \label{sec:probes}}

Molecular observations often provide the best and sometimes the only path to constraining physical disk properties of importance to disk evolution and planet formation. This section reviews chemical probes of disk gas mass and surface density (\ref{sec:mass}), ionization (\ref{sec:ion}), temperature (\ref{sec:temperature}), and dynamics, including planet formation (\ref{sec:dynamics-pf}). In each case we provide a brief overview of how disk chemistry and molecular emission are linked to the disk property or process in question, review the deployed chemical probe(s), and discuss some key results that have emerged or are expected in the near future.

\subsection{Disk Gas Mass and Surface Density \label{sec:mass}}

Disk masses and column densities of both gas and solids determine the planet formation potential of protoplanetary disks. Neither can be directly observed, however, and instead a range of proxies have been developed, each with their own advantages and disadvantages. The most common proxies for disk masses are the mm flux density for the dust component, and CO isotopologue fluxes for the gas counterpart \citep[][and references therein]{Andrews20,Miotello22}. The utility of the CO gas mass probe depends on the CO abundance in disks, which is fundamentally a chemical problem. The disk CO abundance structure has been modeled with different levels of sophistication  including CO freeze-out and photodissociation \citep[e.g.][]{Williams14},  self-shielding and isotope-selective photodissociation \citep{Miotello14}, C and O isotopic fractionation chemistry \citep{Miotello16}, and source specific thermo-chemical models \citep{vanderMarel15,vanderMarel16,Woitke19, Zhang21}. When the CO model grids are compared with resolved or disk-integrated CO isotopologue fluxes, the extracted gas-to-dust ratio is frequently one to two orders of magnitude below the typical ISM ratio of 100 \citep{Miotello17}, which suggests that most models overestimate the CO abundance. Furthermore, in a few cases gas masses can be estimated using HD observations, and these reveal substantial CO depletion compared to the ISM for T Tauri disks \citep{Bergin13,McClure16}, while CO abundances may be more similar to the ISM in Herbig disks  \citep{Kama20}. 

\begin{figure}[h]
\includegraphics[width=5in]{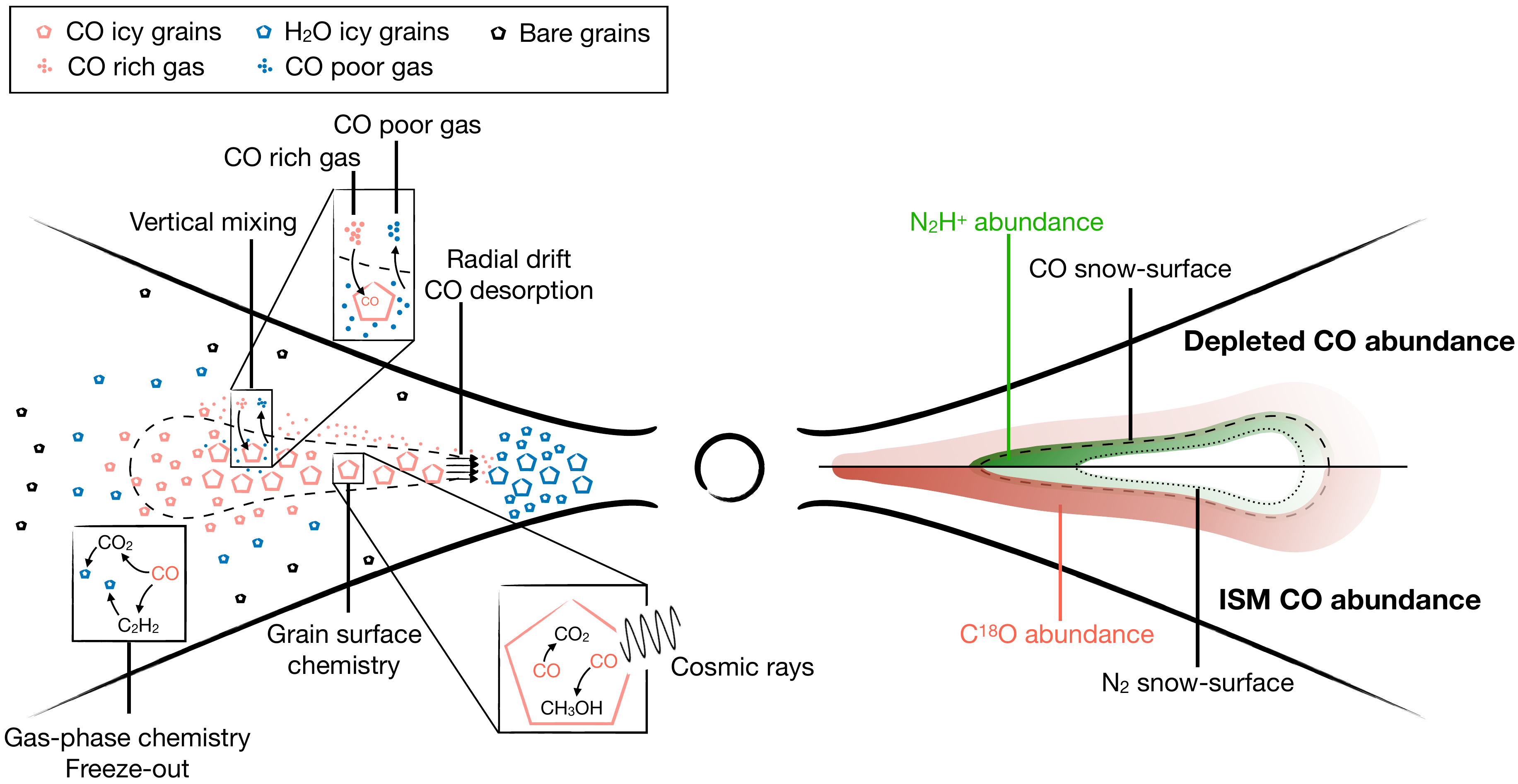}
\caption{Left: main physico-chemical processes reducing the CO gas abundance in protoplanetary disks, highlighting the three main concurring processes: vertical mixing and grain growth, gas phase chemistry and freeze-out, and grain surface chemistry. Right: C$^{18}$O (red) and N$_2$H$^+$ (green) abundances for depleted and ISM-like CO gas abundances, with the N$_2$H$^+$ being abundant between the CO and N$_2$ snow surfaces.}
\label{fig:CO}
\end{figure}

The observational evidence for CO depletion from the gas phase has triggered extensive modeling to understand its physico-chemical cause(s). Figure~\ref{fig:CO} illustrates some of the most commonly proposed scenarios: 1) Gas and dust vertical mixing, driven by large scale motions or turbulence, can remove gaseous CO from the disk upper layers by freezing it onto large dust grains in the disk midplane \citep{Kama16,Schwarz16,Powell22,Furuya22b}. The same mechanism has been invoked to explain the low water abundance in the disk upper layers \citep{Meijerink09,Hogerheijde11,Du15}. 2) Gas-phase reactions can convert CO into less volatile species, such as CO$_2$ \citep{Aikawa99} and small hydrocarbons \citep{Aikawa99,yu17}, which subsequently freeze-out onto dust grains. 3) Frozen-out CO may be transformed into molecules with higher binding energy, such as CO$_2$ or CH$_3$OH ice, through grain-surface chemistry \citep{Schwarz18,Bosman18}. 1) and 3) should mainly operate outside of the CO snowline, though radial diffusion may extend 1) somewhat inwards. The evidence that CO depletion is less severe in warm disks where the CO snowline is further out supports these scenarios, and recent models by \citet{Krijt20} and \citet{vanClepper22} show that both are needed to reproduce the derived low CO abundances of some T Tauri disks. These models also indicate that if dust radial drift is very effective, the CO abundance should be enhanced interior to the CO snowline, which has also been observed towards some disks \citep[e.g.][]{Zhang19}. 

\begin{figure}[h]
\includegraphics[width=3.5in]{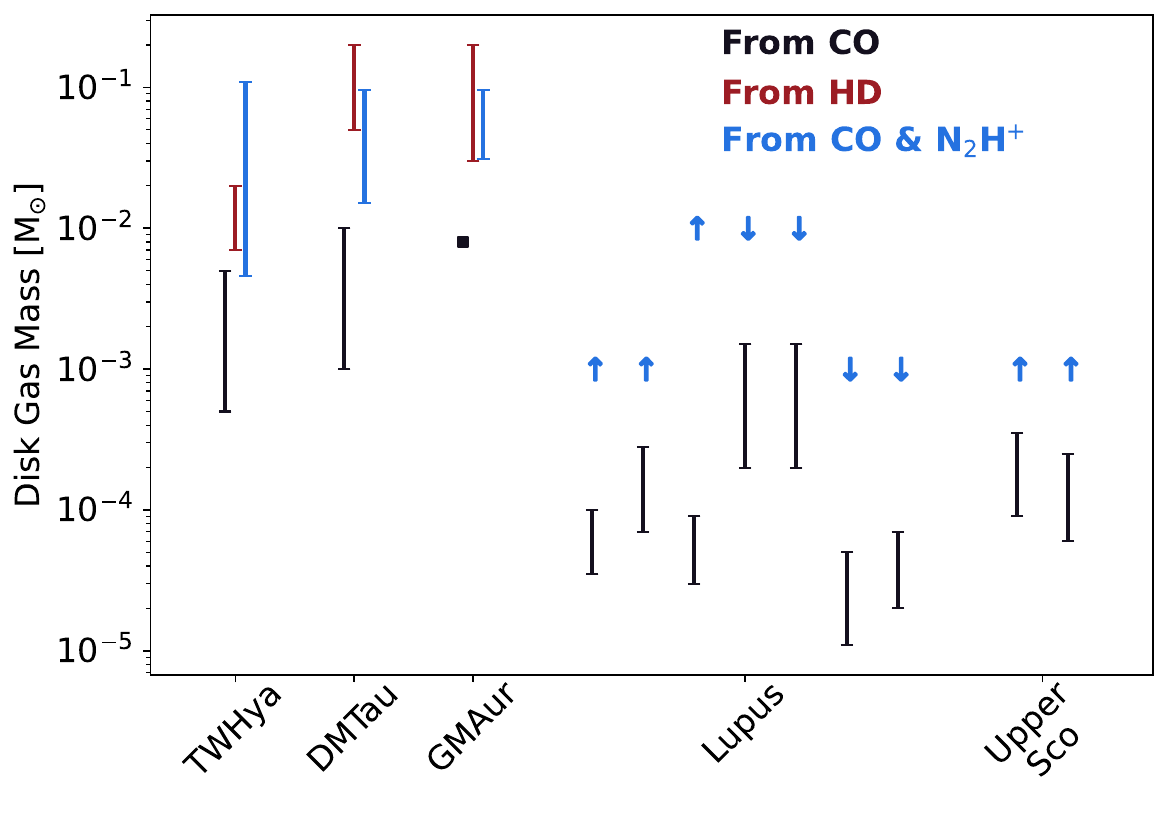}
\caption{Comparison of total H$_2$ gas mass estimates using different molecular tracers: CO (black), HD (red), and N$_2$H$^+$ combined with CO (blue) \citep{Thi10,Miotello16,Miotello17,Zhang19,Anderson19,Zhang21,Trapman22,Anderson22}. }
\label{fig-mass-compare}
\end{figure}

In light of the limitations of CO as a gas mass tracer, the community has explored complementary observational mass diagnostics, and N$_2$H$^+$ has emerged as a promising candidate. N$_2$H$^+$ should correlate with CO depletion (Fig.~\ref{fig:CO}) due to competition between N$_2$ and CO molecules for  H$_3^+$, and N$_2$H$^+$ destruction by gas-phase CO. The combination of N$_2$H$^+$ and CO isotopologues should therefore probe both CO-rich and CO-poor gas and provide better mass estimates than CO-based ones \citep{Trapman22,Anderson22}. Figure \ref{fig-mass-compare} shows that these chemical mass estimates are generally in good agreement with constraints from the HD measurements when available. However,  N$_2$H$^+$ fluxes also depend on ionizing radiation fluxes and the cosmic ray ionization rate, and an additional independent proxy of disk ionization (e.g., HCO$^+$) may be needed to consistently obtain accurate disk gas masses with this approach \citep{vantHoff17,Anderson22}. Other, complementary molecule-based methods to estimate gas masses are also under development. The H$_2$ density could be derived from excitation analysis of molecules not in LTE at a range of disk heights and radii \citep{Teague20}. Finally, locations of snowlines may also be used to derive disk masses due to their dependence on pebble drift if all other aspects of snowline formation are well understood \citep{Powell19}.

\subsection{Disk Ionization}
\label{sec:ion}

The ionization fraction of protoplanetary disks affects their dynamical and chemical evolution \citep[see also discussion in][]{Bergin07}. It regulates the coupling of the disk gas to magnetic field lines and the efficacy of the magneto rotational instability (MRI) as a means of producing turbulence and driving disk evolution, including determining locations of non-turbulent ``dead zones" that are potentially favorable locations for planet formation \citep{Gressel12}. Ionization also sets many chemical time scales. Ion-molecular reaction rates depend directly on ion abundances, and most grain surface and neutral-neutral gas-phase chemistry, include an ion recombination reaction to form the reactants. An example of the latter is the production of H atoms needed to form water on grain surfaces \citep{Cleeves14,Oberg21_Review}.  

\begin{figure}[h]
\includegraphics[width=5in]{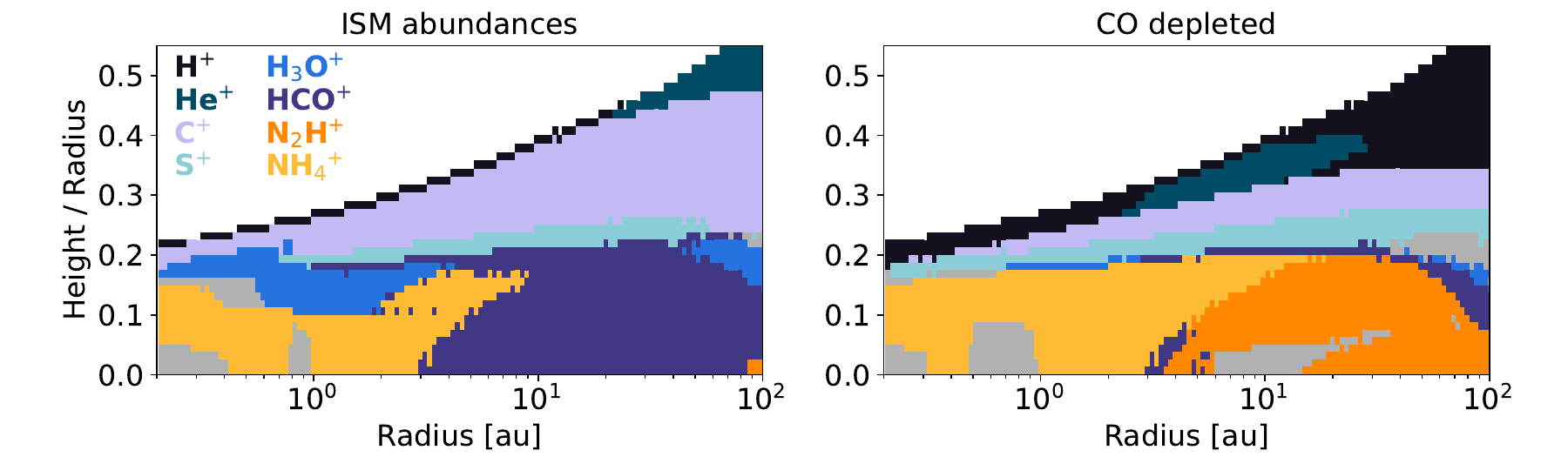}
\caption{Dominant ion species across a T Tauri disk for models with interstellar abundances of C and O (left) and with initial C and O abundances depleted by 100$\times$. Other minor species not listed in the legend are shown in gray. The models \citep{Anderson21, Anderson22} assume a stellar UV spectrum from a low-mass T Tauri star, a total X-ray luminosity of 10$^{30}$~erg~s$^{-1}$, and a cosmic ray ionization rate of 10$^{-18}$~s$^{-1}$ per H$_2$ at the disk surface.}
\label{fig-ions}
\end{figure}

% theory - which ions are where
Disk ionization studies typically have two goals: to determine the ionization fraction and to constrain the main source(s) of ionization (see \S\ref{sec:disk-environ}). The ionization fraction is set by a balance between the ionization and recombination rates, both of which depend on the chemical composition of the disk \citep[see][]{Semenov04,Xu19} as well as on the radiative transfer of the ionizing radiation. Due to the chemical stratification of protoplanetary disks, the main positive charge carrier varies between different disk regions (Fig.~\ref{fig-ions}) and the use of atomic and molecular ions as probes of ionization therefore depends on detailed models and observations of disk chemistry.
Theoretically, atomic ions dominate near the directly irradiated, PDR-like disk surface, while, molecular ions produced via protonation of abundant volatile species (H$_2$O, CO, N$_2$, NH$_3$) by H$_3^+$ dominate in the deeper layers. As shown in Fig. \ref{fig-ions}, the thickness of the atomic ion layer and the region in which molecular ion is the major charge carrier are both highly sensitive to the overall chemical state of a disk: in a disk depleted in C/H and O/H (see \S\ref{sec:ocnsp}), the  C$^+$ region is reduced in favor of H$^+$ and He$^+$, and the HCO$^+$ and H$_3$O$^+$ regions are reduced in favor of N-bearing molecular ions N$_2$H$^+$ and NH$_4^+$. Molecular ion abundances also depend on other elemental abundances, which act as electron donors \citep{Ilgner06}. In addition to atomic and molecular ions, PAHs and grains can be significant charge carriers in the disk \citep[not included in Fig.~\ref{fig-ions}; see e.g.,][]{Thi19}. 

% observational techniques, and observational findings
% KO Constraints on ionization level
Because of the complexity of the atomic and molecular ion structure of disks, there is no consensus on the typical ionization level in different disk environments. Only a handful of disks have been studied in detail with regard to ionization, and the conclusions diverge. Based on a combined analysis of HCO$^+$ and N$_2$H$^+$ emission in TW~Hya, \cite{Cleeves15} found a low ionization level, which constrain 
the cosmic-ray ionization rate to be $\lesssim$0.01$\times$ the interstellar value. By contrast \citet{Aikawa21} found relatively high ionization rates in both the warm molecular layer and midplane of other disks, consistent with standard assumptions about X-ray ionization and relatively high cosmic ray ionization rates. Finally, \citet{Seifert21} found a gradient in the cosmic ray ionization rate across the IM Lup disk, and if this is a general feature it may help to explain some of the diverging results. More studies at high spatial resolution that combine multiple optically thin ionization probes with probes of other disk properties are needed to settle the disk ionization levels across disks.

The relative importance of different sources of ionization should vary across disks \citep{Cleeves15,Rab18}, and between disks around different stars \citep{Walsh15} and situated in different radiation environments \citep{Walsh13}. 
In theory the contributions of  different ionizing sources could be deduced from observations of ions in disks because different ionizing sources impact disk ion abundances in different disk regions. This was e.g. used by \citet{Cleeves15} to extract a relatively limited contribution from cosmic ray ionization, and a larger than expected contribution from X-rays in one disk, suggesting a flaring X-ray stage. X-ray flares may have a time-resolved impact on ionization as seen in \citet{Cleeves17}. The full potential of ionization probes have yet to be realized, however, mainly due to a lack of comprehensive, spatially resolved molecular ion data sets on samples of disks.

% Theory and Constraints on ionization source

% theory - how modeled chemical composition depends on variations in ionization due to star/surrounding environment
%D

%Observations using specific probes have also put some constraints on the main sources of ionizing radiation in disks. 
%The strength of the UV field is constrained using photodissociation products, specifically radicals like CN and C$_2$H \citep[][]{vanTerwisga19,Bosman21_c2h}.

\subsection{Disk Temperature}
\label{sec:temperature}

Detailed disk temperature structures are necessary to correctly interpret molecular line observations, to model planet formation, and to predict the chemical evolution of disks (see also \S\ref{sec:snow_lines} and \S\ref{sec:ocnsp}).  %Qualitatively disk temperature gradients are well understood and predicted by models, but quantitative model predictions have been more challenging due to complex radiative transfer that sensitively depends on the gas and dust density structure, thermal decoupling between dust and gas, and chemical feedback on the gas temperature since gas lines constitute a major cooling mechanism in many disk regions \citep[see \S\ref{sec:disk-environ} and][]{Kamp04,Woitke09,Bruderer12}. 
 Molecular and atomic line observations constitute our best tools to constrain individual disk temperature structures, as well as to benchmark disk models \citep[e.g.][]{Dartois03, Pietu07,Kamp10,Calahan21}. For spectrally well resolved, optically thick lines, the peak brightness temperature ($T_{\rm b}$) is a direct tracer of the local kinetic temperature and optically thick lines (such as the CO ladder in the FIR) can be used to constrain disk gas temperatures by comparing observational data to radiative transfer models \citep[e.g.,][]{Bruderer12,van_der_wiel14}. High spatial resolution observations of optically thick lines provide more direct constraints on the temperature profiles of the emitting layers \citep{Dartois03,Zhang17,Pinte18b}, and by combining optically thick lines  emitting from different disk heights, which can be constrained observationally in mid-inclined disks   (\S\ref{sec:vert_gradients}), it is possible to reconstruct a 2D ($R-z$) map of the disk temperature structure. This has so far mostly been done using CO isotopologues due to required high SNRs for this method \citep[][]{Law21_surf}, but future studies are also expected to make use of other molecules that emit from a larger range of disk layers due to either different chemistry or line excitation, and therefore enable a more complete temperature reconstruction \citep[see e.g.][]{Huang20}. 

For optically thin (or marginally optically thick) lines, single transitions cannot break the degeneracy between column density and excitation temperature, but rotational excitation temperatures can instead be derived with molecular population diagrams using two or more transitions with well characterized energy levels and transition probabilities. The quality of ALMA data allows the construction of molecular population diagrams of a large number of molecules, in several cases also with spatially resolved data. Excitation temperatures have been derived for molecules such as HCN, H$_2$CO, CH$_3$OH, CH$_3$CN, CN, c-C$_3$H$_2$ \citep[e.g.,][]{Teague15,Loomis18b,Pegues20,Teague20,Facchini21,Ilee21,vanderMarel21}. For marginally optically thick lines, it is possible to account for the line intensity saturation and correct for it \citep{Goldsmith99}. For the majority of rotational transitions, their critical density is $<10^8\,$cm$^{-3}$, below the H/H$_2$ density where the lines originate, and an LTE approximation is valid; the excitation temperature therefore directly traces the gas kinetic temperature and can be used as a gas thermometer when the emission location is known. Conversely, whenever the temperature structure is known through other means, the excitation temperature of molecules can be used to locate the vertical layer where the lines originate, anchoring disk chemistry models \citep{Ilee21}. 

In cases where molecular excitation temperatures are challenging to extract, including in the disk midplane, chemical structures can be used to constrain gas and dust temperatures. The most prominent examples are snow-lines and snow-surfaces (see \S\ref{sec:snow_lines} and \S\ref{sec:vert_gradients}): identification of snow-surfaces via chemical markers (as N$_2$H$^+$ for CO, or H$^{13}$CO$^+$ for H$_2$O) uniquely associates a specific region of a disk to the range of sublimation temperatures that laboratory experiments obtain for that given molecule \citep{Fayolle16,Qi19,Zhang21}. Intensity ratios of some molecules have also been associated to specific gas temperatures and could be deployed as thermometers. An example is the HNC/HCN ratio because of  temperature-dependent destruction pathways for the two isomers \citep{Graninger15,Hacar20,Long21}.

The deployment of these different temperature probes, including using them to tune source-specific models, has  confirmed general expectations on temperature structures \citep[e.g.][]{Dartois03,Law21_surf}, but also demonstrated the importance of gas heating in low density disk environments \citep[e.g.][]{Fedele16}, and revealed unexpected thermal sub-structure, such as temperature inversions in outer disk regions, resulting in secondary snowlines of CO \citep{Oberg15,Dutrey17,Cleeves16a,Facchini17}. An analogous thermal inversion can occur in the disk upper layers, where the dust and gas can sometimes thermally de-couple at lower heights than foreseen in typical thermo-chemical models not accounting for settling \citep[e.g.][]{Facchini17}. An open question in disk thermal structure modeling concerns the gas temperature in dust gaps, and hence the chemical evolution in the vicinity of the planets that produce them. A higher penetration of energetic photons should increase both the gas and the dust temperatures \citep{vanderMarel18b,Rab20,Alarcon20a}, but efficient gas-cooling may counteract this in severely dust-depleted gaps
\citep{Facchini18}. Observational constraints on the gas temperature of dust gaps are inconclusive, but in at least two transition disk cavities the gas temperature is enhanced \citep{Leemker22}.%, which may suggest that gap chemistry too occurs at elevated temperatures.

\subsection{Disk Dynamics and Planet Formation } \label{sec:dynamics-pf}

Planet forming disks are highly dynamical environments \citep[\S\ref{sec:context} and recent reviews by][]{DiskDynamics20,Pinte22}, and many of these dynamical timescales are of the same order (or shorter) than typical chemical timescales \citep[e.g.,][]{Semenov11}, complicating disk chemical modeling (\S\ref{sec:models-dyn}).  At the same time, chemistry can affect gas dynamics in disks through its impact on disk ionization and temperature (\S\ref{sec:ion} and \S\ref{sec:temperature}) and a complete chemo-dynamical disk model would need to take this chemical feedback into account. The complex interplay between chemistry and dynamics is one of the major challenges for disk chemistry studies going forward. It is also a potential advantage, however, since it implies that chemical disk structures could be used as probes of disk dynamics and ongoing planet formation.

A set of proposed chemical probes of dynamics exploits the impact of radial advection of volatiles in the gas and/or ice form on inner disk molecular abundances (\S\ref{sec:context}). Efficient inward radial drift of icy pebbles should increase the H$_2$O vapor in the inner disk and produce a sub-stellar C/O ratio. This has tentatively been observed by \citet{Banzatti20} who found an anti-correlation between inner disk H$_2$O line flux and pebble disk size, where a larger disk size implies less pebble drift (probably due to dust-substructure). If instead water is sequestrated in the outer disk due to the efficient transformation of pebbles into boulders, the inner disk should be water poor and the C/O ratio super-stellar.    \citet{Najita13} suggests that this process would explain an observed trend between the HCN-H$_2$O water ratio in the inner disk and disk mass. Upcoming JWST data may be able to distinguish between the two scenarios. Another possible probe of drift is the relative C/H ratios in the inner and outer disk regions \citep{McClure19,Sturm22}. Disk models that include inward accretion flows have also noted large effects on the inner disk organic chemistry \citep{Semenov11,Price20}, but this has not yet been converted into an observational probe. %The chemical composition of dust grains in outer disk regions, and especially signs of thermal processing, has been used to infer outward transport of grains both in our Solar System and in other protoplanetary disks \citep{Bockelee02,Abraham09}. 
 
 Chemical disk structures can also be used to trace time-dependent changes in disk temperature and radiation structures. In an FU Ori outburst, a sudden increase in stellar luminosity and accretion rate moves the icelines of molecules as H$_2$O and CH$_3$OH to large radii. These extended snowlines remain for some time after the outburst and can therefore be used to infer an energetic past. This has been observed in younger disks, but not yet in mature protoplanetary disks \citep{vantHoff18a,Lee19}. In warped disks, a misaligned inner disk can cast shadows onto the outer regions \citep{Facchini18b} and the resulting  azimuthal gradient in the illumination pattern is expected to cause large scale asymmetries in the intensity of molecular lines particularly sensitive to photochemistry \citep{Young21}. Finally, theoretical models predict that heating from massive planets and UV excess caused by high accretion rates during the run-away phase could imprint the thermal and chemical structure of close-by gas, as warm hot-spots detectable with long ALMA integrations \citep{Cleeves15b}. Such features have not yet been detected in the proximity of kinematically-inferred planets \citep{Pinte18b,Casassus19,Izquierdo22}, or directly-detected planets \citep{Facchini21}, but constitute a potential unique probe of planet formation.

In addition to chemical probes, observations of molecular lines at high angular and spatial resolution can be used to directly constrain the gas dynamics of planet forming disks. To date almost all studies make use of the brightest molecular lines, i.e. CO 2--1 or 3--2, without much regard for chemistry \citep[e.g.,][]{Teague19,Casassus19,Izquierdo21}.
%Even when different molecular lines have been used, it is typically because of favorable line excitation properties rather than chemistry, such as when using line widths to measure disk dynamics \citep{Flaherty20,Casassus21}.
There is real potential to further develop such probes, however, to extract density and abundance gradients for different molecular species, by mapping the velocity fields across the disk radial and vertical extent. In the meantime, these kinds of studies have revealed several kinematic structures of relevance to the chemistry of planet formation. In particular there are large scale convective flows co-located with annular substructures in dust continuum \citep{Teague19,Yu21}. These meridional flows show that chemical abundances probed in disk molecular layers can access the planet-forming disk midplane, and deliver chemically evolved gas to the proximity of growing protoplanets \citep{Cridland20}.

\section{LINKING DISK CHEMISTRY AND PLANET COMPOSITIONS \label{sec:planet}}

This review is to a large extent motivated by a close connection between disk chemical structures and processes on the one hand, and the outcome of planet formation on the other. In this final section we review the links between disk chemistry and planet formation and discuss possible directions for future development. 

To begin with snowline locations (\S\ref{sec:snow_lines}) may impact the architectures of planetary systems. Snowlines of different volatiles are predicted to impact the grain coagulation rate, resulting in higher (or lower) rates of planet formation in the vicinity of snowline locations. In the Solar System, the presence of Jupiter at 5~au has long been associated with gas giant formation just outside the water snowline in the Solar nebula \citep[e.g.][]{Stevenson88}. There is also  evidence for a pile-up of gas giant exoplanets around 2~au \citep{Fernandes19}, which may coincide with the location of the water snowline during the relevant disk evolutionary stage. Better exoplanet statistics, as well as observations of snowline locations in samples of disks are needed, however, to establish such links between planet formation and snowline locations with confidence. Currently CO and N$_2$ snowlines have only been observed towards a handful of disks, while H$_2$O snowline constraints are more indirect (and fewer), and no constraints exist for other major snowlines. This small number statistic is made worse by a clear bias towards large and bright disks that are not typical, and it is therefore difficult to extrapolate from existing disk snowline data to snowline locations in `typical' planet-forming disks. High-resolution ALMA data towards larger and less biased disk samples would resolve current uncertainties regarding CO and N$_2$ snowline locations, while the path forward for other snowline determinations is less clear and may require a combination of methods development and new facilities.

Snowlines, or rather planet formation locations with respect to snowlines, should also impact planet compositions (\S\ref{sec:ocnsp}). This idea has been used to constrain where in the Solar nebula Jupiter and other planets formed based on their compositions \citep[e.g.][]{Owen99,Lodders04,Morbidelli16,Oberg19,Bosman19}, and to provide an interpretive framework for exoplanet compositions \citep{Madhusudhan19} with a focus on atmospheric C/O ratios  \citep{Oberg11e}. Indeed atmospheric elemental ratios have great potential to trace exoplanet histories, and while C/O ratios cannot on their own be used to assign an unambiguous planet formation location, C/O combined with C/N, C/H and C/S should yield more informative constraints for gas giants \citep{Oberg11e,Piso16,Hobbs22}, while other elemental ratios may be developed to trace the formation of smaller planets. The success of this method will require a more precise understanding of the distribution of elements in disks than is currently available. This entails better models of the interactions between disk dynamics and molecule condensation and sublimation, as well as laboratory data on ice sublimation in different disk contexts \citep[e.g.][]{Fayolle11,Poptapov18,Simon19,Kruczkiewicz21}. In addition, if {\it in situ} chemistry is important for determining the major elemental carriers, then more complex model development will be needed, as well as laboratory experiments on the relevant chemical transformations. 

Most of the current work linking disk and planet compositions has focused on the formation of giant planets and the elemental composition of their atmospheres, but in the near future the study of Earth analog atmospheres will become feasible. The atmospheres and hydrospheres of rocky planets are shaped by a range of processes, including outgassing of magma, the length of a magma ocean phase, plate tectonics \citep[see e.g.][and references therein]{Lichtenberg22}, and impacts of meteorites and comets. The latter two connects the disk molecular inventories with rocky planet compositions, and comprehensive data sets on the distribution of key organics in asteroid and comet-forming disk environments are therefore needed to predict the prebiotic chemistry on young rocky planets. This will include both innovative observational constraints on disk icy reservoirs, such as those provided by \citet{Booth21-so}, and models and laboratory data on the formation and transformation of organic molecules in disks. In the meantime, comparisons of organics in Solar System comets and a handful of protostellar and protoplanetary disks indicate that exoplanets at least sometimes assemble in a similar chemical environment to the solar nebula \citep{Oberg15Natur, Drozdovskaya19,Oberg21-maps}. In addition, constraining the distribution of life-enabling elements like C, N, O and P across disk solids will be crucial for determining their availability to terrestrial exoplanets \citep{Bergin15}. For Earth-analogs it is also important to explore the links between refractory compositions of both the inner disk solids and planet cores and mantles. The planet core composition is difficult to observe in regular planets, but can be probed using  data from extrasolar planetary bodies that have polluted white dwarf atmospheres \citep{Jura14}. 

Finally, isotopic ratios in gas and solids provide a tool to map out the origins of cometary and planetary volatiles. This has so far been almost exclusively applied to the Solar System to constrain the origin of water on Earth and other planets, as well as in comets and asteroids \citep[see][for reviews]{Ceccarelli14,Altwegg19}. More recently the first isotopic ratio in an exoplanet atmosphere has been reported, potentially unlocking isotopic ratios as a complementary tool to elementary ratios when extracting a planet's formation history \citep{Zhang21-exo}. The deployment of this method requires a detailed understanding of the isotopic composition and fractionation chemistry of disks, however, which is currently incomplete. Additional observations, modeling and experiments are all needed to establish a comprehensive interpretative framework for planetary volatile isotopic compositions.

% Summary Points
\begin{summary}[SUMMARY POINTS]
\begin{enumerate}
\item Observations across the electromagnetic spectrum, from UV to radio wavelengths, are needed to characterize disk solid and gas compositions at the radii relevant to planet formation. 
\item Disk chemical models, anchored in astrochemical computations and laboratory experiments, are essential to estimate the complete chemical compositions of disks and how they evolve over time.
\item Disk gas and solid elemental compositions often deviate from those of the their host stars, which has a direct impact on both the disk organic chemistry, and on the elemental and chemical makeup of planets.
\item The disk chemical structure and evolution is set by an interplay between chemical inheritance and a rich {\it in situ} chemistry, and taking account of both is especially important when considering the major carriers of the volatile elements and the organic inventory at different disk radii.
\item There are multiple lines of evidence for substantial volatile transport and mixing in disks, which chemically connect inner and outer disks as well as disk midplanes and atmospheres.
\item Disk chemistry is closely interlinked with disk temperature, and radiation structures, as well as with disk dynamics, which explains the great potential of molecules as probes of non-chemical disk processes.
\end{enumerate}
\end{summary}

% Future Issues
\begin{issues}[FUTURE ISSUES]
\begin{enumerate}
\item During the past decade ALMA disk chemistry observations have mostly focused on a small and highly biased sample of disks, and statistical samples of outer disk chemical compositions are needed to constrain the chemistry of `typical' planet formation and to link disk and exoplanet compositions.
\item Icy grains constitute the major reservoir of O, C, S and P in disks, and IR observations, informed by laboratory experiments, offer a great opportunity to characterize both their compositions and whether they are inherited intact from the ISM or reformed in the disk. 
\item Establishing clear chemodynamical links between different disk regions will require both the realization of new observational probes, and development of 2D (and eventually 3D) models that combine comprehensive chemical networks with a range of gas and dust dynamics and grain evolution. 
\item The gas and solid compositions that are most relevant for terrestrial planet formation are currently not well constrained outside of the Solar System, which should be addressed by spectroscopic observations at the relevant scales and models that connect the compositions of observable disk layers with the planet-forming midplane.
\item Several outstanding questions in disk chemistry studies, including the locations of key snowlines, the distribution of water, and the organic inventory across most of the planet-forming disk, may only be fully addressed by new observatories provide access to the FIR, and higher sensitivity at (sub)millimeter and radio wavelengths.
\end{enumerate}
\end{issues}

%Disclosure
\section*{DISCLOSURE STATEMENT}
 The authors are not aware of any affiliations, memberships, funding, or financial holdings that might be perceived as affecting the objectivity of this review. 

% Acknowledgements
\section*{ACKNOWLEDGMENTS}

We are grateful to Yuri Aikawa, Inga Kamp, and Jennifer Bergner for insightful comments on an early version of the manuscript. We thank Anne Dutrey, Chunhua Qi, Kamber Schwarz, Hideko Nomura and Charles Law for sharing data that have been used in this review. This work was supported by a grant from the Simons Foundation (686302, KI\"O) an award from the Simons Foundation (321183FY19, KI\"O), and  an NSF AAG Grant (\#1907653, KI\"O). D.E.A. acknowledges support from the
Virginia Initiative on Cosmic Origins (VICO) Postdoctoral
Fellowship and Carnegie Postdoctoral Fellowship. S.F. is funded by the European Union under the European Union's Horizon Europe Research \& Innovation Programme 101076613 (UNVEIL). Views and opinions expressed are however those of the author(s) only and do not necessarily reflect those of the European Union or the European Research Council. Neither the European Union nor the granting authority can be held responsible for them.

\bibliography{mybib}{}

\begin{thebibliography}{}
\expandafter\ifx\csname natexlab\endcsname\relax\def\natexlab#1{#1}\fi

\bibitem[{{Acke} et~al.(2004){Acke}, {van den Ancker}, {Dullemond}, {van
  Boekel} \& {Waters}}]{Acke04}
{Acke} B, {van den Ancker} ME, {Dullemond} CP, {van Boekel} R, {Waters} LBFM.
  2004.
\textit{A\&A} 422:621--626

\bibitem[{{{\'A}d{\'a}mkovics} et~al.(2014){{\'A}d{\'a}mkovics}, {Glassgold} \&
  {Najita}}]{Adamkovics14}
{{\'A}d{\'a}mkovics} M, {Glassgold} AE, {Najita} JR. 2014.
\textit{ApJ} 786:135

\bibitem[{{Ag{\'u}ndez} et~al.(2008){Ag{\'u}ndez}, {Cernicharo} \&
  {Goicoechea}}]{Agundez08}
{Ag{\'u}ndez} M, {Cernicharo} J, {Goicoechea} JR. 2008.
\textit{A\&A} 483:831--837

\bibitem[{{Ag{\'u}ndez} et~al.(2018){Ag{\'u}ndez}, {Roueff}, {Le Petit} \& {Le
  Bourlot}}]{Agundez18}
{Ag{\'u}ndez} M, {Roueff} E, {Le Petit} F, {Le Bourlot} J. 2018.
\textit{A\&A} 616:A19

\bibitem[{{Aikawa} et~al.(2021){Aikawa}, {Cataldi}, {Yamato}, {Zhang}, {Booth}
  et~al.}]{Aikawa21}
{Aikawa} Y, {Cataldi} G, {Yamato} Y, {Zhang} K, {Booth} AS, et~al. 2021.
\textit{arXiv e-prints} :arXiv:2109.06419

\bibitem[{{Aikawa} et~al.(2018){Aikawa}, {Furuya}, {Hincelin} \&
  {Herbst}}]{Aikawa18}
{Aikawa} Y, {Furuya} K, {Hincelin} U, {Herbst} E. 2018.
\textit{ApJ} 855:119

\bibitem[{{Aikawa} \& {Herbst}(1999)}]{Aikawa99}
{Aikawa} Y, {Herbst} E. 1999.
\textit{A\&A} 351:233--246

\bibitem[{{Aikawa} \& {Herbst}(2001)}]{Aikawa01}
{Aikawa} Y, {Herbst} E. 2001.
\textit{A\&A} 371:1107--1117

\bibitem[{{Aikawa} et~al.(2012){Aikawa}, {Kamuro}, {Sakon}, {Itoh}, {Terada}
  et~al.}]{Aikawa12}
{Aikawa} Y, {Kamuro} D, {Sakon} I, {Itoh} Y, {Terada} H, et~al. 2012.
\textit{A\&A} 538:A57

\bibitem[{{Aikawa} et~al.(1996){Aikawa}, {Miyama}, {Nakano} \&
  {Umebayashi}}]{Aikawa96}
{Aikawa} Y, {Miyama} SM, {Nakano} T, {Umebayashi} T. 1996.
\textit{ApJ} 467:684

\bibitem[{{Aikawa} et~al.(2003){Aikawa}, {Momose}, {Thi}, {van Zadelhoff}, {Qi}
  et~al.}]{Aikawa03}
{Aikawa} Y, {Momose} M, {Thi} W, {van Zadelhoff} G, {Qi} C, et~al. 2003.
\textit{PASJ} 55:11--15

\bibitem[{{Aikawa} et~al.(2022){Aikawa}, {Okuzumi} \& {Pontoppidan}}]{Aikawa22}
{Aikawa} Y, {Okuzumi} S, {Pontoppidan} K. 2022.
\textit{arXiv e-prints} :arXiv:2212.14529

\bibitem[{{Aikawa} et~al.(2002){Aikawa}, {van Zadelhoff}, {van Dishoeck} \&
  {Herbst}}]{Aikawa02}
{Aikawa} Y, {van Zadelhoff} GJ, {van Dishoeck} EF, {Herbst} E. 2002.
\textit{A\&A} 386:622--632

\bibitem[{{Akimkin} et~al.(2013){Akimkin}, {Zhukovska}, {Wiebe}, {Semenov},
  {Pavlyuchenkov} et~al.}]{Akimkin13}
{Akimkin} V, {Zhukovska} S, {Wiebe} D, {Semenov} D, {Pavlyuchenkov} Y, et~al.
  2013.
\textit{ApJ} 766:8

\bibitem[{{Alarc{\'o}n} et~al.(2020){Alarc{\'o}n}, {Teague}, {Zhang}, {Bergin}
  \& {Barraza-Alfaro}}]{Alarcon20a}
{Alarc{\'o}n} F, {Teague} R, {Zhang} K, {Bergin} EA, {Barraza-Alfaro} M. 2020.
\textit{ApJ} 905:68

\bibitem[{{Altwegg} et~al.(2019){Altwegg}, {Balsiger} \&
  {Fuselier}}]{Altwegg19}
{Altwegg} K, {Balsiger} H, {Fuselier} SA. 2019.
\textit{ARA\&A} 57:113--155

\bibitem[{{Altwegg} et~al.(2020){Altwegg}, {Balsiger}, {H{\"a}nni}, {Rubin},
  {Schuhmann} et~al.}]{Altwegg20}
{Altwegg} K, {Balsiger} H, {H{\"a}nni} N, {Rubin} M, {Schuhmann} M, et~al.
  2020.
\textit{Nature Astronomy} 4:533--540

\bibitem[{{Anderson} et~al.(2019){Anderson}, {Blake}, {Bergin}, {Zhang},
  {Carpenter} et~al.}]{Anderson19}
{Anderson} DE, {Blake} GA, {Bergin} EA, {Zhang} K, {Carpenter} JM, et~al. 2019.
\textit{ApJ} 881:127

\bibitem[{{Anderson} et~al.(2021){Anderson}, {Blake}, {Cleeves}, {Bergin},
  {Zhang} et~al.}]{Anderson21}
{Anderson} DE, {Blake} GA, {Cleeves} LI, {Bergin} EA, {Zhang} K, et~al. 2021.
\textit{ApJ} 909:55

\bibitem[{{Anderson} et~al.(2022){Anderson}, {Cleeves}, {Blake}, {Bergin},
  {Zhang} et~al.}]{Anderson22}
{Anderson} DE, {Cleeves} LI, {Blake} GA, {Bergin} EA, {Zhang} K, et~al. 2022.
\textit{ApJ} 927:229

\bibitem[{{Andrews}(2015)}]{Andrews15}
{Andrews} SM. 2015.
\textit{PASP} 127:961

\bibitem[{{Andrews}(2020)}]{Andrews20}
{Andrews} SM. 2020.
\textit{ARA\&A} 58:483--528

\bibitem[{{Andrews} et~al.(2018){Andrews}, {Huang}, {P{\'e}rez}, {Isella},
  {Dullemond} et~al.}]{Andrews18}
{Andrews} SM, {Huang} J, {P{\'e}rez} LM, {Isella} A, {Dullemond} CP, et~al.
  2018.
\textit{ApJL} 869:L41

\bibitem[{{Ansdell} et~al.(2016){Ansdell}, {Williams}, {van der Marel},
  {Carpenter}, {Guidi} et~al.}]{Ansdell16}
{Ansdell} M, {Williams} JP, {van der Marel} N, {Carpenter} JM, {Guidi} G,
  et~al. 2016.
\textit{ApJ} 828:46

\bibitem[{{Antonellini} et~al.(2016){Antonellini}, {Kamp}, {Lahuis}, {Woitke},
  {Thi} et~al.}]{Antonellini16}
{Antonellini} S, {Kamp} I, {Lahuis} F, {Woitke} P, {Thi} WF, et~al. 2016.
\textit{A\&A} 585:A61

\bibitem[{{Ardila} et~al.(2013){Ardila}, {Herczeg}, {Gregory}, {Ingleby},
  {France} et~al.}]{Ardila2013}
{Ardila} DR, {Herczeg} GJ, {Gregory} SG, {Ingleby} L, {France} K, et~al. 2013.
\textit{ApJs} 207:1

\bibitem[{{Armitage} et~al.(2001){Armitage}, {Livio} \& {Pringle}}]{Armitage01}
{Armitage} PJ, {Livio} M, {Pringle} JE. 2001.
\textit{MNRAS} 324:705--711

\bibitem[{{Arulanantham} et~al.(2021){Arulanantham}, {France}, {Hoadley},
  {Schneider}, {Espaillat} et~al.}]{Arulanatham21}
{Arulanantham} N, {France} K, {Hoadley} K, {Schneider} PC, {Espaillat} CC,
  et~al. 2021.
\textit{AJ} 162:185

\bibitem[{{Bai} \& {Stone}(2013)}]{Bai13}
{Bai} XN, {Stone} JM. 2013.
\textit{ApJ} 769:76

\bibitem[{{Ballering} et~al.(2021){Ballering}, {Cleeves} \&
  {Anderson}}]{Ballering21}
{Ballering} NP, {Cleeves} LI, {Anderson} DE. 2021.
\textit{ApJ} 920:115

\bibitem[{{Banzatti} et~al.(2012){Banzatti}, {Meyer}, {Bruderer}, {Geers},
  {Pascucci} et~al.}]{Banzatti12}
{Banzatti} A, {Meyer} MR, {Bruderer} S, {Geers} V, {Pascucci} I, et~al. 2012.
\textit{ApJ} 745:90

\bibitem[{{Banzatti} et~al.(2020){Banzatti}, {Pascucci}, {Bosman}, {Pinilla},
  {Salyk} et~al.}]{Banzatti20}
{Banzatti} A, {Pascucci} I, {Bosman} AD, {Pinilla} P, {Salyk} C, et~al. 2020.
\textit{ApJ} 903:124

\bibitem[{{Banzatti} et~al.(2015){Banzatti}, {Pinilla}, {Ricci}, {Pontoppidan},
  {Birnstiel} \& {Ciesla}}]{Banzatti15}
{Banzatti} A, {Pinilla} P, {Ricci} L, {Pontoppidan} KM, {Birnstiel} T, {Ciesla}
  F. 2015.
\textit{ApJL} 815:L15

\bibitem[{{Banzatti} et~al.(2022){Banzatti}, {Pontoppidan}, {P{\'e}rez
  Ch{\'a}vez}, {Diehl}, {Salyk} et~al.}]{Banzatti22}
{Banzatti} A, {Pontoppidan} KM, {P{\'e}rez Ch{\'a}vez} J, {Diehl} L, {Salyk} C,
  et~al. 2022.
\textit{arXiv e-prints} :arXiv:2209.08216

\bibitem[{{Bar-Nun} et~al.(1985){Bar-Nun}, {Herman}, {Laufer} \&
  {Rappaport}}]{Bar-Nun85}
{Bar-Nun} A, {Herman} G, {Laufer} D, {Rappaport} ML. 1985.
\textit{Icarus} 63:317--332

\bibitem[{{Barenfeld} et~al.(2016){Barenfeld}, {Carpenter}, {Ricci} \&
  {Isella}}]{Barenfeld16}
{Barenfeld} SA, {Carpenter} JM, {Ricci} L, {Isella} A. 2016.
\textit{ApJ} 827:142

\bibitem[{{Bergin} et~al.(2007){Bergin}, {Aikawa}, {Blake} \& {van
  Dishoeck}}]{Bergin07}
{Bergin} EA, {Aikawa} Y, {Blake} GA, {van Dishoeck} EF. 2007.
\textit{{The Chemical Evolution of Protoplanetary Disks}}. In
  \textit{Protostars and Planets V}, eds. B~{Reipurth}, D~{Jewitt}, K~{Keil}

\bibitem[{{Bergin} et~al.(2023){Bergin}, {Alexander}, {Drozdovskaya},
  {Gounelle} \& {Pfalzner}}]{Bergin23}
{Bergin} EA, {Alexander} C, {Drozdovskaya} M, {Gounelle} M, {Pfalzner} S. 2023.
\textit{arXiv e-prints} :arXiv:2301.05212

\bibitem[{{Bergin} et~al.(2015){Bergin}, {Blake}, {Ciesla}, {Hirschmann} \&
  {Li}}]{Bergin15}
{Bergin} EA, {Blake} GA, {Ciesla} F, {Hirschmann} MM, {Li} J. 2015.
\textit{Proceedings of the National Academy of Science} 112:8965--8970

\bibitem[{{Bergin} et~al.(2013){Bergin}, {Cleeves}, {Gorti}, {Zhang}, {Blake}
  et~al.}]{Bergin13}
{Bergin} EA, {Cleeves} LI, {Gorti} U, {Zhang} K, {Blake} GA, et~al. 2013.
\textit{Nature} 493:644--646

\bibitem[{{Bergin} et~al.(1995){Bergin}, {Langer} \& {Goldsmith}}]{Bergin95}
{Bergin} EA, {Langer} WD, {Goldsmith} PF. 1995.
\textit{ApJ} 441:222--243

\bibitem[{{Bergner} \& {Ciesla}(2021)}]{Bergner21-model}
{Bergner} JB, {Ciesla} F. 2021.
\textit{ApJ} 919:45

\bibitem[{{Bergner} et~al.(2018){Bergner}, {Guzm{\'a}n}, {{\"O}berg}, {Loomis}
  \& {Pegues}}]{Bergner18}
{Bergner} JB, {Guzm{\'a}n} VG, {{\"O}berg} KI, {Loomis} RA, {Pegues} J. 2018.
\textit{ApJ} 857:69

\bibitem[{{Bergner} et~al.(2020){Bergner}, {{\"O}berg}, {Bergin}, {Andrews},
  {Blake} et~al.}]{Bergner20}
{Bergner} JB, {{\"O}berg} KI, {Bergin} EA, {Andrews} SM, {Blake} GA, et~al.
  2020.
\textit{ApJ} 898:97

\bibitem[{{Bergner} et~al.(2019){Bergner}, {{\"O}berg}, {Bergin}, {Loomis},
  {Pegues} \& {Qi}}]{Bergner19}
{Bergner} JB, {{\"O}berg} KI, {Bergin} EA, {Loomis} RA, {Pegues} J, {Qi} C.
  2019.
\textit{ApJ} 876:25

\bibitem[{{Bergner} et~al.(2021){Bergner}, {{\"O}berg}, {Guzm{\'a}n}, {Law},
  {Loomis} et~al.}]{Bergner21}
{Bergner} JB, {{\"O}berg} KI, {Guzm{\'a}n} VV, {Law} CJ, {Loomis} RA, et~al.
  2021.
\textit{ApJs} 257:11

\bibitem[{{Bernstein} et~al.(2002){Bernstein}, {Dworkin}, {Sandford}, {Cooper}
  \& {Allamandola}}]{Bernstein02}
{Bernstein} MP, {Dworkin} JP, {Sandford} SA, {Cooper} GW, {Allamandola} LJ.
  2002.
\textit{Nature} 416:401--403

\bibitem[{{Bethell} \& {Bergin}(2011)}]{Bethell11a}
{Bethell} TJ, {Bergin} EA. 2011.
\textit{ApJ} 739:78

\bibitem[{{Bjerkeli} et~al.(2016){Bjerkeli}, {J{\o}rgensen}, {Bergin},
  {Frimann}, {Harsono} et~al.}]{Bjerkeli16}
{Bjerkeli} P, {J{\o}rgensen} JK, {Bergin} EA, {Frimann} S, {Harsono} D, et~al.
  2016.
\textit{A\&A} 595:A39

\bibitem[{{Blevins} et~al.(2016){Blevins}, {Pontoppidan}, {Banzatti}, {Zhang},
  {Najita} et~al.}]{Blevins16}
{Blevins} SM, {Pontoppidan} KM, {Banzatti} A, {Zhang} K, {Najita} JR, et~al.
  2016.
\textit{ApJ} 818:22

\bibitem[{{Bockel{\'e}e-Morvan} et~al.(2015){Bockel{\'e}e-Morvan}, {Calmonte},
  {Charnley}, {Duprat}, {Engrand} et~al.}]{Bockelee15}
{Bockel{\'e}e-Morvan} D, {Calmonte} U, {Charnley} S, {Duprat} J, {Engrand} C,
  et~al. 2015.
\textit{SSRv} 197:47--83

\bibitem[{{Boogert} et~al.(2015){Boogert}, {Gerakines} \&
  {Whittet}}]{Boogert15}
{Boogert} ACA, {Gerakines} PA, {Whittet} DCB. 2015.
\textit{ARA\&A} 53:541--581

\bibitem[{{Booth} et~al.(2021{\natexlab{a}}){Booth}, {van der Marel},
  {Leemker}, {van Dishoeck} \& {Ohashi}}]{Booth21-so}
{Booth} AS, {van der Marel} N, {Leemker} M, {van Dishoeck} EF, {Ohashi} S.
  2021{\natexlab{a}}.
\textit{A\&A} 651:L6

\bibitem[{{Booth} et~al.(2019){Booth}, {Walsh}, {Ilee}, {Notsu}, {Qi}
  et~al.}]{Booth19}
{Booth} AS, {Walsh} C, {Ilee} JD, {Notsu} S, {Qi} C, et~al. 2019.
\textit{ApJL} 882:L31

\bibitem[{{Booth} et~al.(2021{\natexlab{b}}){Booth}, {Walsh}, {Terwisscha van
  Scheltinga}, {van Dishoeck}, {Ilee} et~al.}]{Booth21}
{Booth} AS, {Walsh} C, {Terwisscha van Scheltinga} J, {van Dishoeck} EF, {Ilee}
  JD, et~al. 2021{\natexlab{b}}.
\textit{Nature Astronomy}

\bibitem[{{Booth} \& {Ilee}(2019)}]{Booth_R19}
{Booth} RA, {Ilee} JD. 2019.
\textit{MNRAS} 487:3998--4011

\bibitem[{{Bosman} et~al.(2021{\natexlab{a}}){Bosman}, {Alarc{\'o}n}, {Bergin},
  {Zhang}, {van't Hoff} et~al.}]{Bosman21_VII}
{Bosman} AD, {Alarc{\'o}n} F, {Bergin} EA, {Zhang} K, {van't Hoff} MLR, et~al.
  2021{\natexlab{a}}.
\textit{ApJs} 257:7

\bibitem[{{Bosman} et~al.(2021{\natexlab{b}}){Bosman}, {Alarc{\'o}n}, {Zhang}
  \& {Bergin}}]{Bosman21_c2h}
{Bosman} AD, {Alarc{\'o}n} F, {Zhang} K, {Bergin} EA. 2021{\natexlab{b}}.
\textit{ApJ} 910:3

\bibitem[{{Bosman} et~al.(2021{\natexlab{c}}){Bosman}, {Bergin}, {Loomis},
  {Andrews}, {van't Hoff} et~al.}]{Bosman21_XV}
{Bosman} AD, {Bergin} EA, {Loomis} RA, {Andrews} SM, {van't Hoff} MLR, et~al.
  2021{\natexlab{c}}.
\textit{ApJs} 257:15

\bibitem[{{Bosman} et~al.(2017){Bosman}, {Bruderer} \& {van
  Dishoeck}}]{Bosman17}
{Bosman} AD, {Bruderer} S, {van Dishoeck} EF. 2017.
\textit{A\&A} 601:A36

\bibitem[{{Bosman} et~al.(2019){Bosman}, {Cridland} \& {Miguel}}]{Bosman19}
{Bosman} AD, {Cridland} AJ, {Miguel} Y. 2019.
\textit{A\&A} 632:L11

\bibitem[{{Bosman} et~al.(2018){Bosman}, {Walsh} \& {van Dishoeck}}]{Bosman18}
{Bosman} AD, {Walsh} C, {van Dishoeck} EF. 2018.
\textit{A\&A} 618:A182

\bibitem[{{Bouwman} et~al.(2008){Bouwman}, {Henning}, {Hillenbrand}, {Meyer},
  {Pascucci} et~al.}]{Bouwman08}
{Bouwman} J, {Henning} T, {Hillenbrand} LA, {Meyer} MR, {Pascucci} I, et~al.
  2008.
\textit{ApJ} 683:479--498

\bibitem[{{Brinch} \& {Hogerheijde}(2010)}]{Brinch10}
{Brinch} C, {Hogerheijde} MR. 2010.
\textit{A\&A} 523:A25

\bibitem[{{Brittain} et~al.(2007){Brittain}, {Simon}, {Najita} \&
  {Rettig}}]{Brittain07}
{Brittain} SD, {Simon} T, {Najita} JR, {Rettig} TW. 2007.
\textit{ApJ} 659:685--704

\bibitem[{{Bruderer}(2013)}]{Bruderer13}
{Bruderer} S. 2013.
\textit{A\&A} 559:A46

\bibitem[{{Bruderer} et~al.(2015){Bruderer}, {Harsono} \& {van
  Dishoeck}}]{Bruderer15}
{Bruderer} S, {Harsono} D, {van Dishoeck} EF. 2015.
\textit{A\&A} 575:A94

\bibitem[{{Bruderer} et~al.(2012){Bruderer}, {van Dishoeck}, {Doty} \&
  {Herczeg}}]{Bruderer12}
{Bruderer} S, {van Dishoeck} EF, {Doty} SD, {Herczeg} GJ. 2012.
\textit{A\&A} 541:A91

\bibitem[{{Brunken} et~al.(2022){Brunken}, {Booth}, {Leemker}, {Nazari}, {van
  der Marel} \& {van Dishoeck}}]{Brunken22}
{Brunken} NGC, {Booth} AS, {Leemker} M, {Nazari} P, {van der Marel} N, {van
  Dishoeck} EF. 2022.
\textit{A\&A} 659:A29

\bibitem[{{Calahan} et~al.(2021){Calahan}, {Bergin}, {Zhang}, {Schwarz},
  {{\"O}berg} et~al.}]{Calahan21}
{Calahan} JK, {Bergin} EA, {Zhang} K, {Schwarz} KR, {{\"O}berg} KI, et~al.
  2021.
\textit{ApJs} 257:17

\bibitem[{{Canta} et~al.(2021){Canta}, {Teague}, {Le Gal} \&
  {{\"O}berg}}]{Canta21}
{Canta} A, {Teague} R, {Le Gal} R, {{\"O}berg} KI. 2021.
\textit{ApJ} 922:62

\bibitem[{{Carr} \& {Najita}(2008)}]{Carr08}
{Carr} JS, {Najita} JR. 2008.
\textit{Science} 319:1504--

\bibitem[{{Carr} \& {Najita}(2011)}]{Carr11}
{Carr} JS, {Najita} JR. 2011.
\textit{ApJ} 733:102

\bibitem[{Carr et~al.(2018)Carr, Najita \& Salyk}]{Carr18}
Carr JS, Najita JR, Salyk C. 2018.
\textit{RNAAS} 2:169

\bibitem[{{Casassus} et~al.(2021){Casassus}, {Christiaens}, {C{\'a}rcamo},
  {P{\'e}rez}, {Weber} et~al.}]{Casassus21}
{Casassus} S, {Christiaens} V, {C{\'a}rcamo} M, {P{\'e}rez} S, {Weber} P,
  et~al. 2021.
\textit{MNRAS} 507:3789--3809

\bibitem[{{Casassus} \& {P{\'e}rez}(2019)}]{Casassus19}
{Casassus} S, {P{\'e}rez} S. 2019.
\textit{ApJL} 883:L41

\bibitem[{{Cataldi} et~al.(2021){Cataldi}, {Yamato}, {Aikawa}, {Bergner},
  {Furuya} et~al.}]{Cataldi21}
{Cataldi} G, {Yamato} Y, {Aikawa} Y, {Bergner} JB, {Furuya} K, et~al. 2021.
\textit{ApJs} 257:10

\bibitem[{{Cazzoletti} et~al.(2018){Cazzoletti}, {van Dishoeck}, {Visser},
  {Facchini} \& {Bruderer}}]{Cazzoletti18}
{Cazzoletti} P, {van Dishoeck} EF, {Visser} R, {Facchini} S, {Bruderer} S.
  2018.
\textit{A\&A} 609:A93

\bibitem[{{Ceccarelli} et~al.(2014){Ceccarelli}, {Caselli},
  {Bockel{\'e}e-Morvan}, {Mousis}, {Pizzarello} et~al.}]{Ceccarelli14}
{Ceccarelli} C, {Caselli} P, {Bockel{\'e}e-Morvan} D, {Mousis} O, {Pizzarello}
  S, et~al. 2014.
\textit{{Deuterium Fractionation: The Ariadne's Thread from the Precollapse
  Phase to Meteorites and Comets Today}}. In \textit{Protostars and Planets
  VI}, eds. H~{Beuther}, RS~{Klessen}, CP~{Dullemond}, T~{Henning}

\bibitem[{{Chapillon} et~al.(2012){Chapillon}, {Dutrey}, {Guilloteau},
  {Pi{\'e}tu}, {Wakelam} et~al.}]{Chapillon12}
{Chapillon} E, {Dutrey} A, {Guilloteau} S, {Pi{\'e}tu} V, {Wakelam} V, et~al.
  2012.
\textit{ApJ} 756:58

\bibitem[{{Chiang} \& {Youdin}(2010)}]{Chiang10}
{Chiang} E, {Youdin} AN. 2010.
\textit{Annual Review of Earth and Planetary Sciences} 38:493--522

\bibitem[{{Chiang} \& {Goldreich}(1997)}]{Chiang97}
{Chiang} EI, {Goldreich} P. 1997.
\textit{ApJ} 490:368

\bibitem[{{Chuang} et~al.(2017){Chuang}, {Fedoseev}, {Qasim}, {Ioppolo}, {van
  Dishoeck} \& {Linnartz}}]{Chuang17}
{Chuang} KJ, {Fedoseev} G, {Qasim} D, {Ioppolo} S, {van Dishoeck} EF,
  {Linnartz} H. 2017.
\textit{MNRAS} 467:2552--2565

\bibitem[{{Ciesla} \& {Cuzzi}(2006)}]{Ciesla06}
{Ciesla} FJ, {Cuzzi} JN. 2006.
\textit{Icarus} 181:178--204

\bibitem[{{Ciesla} \& {Sandford}(2012)}]{Ciesla12}
{Ciesla} FJ, {Sandford} SA. 2012.
\textit{Science} 336:452

\bibitem[{{Cieza} et~al.(2016){Cieza}, {Casassus}, {Tobin}, {Bos}, {Williams}
  et~al.}]{Cieza16}
{Cieza} LA, {Casassus} S, {Tobin} J, {Bos} SP, {Williams} JP, et~al. 2016.
\textit{Nature} 535:258--261

\bibitem[{{Cleeves}(2016)}]{Cleeves16a}
{Cleeves} LI. 2016.
\textit{ApJL} 816:L21

\bibitem[{{Cleeves} et~al.(2014){Cleeves}, {Bergin}, {Alexander}, {Du},
  {Graninger} et~al.}]{Cleeves14}
{Cleeves} LI, {Bergin} EA, {Alexander} CMO, {Du} F, {Graninger} D, et~al. 2014.
\textit{Science} 345:1590--1593

\bibitem[{{Cleeves} et~al.(2015{\natexlab{a}}){Cleeves}, {Bergin} \&
  {Harries}}]{Cleeves15b}
{Cleeves} LI, {Bergin} EA, {Harries} TJ. 2015{\natexlab{a}}.
\textit{ApJ} 807:2

\bibitem[{{Cleeves} et~al.(2017){Cleeves}, {Bergin}, {{\"O}berg}, {Andrews},
  {Wilner} \& {Loomis}}]{Cleeves17}
{Cleeves} LI, {Bergin} EA, {{\"O}berg} KI, {Andrews} S, {Wilner} D, {Loomis} R.
  2017.
\textit{ApJL} 843:L3

\bibitem[{{Cleeves} et~al.(2015{\natexlab{b}}){Cleeves}, {Bergin}, {Qi},
  {Adams} \& {{\"O}berg}}]{Cleeves15}
{Cleeves} LI, {Bergin} EA, {Qi} C, {Adams} FC, {{\"O}berg} KI.
  2015{\natexlab{b}}.
\textit{ApJ} 799:204

\bibitem[{{Cleeves} et~al.(2018){Cleeves}, {{\"O}berg}, {Wilner}, {Huang},
  {Loomis} et~al.}]{Cleeves18}
{Cleeves} LI, {{\"O}berg} KI, {Wilner} DJ, {Huang} J, {Loomis} RA, et~al. 2018.
\textit{ApJ} 865:155

\bibitem[{{Collings} et~al.(2004){Collings}, {Anderson}, {Chen}, {Dever},
  {Viti} et~al.}]{Collings04}
{Collings} MP, {Anderson} MA, {Chen} R, {Dever} JW, {Viti} S, et~al. 2004.
\textit{MNRAS} 354:1133--1140

\bibitem[{{Collings} et~al.(2003){Collings}, {Dever}, {Fraser} \&
  {McCoustra}}]{Collings03}
{Collings} MP, {Dever} JW, {Fraser} HJ, {McCoustra} MRS. 2003.
\textit{Ap\&SS} 285:633--659

\bibitem[{{Cridland} et~al.(2020){Cridland}, {van Dishoeck}, {Alessi} \&
  {Pudritz}}]{Cridland20}
{Cridland} AJ, {van Dishoeck} EF, {Alessi} M, {Pudritz} RE. 2020.
\textit{A\&A} 642:A229

\bibitem[{{Cuzzi} \& {Zahnle}(2004)}]{Cuzzi04}
{Cuzzi} JN, {Zahnle} KJ. 2004.
\textit{ApJ} 614:490--496

\bibitem[{{D'Alessio} et~al.(2006){D'Alessio}, {Calvet}, {Hartmann},
  {Franco-Hern{\'a}ndez} \& {Serv{\'{\i}}n}}]{dAlessio06}
{D'Alessio} P, {Calvet} N, {Hartmann} L, {Franco-Hern{\'a}ndez} R,
  {Serv{\'{\i}}n} H. 2006.
\textit{ApJ} 638:314--335

\bibitem[{{D'Alessio} et~al.(1999){D'Alessio}, {Calvet}, {Hartmann}, {Lizano}
  \& {Cant{\'o}}}]{dAlessio99}
{D'Alessio} P, {Calvet} N, {Hartmann} L, {Lizano} S, {Cant{\'o}} J. 1999.
\textit{ApJ} 527:893--909

\bibitem[{{Dartois} et~al.(2003){Dartois}, {Dutrey} \&
  {Guilloteau}}]{Dartois03}
{Dartois} E, {Dutrey} A, {Guilloteau} S. 2003.
\textit{A\&A} 399:773--787

\bibitem[{{Disk Dynamics Collaboration} et~al.(2020){Disk Dynamics
  Collaboration}, {Armitage}, {Bae}, {Benisty}, {Bergin}
  et~al.}]{DiskDynamics20}
{Disk Dynamics Collaboration}, {Armitage} PJ, {Bae} J, {Benisty} M, {Bergin}
  EA, et~al. 2020.
\textit{arXiv e-prints} :arXiv:2009.04345

\bibitem[{{Drake} et~al.(2005){Drake}, {Testa} \& {Hartmann}}]{Drake2005}
{Drake} JJ, {Testa} P, {Hartmann} L. 2005.
\textit{ApJL} 627:L149--L152

\bibitem[{{Drazkowska} et~al.(2022){Drazkowska}, {Bitsch}, {Lambrechts},
  {Mulders}, {Harsono} et~al.}]{Drazkowska22}
{Drazkowska} J, {Bitsch} B, {Lambrechts} M, {Mulders} GD, {Harsono} D, et~al.
  2022.
\textit{arXiv e-prints} :arXiv:2203.09759

\bibitem[{{Drozdovskaya} et~al.(2019){Drozdovskaya}, {van Dishoeck}, {Rubin},
  {J{\o}rgensen} \& {Altwegg}}]{Drozdovskaya19}
{Drozdovskaya} MN, {van Dishoeck} EF, {Rubin} M, {J{\o}rgensen} JK, {Altwegg}
  K. 2019.
\textit{MNRAS} 490:50--79

\bibitem[{{Drozdovskaya} et~al.(2016){Drozdovskaya}, {Walsh}, {van Dishoeck},
  {Furuya}, {Marboeuf} et~al.}]{Drozdovskaya16}
{Drozdovskaya} MN, {Walsh} C, {van Dishoeck} EF, {Furuya} K, {Marboeuf} U,
  et~al. 2016.
\textit{MNRAS} 462:977--993

\bibitem[{{Du} \& {Bergin}(2014)}]{Du14}
{Du} F, {Bergin} EA. 2014.
\textit{ApJ} 792:2

\bibitem[{{Du} et~al.(2015){Du}, {Bergin} \& {Hogerheijde}}]{Du15}
{Du} F, {Bergin} EA, {Hogerheijde} MR. 2015.
\textit{ApJL} 807:L32

\bibitem[{{Dutrey} et~al.(1996){Dutrey}, {Guilloteau}, {Duvert}, {Prato},
  {Simon} et~al.}]{Dutrey96}
{Dutrey} A, {Guilloteau} S, {Duvert} G, {Prato} L, {Simon} M, et~al. 1996.
\textit{A\&A} 309:493--504

\bibitem[{{Dutrey} et~al.(2008){Dutrey}, {Guilloteau}, {Pi{\'e}tu},
  {Chapillon}, {Gueth} et~al.}]{Dutrey08}
{Dutrey} A, {Guilloteau} S, {Pi{\'e}tu} V, {Chapillon} E, {Gueth} F, et~al.
  2008.
\textit{A\&A} 490:L15--L18

\bibitem[{{Dutrey} et~al.(2017){Dutrey}, {Guilloteau}, {Pi{\'e}tu},
  {Chapillon}, {Wakelam} et~al.}]{Dutrey17}
{Dutrey} A, {Guilloteau} S, {Pi{\'e}tu} V, {Chapillon} E, {Wakelam} V, et~al.
  2017.
\textit{A\&A} 607:A130

\bibitem[{{Dutrey} et~al.(2007){Dutrey}, {Henning}, {Guilloteau}, {Semenov},
  {Pi{\'e}tu} et~al.}]{Dutrey07}
{Dutrey} A, {Henning} T, {Guilloteau} S, {Semenov} D, {Pi{\'e}tu} V, et~al.
  2007.
\textit{A\&A} 464:615--623

\bibitem[{{Dutrey} et~al.(2014){Dutrey}, {Semenov}, {Chapillon}, {Gorti},
  {Guilloteau} et~al.}]{Dutrey14}
{Dutrey} A, {Semenov} D, {Chapillon} E, {Gorti} U, {Guilloteau} S, et~al. 2014.
\textit{{Physical and Chemical Structure of Planet-Forming Disks Probed by
  Millimeter Observations and Modeling}}. In \textit{Protostars and Planets
  VI}, eds. H~{Beuther}, RS~{Klessen}, CP~{Dullemond}, T~{Henning}

\bibitem[{{Duval} et~al.(2022){Duval}, {Bosman} \& {Bergin}}]{Duval22}
{Duval} SE, {Bosman} AD, {Bergin} EA. 2022.
\textit{ApJL} 934:L25

\bibitem[{{Eistrup} \& {Henning}(2022)}]{Eistrup22}
{Eistrup} C, {Henning} T. 2022.
\textit{arXiv e-prints} :arXiv:2208.07390

\bibitem[{{Eistrup} et~al.(2016){Eistrup}, {Walsh} \& {van
  Dishoeck}}]{Eistrup16}
{Eistrup} C, {Walsh} C, {van Dishoeck} EF. 2016.
\textit{A\&A} 595:A83

\bibitem[{{Endres} et~al.(2016){Endres}, {Schlemmer}, {Schilke}, {Stutzki} \&
  {M{\"u}ller}}]{Endres16}
{Endres} CP, {Schlemmer} S, {Schilke} P, {Stutzki} J, {M{\"u}ller} HSP. 2016.
\textit{Journal of Molecular Spectroscopy} 327:95--104

\bibitem[{{Facchini} et~al.(2017){Facchini}, {Birnstiel}, {Bruderer} \& {van
  Dishoeck}}]{Facchini17}
{Facchini} S, {Birnstiel} T, {Bruderer} S, {van Dishoeck} EF. 2017.
\textit{A\&A} 605:A16

\bibitem[{{Facchini} et~al.(2018{\natexlab{a}}){Facchini}, {Juh{\'a}sz} \&
  {Lodato}}]{Facchini18b}
{Facchini} S, {Juh{\'a}sz} A, {Lodato} G. 2018{\natexlab{a}}.
\textit{MNRAS} 473:4459--4475

\bibitem[{{Facchini} et~al.(2016){Facchini}, {Manara}, {Schneider}, {Clarke},
  {Bouvier} et~al.}]{Facchini16}
{Facchini} S, {Manara} CF, {Schneider} PC, {Clarke} CJ, {Bouvier} J, et~al.
  2016.
\textit{A\&A} 596:A38

\bibitem[{{Facchini} et~al.(2018{\natexlab{b}}){Facchini}, {Pinilla}, {van
  Dishoeck} \& {de Juan Ovelar}}]{Facchini18}
{Facchini} S, {Pinilla} P, {van Dishoeck} EF, {de Juan Ovelar} M.
  2018{\natexlab{b}}.
\textit{A\&A} 612:A104

\bibitem[{{Facchini} et~al.(2021){Facchini}, {Teague}, {Bae}, {Benisty},
  {Keppler} \& {Isella}}]{Facchini21}
{Facchini} S, {Teague} R, {Bae} J, {Benisty} M, {Keppler} M, {Isella} A. 2021.
\textit{AJ} 162:99

\bibitem[{{Faure} et~al.(2015){Faure}, {Quirico}, {Faure}, {Schmitt},
  {Theul{\'e}} \& {Marboeuf}}]{Faure15b}
{Faure} M, {Quirico} E, {Faure} A, {Schmitt} B, {Theul{\'e}} P, {Marboeuf} U.
  2015.
\textit{Icarus} 261:14--30

\bibitem[{{Fayolle} et~al.(2016){Fayolle}, {Balfe}, {Loomis}, {Bergner},
  {Graninger} et~al.}]{Fayolle16}
{Fayolle} EC, {Balfe} J, {Loomis} R, {Bergner} J, {Graninger} D, et~al. 2016.
\textit{ApJL} 816:L28

\bibitem[{{Fayolle} et~al.(2011){Fayolle}, {Bertin}, {Romanzin}, {Michaut},
  {{\"O}berg} et~al.}]{Fayolle11}
{Fayolle} EC, {Bertin} M, {Romanzin} C, {Michaut} X, {{\"O}berg} KI, et~al.
  2011.
\textit{ApJL} 739:L36

\bibitem[{{Fedele} et~al.(2013){Fedele}, {Bruderer}, {van Dishoeck}, {Carr},
  {Herczeg} et~al.}]{Fedele13}
{Fedele} D, {Bruderer} S, {van Dishoeck} EF, {Carr} J, {Herczeg} GJ, et~al.
  2013.
\textit{A\&A} 559:A77

\bibitem[{{Fedele} \& {Favre}(2020)}]{Fedele20}
{Fedele} D, {Favre} C. 2020.
\textit{A\&A} 638:A110

\bibitem[{{Fedele} et~al.(2016){Fedele}, {van Dishoeck}, {Kama}, {Bruderer} \&
  {Hogerheijde}}]{Fedele16}
{Fedele} D, {van Dishoeck} EF, {Kama} M, {Bruderer} S, {Hogerheijde} MR. 2016.
\textit{A\&A} 591:A95

\bibitem[{{Fernandes} et~al.(2019){Fernandes}, {Mulders}, {Pascucci},
  {Mordasini} \& {Emsenhuber}}]{Fernandes19}
{Fernandes} RB, {Mulders} GD, {Pascucci} I, {Mordasini} C, {Emsenhuber} A.
  2019.
\textit{ApJ} 874:81

\bibitem[{{France} et~al.(2012){France}, {Schindhelm}, {Herczeg}, {Brown},
  {Abgrall} et~al.}]{France2012}
{France} K, {Schindhelm} R, {Herczeg} GJ, {Brown} A, {Abgrall} H, et~al. 2012.
\textit{ApJ} 756:171

\bibitem[{{Fuchs} et~al.(2009){Fuchs}, {Cuppen}, {Ioppolo}, {Romanzin},
  {Bisschop} et~al.}]{Fuchs09}
{Fuchs} GW, {Cuppen} HM, {Ioppolo} S, {Romanzin} C, {Bisschop} SE, et~al. 2009.
\textit{A\&A} 505:629--639

\bibitem[{{Fuente} et~al.(2010){Fuente}, {Cernicharo}, {Ag{\'u}ndez},
  {Bern{\'e}}, {Goicoechea} et~al.}]{Fuente10}
{Fuente} A, {Cernicharo} J, {Ag{\'u}ndez} M, {Bern{\'e}} O, {Goicoechea} JR,
  et~al. 2010.
\textit{A\&A} 524:A19

\bibitem[{{Furuya} \& {Aikawa}(2014)}]{Furuya14}
{Furuya} K, {Aikawa} Y. 2014.
\textit{ApJ} 790:97

\bibitem[{{Furuya} et~al.(2022{\natexlab{a}}){Furuya}, {Lee} \&
  {Nomura}}]{Furuya22b}
{Furuya} K, {Lee} S, {Nomura} H. 2022{\natexlab{a}}.
\textit{ApJ} 938:29

\bibitem[{{Furuya} et~al.(2022{\natexlab{b}}){Furuya}, {Tsukagoshi}, {Qi},
  {Nomura}, {Cleeves} et~al.}]{Furuya22}
{Furuya} K, {Tsukagoshi} T, {Qi} C, {Nomura} H, {Cleeves} LI, et~al.
  2022{\natexlab{b}}.
\textit{ApJ} 926:148

\bibitem[{{Gavino} et~al.(2021){Gavino}, {Dutrey}, {Wakelam}, {Guilloteau},
  {Kobus} et~al.}]{Gavino21}
{Gavino} S, {Dutrey} A, {Wakelam} V, {Guilloteau} S, {Kobus} J, et~al. 2021.
\textit{A\&A} 654:A65

\bibitem[{{Geers} et~al.(2006){Geers}, {Augereau}, {Pontoppidan}, {Dullemond},
  {Visser} et~al.}]{Geers06}
{Geers} VC, {Augereau} JC, {Pontoppidan} KM, {Dullemond} CP, {Visser} R, et~al.
  2006.
\textit{A\&A} 459:545--556

\bibitem[{{Gibb} et~al.(2007){Gibb}, {Van Brunt}, {Brittain} \&
  {Rettig}}]{Gibb07}
{Gibb} EL, {Van Brunt} KA, {Brittain} SD, {Rettig} TW. 2007.
\textit{ApJ} 660:1572--1579

\bibitem[{{Glassgold} et~al.(2009){Glassgold}, {Meijerink} \&
  {Najita}}]{Glassgold09}
{Glassgold} AE, {Meijerink} R, {Najita} JR. 2009.
\textit{ApJ} 701:142--153

\bibitem[{{Glassgold} et~al.(2004){Glassgold}, {Najita} \&
  {Igea}}]{Glassgold04}
{Glassgold} AE, {Najita} J, {Igea} J. 2004.
\textit{ApJ} 615:972--990

\bibitem[{{Goldsmith} \& {Langer}(1999)}]{Goldsmith99}
{Goldsmith} PF, {Langer} WD. 1999.
\textit{ApJ} 517:209--225

\bibitem[{{Gordon} et~al.(2022){Gordon}, {Rothman}, {Hargreaves}, {Hashemi},
  {Karlovets} et~al.}]{Gordon22}
{Gordon} IE, {Rothman} LS, {Hargreaves} RJ, {Hashemi} R, {Karlovets} EV, et~al.
  2022.
\textit{JQSRT} 277:107949

\bibitem[{{Graninger} et~al.(2015){Graninger}, {{\"O}berg}, {Qi} \&
  {Kastner}}]{Graninger15}
{Graninger} D, {{\"O}berg} KI, {Qi} C, {Kastner} J. 2015.
\textit{ApJL} 807:L15

\bibitem[{{Grant} et~al.(2022){Grant}, {van Dishoeck}, {Tabone}, {Gasman},
  {Henning} et~al.}]{Grant22}
{Grant} SL, {van Dishoeck} EF, {Tabone} B, {Gasman} D, {Henning} T, et~al.
  2022.
\textit{arXiv e-prints} :arXiv:2212.08047

\bibitem[{{Gressel} et~al.(2012){Gressel}, {Nelson} \& {Turner}}]{Gressel12}
{Gressel} O, {Nelson} RP, {Turner} NJ. 2012.
\textit{MNRAS} 422:1140--1159

\bibitem[{{Guilloteau} et~al.(2013){Guilloteau}, {Di Folco}, {Dutrey}, {Simon},
  {Grosso} \& {Pi{\'e}tu}}]{Guilloteau13}
{Guilloteau} S, {Di Folco} E, {Dutrey} A, {Simon} M, {Grosso} N, {Pi{\'e}tu} V.
  2013.
\textit{A\&A} 549:A92

\bibitem[{{Guilloteau} et~al.(2016){Guilloteau}, {Reboussin}, {Dutrey},
  {Chapillon}, {Wakelam} et~al.}]{Guilloteau16}
{Guilloteau} S, {Reboussin} L, {Dutrey} A, {Chapillon} E, {Wakelam} V, et~al.
  2016.
\textit{A\&A} 592:A124

\bibitem[{{Gundlach} \& {Blum}(2015)}]{Gundlach15}
{Gundlach} B, {Blum} J. 2015.
\textit{ApJ} 798:34

\bibitem[{{G{\"u}nther} et~al.(2018){G{\"u}nther}, {Birnstiel}, {Huenemoerder},
  {Principe}, {Schneider} et~al.}]{Gunther2018}
{G{\"u}nther} HM, {Birnstiel} T, {Huenemoerder} DP, {Principe} DA, {Schneider}
  PC, et~al. 2018.
\textit{AJ} 156:56

\bibitem[{{Guzm{\'a}n} et~al.(2021){Guzm{\'a}n}, {Bergner}, {Law}, {{\"O}berg},
  {Walsh} et~al.}]{Guzman21}
{Guzm{\'a}n} VV, {Bergner} JB, {Law} CJ, {{\"O}berg} KI, {Walsh} C, et~al.
  2021.
\textit{ApJs} 257:6

\bibitem[{{Guzm{\'a}n} et~al.(2018){Guzm{\'a}n}, {Huang}, {Andrews}, {Isella},
  {P{\'e}rez} et~al.}]{Guzman18}
{Guzm{\'a}n} VV, {Huang} J, {Andrews} SM, {Isella} A, {P{\'e}rez} LM, et~al.
  2018.
\textit{ApJL} 869:L48

\bibitem[{{Guzm{\'a}n} et~al.(2017){Guzm{\'a}n}, {{\"O}berg}, {Huang}, {Loomis}
  \& {Qi}}]{Guzman17}
{Guzm{\'a}n} VV, {{\"O}berg} KI, {Huang} J, {Loomis} R, {Qi} C. 2017.
\textit{ApJ} 836:30

\bibitem[{{Hacar} et~al.(2020){Hacar}, {Bosman} \& {van Dishoeck}}]{Hacar20}
{Hacar} A, {Bosman} AD, {van Dishoeck} EF. 2020.
\textit{A\&A} 635:A4

\bibitem[{{Hartmann} et~al.(1998){Hartmann}, {Calvet}, {Gullbring} \&
  {D'Alessio}}]{Hartmann98}
{Hartmann} L, {Calvet} N, {Gullbring} E, {D'Alessio} P. 1998.
\textit{ApJ} 495:385

\bibitem[{{Heays} et~al.(2014){Heays}, {Visser}, {Gredel}, {Ubachs}, {Lewis}
  et~al.}]{Heays14}
{Heays} AN, {Visser} R, {Gredel} R, {Ubachs} W, {Lewis} BR, et~al. 2014.
\textit{A\&A} 562:A61

\bibitem[{{Henning} \& {Semenov}(2013)}]{Henning13}
{Henning} T, {Semenov} D. 2013.
\textit{Chemical Reviews} 113:9016--9042

\bibitem[{{Henning} et~al.(2010){Henning}, {Semenov}, {Guilloteau}, {Dutrey},
  {Hersant} et~al.}]{Henning10}
{Henning} T, {Semenov} D, {Guilloteau} S, {Dutrey} A, {Hersant} F, et~al. 2010.
\textit{ApJ} 714:1511--1520

\bibitem[{{Hersant} et~al.(2009){Hersant}, {Wakelam}, {Dutrey}, {Guilloteau} \&
  {Herbst}}]{Hersant09}
{Hersant} F, {Wakelam} V, {Dutrey} A, {Guilloteau} S, {Herbst} E. 2009.
\textit{A\&A} 493:L49--L52

\bibitem[{{Hidaka} et~al.(2004){Hidaka}, {Watanabe}, {Shiraki}, {Nagaoka} \&
  {Kouchi}}]{Hidaka04}
{Hidaka} H, {Watanabe} N, {Shiraki} T, {Nagaoka} A, {Kouchi} A. 2004.
\textit{ApJ} 614:1124--1131

\bibitem[{{Hily-Blant} et~al.(2017){Hily-Blant}, {Magalhaes}, {Kastner},
  {Faure}, {Forveille} \& {Qi}}]{Hily-Blant17}
{Hily-Blant} P, {Magalhaes} V, {Kastner} J, {Faure} A, {Forveille} T, {Qi} C.
  2017.
\textit{A\&A} 603:L6

\bibitem[{{Hily-Blant} et~al.(2019){Hily-Blant}, {Magalhaes de Souza},
  {Kastner} \& {Forveille}}]{Hily-Blant19}
{Hily-Blant} P, {Magalhaes de Souza} V, {Kastner} J, {Forveille} T. 2019.
\textit{A\&A} 632:L12

\bibitem[{{Hobbs} et~al.(2022){Hobbs}, {Shorttle}, {Madhusudhan} \&
  {Nikku}}]{Hobbs22}
{Hobbs} R, {Shorttle} O, {Madhusudhan}, {Nikku}. 2022.
\textit{MNRAS}

\bibitem[{{Hogerheijde} et~al.(2011){Hogerheijde}, {Bergin}, {Brinch},
  {Cleeves}, {Fogel} et~al.}]{Hogerheijde11}
{Hogerheijde} MR, {Bergin} EA, {Brinch} C, {Cleeves} LI, {Fogel} JKJ, et~al.
  2011.
\textit{Science} 334:338--

\bibitem[{{Hogerheijde} \& {van der Tak}(2000)}]{Hogerheijde00}
{Hogerheijde} MR, {van der Tak} FFS. 2000.
\textit{A\&A} 362:697--710

\bibitem[{{Honda} et~al.(2009){Honda}, {Inoue}, {Fukagawa}, {Oka}, {Nakamoto}
  et~al.}]{Honda09}
{Honda} M, {Inoue} AK, {Fukagawa} M, {Oka} A, {Nakamoto} T, et~al. 2009.
\textit{ApJL} 690:L110--L113

\bibitem[{{Huang} et~al.(2018{\natexlab{a}}){Huang}, {Andrews}, {Cleeves},
  {{\"O}berg}, {Wilner} et~al.}]{Huang18}
{Huang} J, {Andrews} SM, {Cleeves} LI, {{\"O}berg} KI, {Wilner} DJ, et~al.
  2018{\natexlab{a}}.
\textit{ApJ} 852:122

\bibitem[{{Huang} et~al.(2018{\natexlab{b}}){Huang}, {Andrews}, {Dullemond},
  {Isella}, {P{\'e}rez} et~al.}]{Huang18b}
{Huang} J, {Andrews} SM, {Dullemond} CP, {Isella} A, {P{\'e}rez} LM, et~al.
  2018{\natexlab{b}}.
\textit{ApJL} 869:L42

\bibitem[{{Huang} et~al.(2020){Huang}, {Andrews}, {Dullemond}, {{\"O}berg},
  {Qi} et~al.}]{Huang20}
{Huang} J, {Andrews} SM, {Dullemond} CP, {{\"O}berg} KI, {Qi} C, et~al. 2020.
\textit{ApJ} 891:48

\bibitem[{{Huang} et~al.(2021){Huang}, {Bergin}, {{\"O}berg}, {Andrews},
  {Teague} et~al.}]{Huang21}
{Huang} J, {Bergin} EA, {{\"O}berg} KI, {Andrews} SM, {Teague} R, et~al. 2021.
\textit{ApJS} 257:19

\bibitem[{{Huang} \& {{\"O}berg}(2015)}]{Huang15}
{Huang} J, {{\"O}berg} KI. 2015.
\textit{ApJL} 809:L26

\bibitem[{{Huang} et~al.(2017){Huang}, {{\"O}berg}, {Qi}, {Aikawa}, {Andrews}
  et~al.}]{Huang17}
{Huang} J, {{\"O}berg} KI, {Qi} C, {Aikawa} Y, {Andrews} SM, et~al. 2017.
\textit{ApJ} 835:231

\bibitem[{{Ilee} et~al.(2021){Ilee}, {Walsh}, {Booth}, {Aikawa}, {Andrews}
  et~al.}]{Ilee21}
{Ilee} JD, {Walsh} C, {Booth} AS, {Aikawa} Y, {Andrews} SM, et~al. 2021.
\textit{ApJs} 257:9

\bibitem[{{Ilgner} et~al.(2004){Ilgner}, {Henning}, {Markwick} \&
  {Millar}}]{Ilgner04}
{Ilgner} M, {Henning} T, {Markwick} AJ, {Millar} TJ. 2004.
\textit{A\&A} 415:643--659

\bibitem[{{Ilgner} \& {Nelson}(2006)}]{Ilgner06}
{Ilgner} M, {Nelson} RP. 2006.
\textit{A\&A} 445:205--222

\bibitem[{{Ioppolo} et~al.(2021){Ioppolo}, {Fedoseev}, {Chuang}, {Cuppen},
  {Clements} et~al.}]{Ioppolo21}
{Ioppolo} S, {Fedoseev} G, {Chuang} KJ, {Cuppen} HM, {Clements} AR, et~al.
  2021.
\textit{Nature Astronomy} 5:197--205

\bibitem[{{Izquierdo} et~al.(2022){Izquierdo}, {Facchini}, {Rosotti}, {van
  Dishoeck} \& {Testi}}]{Izquierdo22}
{Izquierdo} AF, {Facchini} S, {Rosotti} GP, {van Dishoeck} EF, {Testi} L. 2022.
\textit{ApJ} 928:2

\bibitem[{{Izquierdo} et~al.(2021){Izquierdo}, {Testi}, {Facchini}, {Rosotti}
  \& {van Dishoeck}}]{Izquierdo21}
{Izquierdo} AF, {Testi} L, {Facchini} S, {Rosotti} GP, {van Dishoeck} EF. 2021.
\textit{A\&A} 650:A179

\bibitem[{{Jiang} et~al.(2022){Jiang}, {Zhu} \& {Ormel}}]{Jiang22}
{Jiang} H, {Zhu} W, {Ormel} CW. 2022.
\textit{ApJL} 924:L31

\bibitem[{{Jonkheid} et~al.(2004){Jonkheid}, {Faas}, {van Zadelhoff} \& {van
  Dishoeck}}]{Jonkheid04}
{Jonkheid} B, {Faas} FGA, {van Zadelhoff} G, {van Dishoeck} EF. 2004.
\textit{A\&A} 428:511--521

\bibitem[{{J{\o}rgensen} et~al.(2020){J{\o}rgensen}, {Belloche} \&
  {Garrod}}]{Jorgensen20}
{J{\o}rgensen} JK, {Belloche} A, {Garrod} RT. 2020.
\textit{ARA\&A} 58:727--778

\bibitem[{{Jura} \& {Young}(2014)}]{Jura14}
{Jura} M, {Young} ED. 2014.
\textit{Annual Review of Earth and Planetary Sciences} 42:45--67

\bibitem[{{Kama} et~al.(2016){Kama}, {Bruderer}, {Carney}, {Hogerheijde}, {van
  Dishoeck} et~al.}]{Kama16}
{Kama} M, {Bruderer} S, {Carney} M, {Hogerheijde} M, {van Dishoeck} EF, et~al.
  2016.
\textit{A\&A} 588:A108

\bibitem[{{Kama} et~al.(2019){Kama}, {Shorttle}, {Jermyn}, {Folsom}, {Furuya}
  et~al.}]{Kama19}
{Kama} M, {Shorttle} O, {Jermyn} AS, {Folsom} CP, {Furuya} K, et~al. 2019.
\textit{ApJ} 885:114

\bibitem[{{Kama} et~al.(2020){Kama}, {Trapman}, {Fedele}, {Bruderer},
  {Hogerheijde} et~al.}]{Kama20}
{Kama} M, {Trapman} L, {Fedele} D, {Bruderer} S, {Hogerheijde} MR, et~al. 2020.
\textit{A\&A} 634:A88

\bibitem[{{Kamp} \& {Dullemond}(2004)}]{Kamp04}
{Kamp} I, {Dullemond} CP. 2004.
\textit{ApJ} 615:991--999

\bibitem[{{Kamp} et~al.(2013){Kamp}, {Thi}, {Meeus}, {Woitke}, {Pinte}
  et~al.}]{Kamp13}
{Kamp} I, {Thi} WF, {Meeus} G, {Woitke} P, {Pinte} C, et~al. 2013.
\textit{A\&A} 559:A24

\bibitem[{{Kamp} et~al.(2017){Kamp}, {Thi}, {Woitke}, {Rab}, {Bouma} \&
  {M{\'e}nard}}]{Kamp17}
{Kamp} I, {Thi} WF, {Woitke} P, {Rab} C, {Bouma} S, {M{\'e}nard} F. 2017.
\textit{A\&A} 607:A41

\bibitem[{{Kamp} et~al.(2010){Kamp}, {Tilling}, {Woitke}, {Thi} \&
  {Hogerheijde}}]{Kamp10}
{Kamp} I, {Tilling} I, {Woitke} P, {Thi} WF, {Hogerheijde} M. 2010.
\textit{A\&A} 510:A18

\bibitem[{{Kastner} et~al.(2014){Kastner}, {Hily-Blant}, {Rodriguez}, {Punzi}
  \& {Forveille}}]{Kastner14}
{Kastner} JH, {Hily-Blant} P, {Rodriguez} DR, {Punzi} K, {Forveille} T. 2014.
\textit{ApJ} 793:55

\bibitem[{{Kastner} et~al.(1997){Kastner}, {Zuckerman}, {Weintraub} \&
  {Forveille}}]{Kastner97}
{Kastner} JH, {Zuckerman} B, {Weintraub} DA, {Forveille} T. 1997.
\textit{Science} 277:67--71

\bibitem[{{Kress} et~al.(2010){Kress}, {Tielens} \& {Frenklach}}]{Kress10}
{Kress} ME, {Tielens} AGGM, {Frenklach} M. 2010.
\textit{Advances in Space Research} 46:44--49

\bibitem[{{Krijt} et~al.(2020){Krijt}, {Bosman}, {Zhang}, {Schwarz}, {Ciesla}
  \& {Bergin}}]{Krijt20}
{Krijt} S, {Bosman} AD, {Zhang} K, {Schwarz} KR, {Ciesla} FJ, {Bergin} EA.
  2020.
\textit{ApJ} 899:134

\bibitem[{{Krijt} et~al.(2016){Krijt}, {Ciesla} \& {Bergin}}]{Krijt16}
{Krijt} S, {Ciesla} FJ, {Bergin} EA. 2016.
\textit{ApJ} 833:285

\bibitem[{{Krijt} et~al.(2018){Krijt}, {Schwarz}, {Bergin} \&
  {Ciesla}}]{Krijt18}
{Krijt} S, {Schwarz} KR, {Bergin} EA, {Ciesla} FJ. 2018.
\textit{ApJ} 864:78

\bibitem[{{Kruczkiewicz} et~al.(2021){Kruczkiewicz}, {Vitorino}, {Congiu},
  {Theul{\'e}} \& {Dulieu}}]{Kruczkiewicz21}
{Kruczkiewicz} F, {Vitorino} J, {Congiu} E, {Theul{\'e}} P, {Dulieu} F. 2021.
\textit{A\&A} 652:A29

\bibitem[{{Lahuis} et~al.(2007){Lahuis}, {van Dishoeck}, {Blake}, {Evans},
  {Kessler-Silacci} \& {Pontoppidan}}]{Lahuis07}
{Lahuis} F, {van Dishoeck} EF, {Blake} GA, {Evans} Neal~J. I, {Kessler-Silacci}
  JE, {Pontoppidan} KM. 2007.
\textit{ApJ} 665:492--511

\bibitem[{{Law} et~al.(2022){Law}, {Crystian}, {Teague}, {{\"O}berg}, {Rich}
  et~al.}]{Law22}
{Law} CJ, {Crystian} S, {Teague} R, {{\"O}berg} KI, {Rich} EA, et~al. 2022.
\textit{ApJ} 932:114

\bibitem[{{Law} et~al.(2021{\natexlab{a}}){Law}, {Loomis}, {Teague},
  {{\"O}berg}, {Czekala} et~al.}]{Law21_radprof}
{Law} CJ, {Loomis} RA, {Teague} R, {{\"O}berg} KI, {Czekala} I, et~al.
  2021{\natexlab{a}}.
\textit{ApJs} 257:3

\bibitem[{{Law} et~al.(2021{\natexlab{b}}){Law}, {Teague}, {Loomis}, {Bae},
  {{\"O}berg} et~al.}]{Law21_surf}
{Law} CJ, {Teague} R, {Loomis} RA, {Bae} J, {{\"O}berg} KI, et~al.
  2021{\natexlab{b}}.
\textit{ApJs} 257:4

\bibitem[{{Le Gal} et~al.(2019{\natexlab{a}}){Le Gal}, {Brady}, {{\"O}berg},
  {Roueff} \& {Le Petit}}]{LeGal19b}
{Le Gal} R, {Brady} MT, {{\"O}berg} KI, {Roueff} E, {Le Petit} F.
  2019{\natexlab{a}}.
\textit{ApJ} 886:86

\bibitem[{{Le Gal} et~al.(2019{\natexlab{b}}){Le Gal}, {{\"O}berg}, {Loomis},
  {Pegues} \& {Bergner}}]{LeGal19}
{Le Gal} R, {{\"O}berg} KI, {Loomis} RA, {Pegues} J, {Bergner} JB.
  2019{\natexlab{b}}.
\textit{ApJ} 876:72

\bibitem[{{Le Gal} et~al.(2021){Le Gal}, {{\"O}berg}, {Teague}, {Loomis}, {Law}
  et~al.}]{LeGal21}
{Le Gal} R, {{\"O}berg} KI, {Teague} R, {Loomis} RA, {Law} CJ, et~al. 2021.
\textit{ApJs} 257:12

\bibitem[{{Lee} et~al.(2019){Lee}, {Lee}, {Baek}, {Aikawa}, {Cieza}
  et~al.}]{Lee19}
{Lee} JE, {Lee} S, {Baek} G, {Aikawa} Y, {Cieza} L, et~al. 2019.
\textit{Nature Astronomy} 3:314--319

\bibitem[{{Leemker} et~al.(2022){Leemker}, {Booth}, {van Dishoeck},
  {P{\'e}rez-S{\'a}nchez}, {Szul{\'a}gyi} et~al.}]{Leemker22}
{Leemker} M, {Booth} AS, {van Dishoeck} EF, {P{\'e}rez-S{\'a}nchez} AF,
  {Szul{\'a}gyi} J, et~al. 2022.
\textit{A\&A} 663:A23

\bibitem[{{Leemker} et~al.(2021){Leemker}, {van't Hoff}, {Trapman}, {van
  Gelder}, {Hogerheijde} et~al.}]{Leemker21}
{Leemker} M, {van't Hoff} MLR, {Trapman} L, {van Gelder} ML, {Hogerheijde} MR,
  et~al. 2021.
\textit{A\&A} 646:A3

\bibitem[{{Lichtenberg} et~al.(2022){Lichtenberg}, {Schaefer}, {Nakajima} \&
  {Fischer}}]{Lichtenberg22}
{Lichtenberg} T, {Schaefer} LK, {Nakajima} M, {Fischer} RA. 2022.
\textit{arXiv e-prints} :arXiv:2203.10023

\bibitem[{{Lodders}(2003)}]{Lodders03}
{Lodders} K. 2003.
\textit{ApJ} 591:1220--1247

\bibitem[{{Lodders}(2004)}]{Lodders04}
{Lodders} K. 2004.
\textit{ApJ} 611:587--597

\bibitem[{{Long} et~al.(2021){Long}, {Bosman}, {Cazzoletti}, {van Dishoeck},
  {{\"O}berg} et~al.}]{Long21}
{Long} F, {Bosman} AD, {Cazzoletti} P, {van Dishoeck} EF, {{\"O}berg} KI,
  et~al. 2021.
\textit{A\&A} 647:A118

\bibitem[{{Long} et~al.(2017){Long}, {Herczeg}, {Pascucci}, {Drabek-Maunder},
  {Mohanty} et~al.}]{Long17}
{Long} F, {Herczeg} GJ, {Pascucci} I, {Drabek-Maunder} E, {Mohanty} S, et~al.
  2017.
\textit{ApJ} 844:99

\bibitem[{{Long} et~al.(2018){Long}, {Pinilla}, {Herczeg}, {Harsono},
  {Dipierro} et~al.}]{Long18}
{Long} F, {Pinilla} P, {Herczeg} GJ, {Harsono} D, {Dipierro} G, et~al. 2018.
\textit{ApJ} 869:17

\bibitem[{{Loomis} et~al.(2018{\natexlab{a}}){Loomis}, {Cleeves}, {{\"O}berg},
  {Aikawa}, {Bergner} et~al.}]{Loomis18b}
{Loomis} RA, {Cleeves} LI, {{\"O}berg} KI, {Aikawa} Y, {Bergner} J, et~al.
  2018{\natexlab{a}}.
\textit{ApJ} 859:131

\bibitem[{{Loomis} et~al.(2020){Loomis}, {{\"O}berg}, {Andrews}, {Bergin},
  {Bergner} et~al.}]{Loomis20}
{Loomis} RA, {{\"O}berg} KI, {Andrews} SM, {Bergin} E, {Bergner} J, et~al.
  2020.
\textit{ApJ} 893:101

\bibitem[{{Loomis} et~al.(2018{\natexlab{b}}){Loomis}, {{\"O}berg}, {Andrews},
  {Walsh}, {Czekala} et~al.}]{Loomis18a}
{Loomis} RA, {{\"O}berg} KI, {Andrews} SM, {Walsh} C, {Czekala} I, et~al.
  2018{\natexlab{b}}.
\textit{AJ} 155:182

\bibitem[{{Lunine} \& {Stevenson}(1985)}]{Lunine85}
{Lunine} JI, {Stevenson} DJ. 1985.
\textit{ApJs} 58:493--531

\bibitem[{{Lynden-Bell} \& {Pringle}(1974)}]{LyndenBell74}
{Lynden-Bell} D, {Pringle} JE. 1974.
\textit{MNRAS} 168:603--637

\bibitem[{{Lyons} \& {Young}(2005)}]{Lyons05}
{Lyons} JR, {Young} ED. 2005.
\textit{{Photochemical Speciation of Oxygen Isotopes in the Solar Nebula}}. In
  \textit{Chondrites and the Protoplanetary Disk}, eds. AN~{Krot}, ERD {Scott},
  B~{Reipurth}, vol. 341 of \textit{Astronomical Society of the Pacific
  Conference Series}

\bibitem[{{Madhusudhan}(2019)}]{Madhusudhan19}
{Madhusudhan} N. 2019.
\textit{ARA\&A} 57:617--663

\bibitem[{{Mandell} et~al.(2012){Mandell}, {Bast}, {van Dishoeck}, {Blake},
  {Salyk} et~al.}]{Mandell12}
{Mandell} AM, {Bast} J, {van Dishoeck} EF, {Blake} GA, {Salyk} C, et~al. 2012.
\textit{ApJ} 747:92

\bibitem[{{McClure}(2019)}]{McClure19}
{McClure} MK. 2019.
\textit{A\&A} 632:A32

\bibitem[{{McClure} et~al.(2016){McClure}, {Bergin}, {Cleeves}, {van Dishoeck},
  {Blake} et~al.}]{McClure16}
{McClure} MK, {Bergin} EA, {Cleeves} LI, {van Dishoeck} EF, {Blake} GA, et~al.
  2016.
\textit{ApJ} 831:167

\bibitem[{{McElroy} et~al.(2013){McElroy}, {Walsh}, {Markwick}, {Cordiner},
  {Smith} \& {Millar}}]{McElroy13}
{McElroy} D, {Walsh} C, {Markwick} AJ, {Cordiner} MA, {Smith} K, {Millar} TJ.
  2013.
\textit{A\&A} 550:A36

\bibitem[{{McGuire}(2022)}]{McGuire22}
{McGuire} BA. 2022.
\textit{ApJs} 259:30

\bibitem[{{Meeus} et~al.(2012){Meeus}, {Montesinos}, {Mendigut{\'\i}a}, {Kamp},
  {Thi} et~al.}]{Meeus12}
{Meeus} G, {Montesinos} B, {Mendigut{\'\i}a} I, {Kamp} I, {Thi} WF, et~al.
  2012.
\textit{A\&A} 544:A78

\bibitem[{{Meeus} et~al.(2013){Meeus}, {Salyk}, {Bruderer}, {Fedele},
  {Maaskant} et~al.}]{Meeus13}
{Meeus} G, {Salyk} C, {Bruderer} S, {Fedele} D, {Maaskant} K, et~al. 2013.
\textit{A\&A} 559:A84

\bibitem[{{Meijerink} et~al.(2009){Meijerink}, {Pontoppidan}, {Blake},
  {Poelman} \& {Dullemond}}]{Meijerink09}
{Meijerink} R, {Pontoppidan} KM, {Blake} GA, {Poelman} DR, {Dullemond} CP.
  2009.
\textit{ApJ} 704:1471--1481

\bibitem[{{Min} et~al.(2016){Min}, {Bouwman}, {Dominik}, {Waters},
  {Pontoppidan} et~al.}]{Min16}
{Min} M, {Bouwman} J, {Dominik} C, {Waters} LBFM, {Pontoppidan} KM, et~al.
  2016.
\textit{A\&A} 593:A11

\bibitem[{{Miotello} et~al.(2014){Miotello}, {Bruderer} \& {van
  Dishoeck}}]{Miotello14}
{Miotello} A, {Bruderer} S, {van Dishoeck} EF. 2014.
\textit{A\&A} 572:A96

\bibitem[{{Miotello} et~al.(2019){Miotello}, {Facchini}, {van Dishoeck},
  {Cazzoletti}, {Testi} et~al.}]{Miotello19}
{Miotello} A, {Facchini} S, {van Dishoeck} EF, {Cazzoletti} P, {Testi} L,
  et~al. 2019.
\textit{A\&A} 631:A69

\bibitem[{{Miotello} et~al.(2022){Miotello}, {Kamp}, {Birnstiel}, {Cleeves} \&
  {Kataoka}}]{Miotello22}
{Miotello} A, {Kamp} I, {Birnstiel} T, {Cleeves} LI, {Kataoka} A. 2022.
\textit{arXiv e-prints} :arXiv:2203.09818

\bibitem[{{Miotello} et~al.(2016){Miotello}, {van Dishoeck}, {Kama} \&
  {Bruderer}}]{Miotello16}
{Miotello} A, {van Dishoeck} EF, {Kama} M, {Bruderer} S. 2016.
\textit{A\&A} 594:A85

\bibitem[{{Miotello} et~al.(2017){Miotello}, {van Dishoeck}, {Williams},
  {Ansdell}, {Guidi} et~al.}]{Miotello17}
{Miotello} A, {van Dishoeck} EF, {Williams} JP, {Ansdell} M, {Guidi} G, et~al.
  2017.
\textit{A\&A} 599:A113

\bibitem[{{Molyarova} et~al.(2018){Molyarova}, {Akimkin}, {Semenov},
  {{\'A}brah{\'a}m}, {Henning} et~al.}]{Molyarova18}
{Molyarova} T, {Akimkin} V, {Semenov} D, {{\'A}brah{\'a}m} P, {Henning} T,
  et~al. 2018.
\textit{ApJ} 866:46

\bibitem[{{Morbidelli} et~al.(2016){Morbidelli}, {Bitsch}, {Crida}, {Gounelle},
  {Guillot} et~al.}]{Morbidelli16}
{Morbidelli} A, {Bitsch} B, {Crida} A, {Gounelle} M, {Guillot} T, et~al. 2016.
\textit{Icarus} 267:368--376

\bibitem[{{Mu{\~n}oz Caro} et~al.(2002){Mu{\~n}oz Caro}, {Meierhenrich},
  {Schutte}, {Barbier}, {Arcones Segovia} et~al.}]{MunozCaro02}
{Mu{\~n}oz Caro} GM, {Meierhenrich} UJ, {Schutte} WA, {Barbier} B, {Arcones
  Segovia} A, et~al. 2002.
\textit{Nature} 416:403--406

\bibitem[{{M{\"u}ller} et~al.(2005){M{\"u}ller}, {Schl{\"o}der}, {Stutzki} \&
  {Winnewisser}}]{Muller05}
{M{\"u}ller} HSP, {Schl{\"o}der} F, {Stutzki} J, {Winnewisser} G. 2005.
\textit{Journal of Molecular Structure} 742:215--227

\bibitem[{{M{\"u}ller} et~al.(2001){M{\"u}ller}, {Thorwirth}, {Roth} \&
  {Winnewisser}}]{Muller01}
{M{\"u}ller} HSP, {Thorwirth} S, {Roth} DA, {Winnewisser} G. 2001.
\textit{A\&A} 370:L49--L52

\bibitem[{{Mumma} \& {Charnley}(2011)}]{Mumma11}
{Mumma} MJ, {Charnley} SB. 2011.
\textit{ARA\&A} 49:471--524

\bibitem[{{Najita} et~al.(2003){Najita}, {Carr} \& {Mathieu}}]{Najita03}
{Najita} J, {Carr} JS, {Mathieu} RD. 2003.
\textit{ApJ} 589:931--952

\bibitem[{{Najita} \& {{\'A}d{\'a}mkovics}(2017)}]{Najita17}
{Najita} JR, {{\'A}d{\'a}mkovics} M. 2017.
\textit{ApJ} 847:6

\bibitem[{{Najita} et~al.(2011){Najita}, {{\'A}d{\'a}mkovics} \&
  {Glassgold}}]{Najita11}
{Najita} JR, {{\'A}d{\'a}mkovics} M, {Glassgold} AE. 2011.
\textit{ApJ} 743:147

\bibitem[{{Najita} et~al.(2021){Najita}, {Carr}, {Brittain}, {Lacy}, {Richter}
  \& {Doppmann}}]{Najita21}
{Najita} JR, {Carr} JS, {Brittain} SD, {Lacy} JH, {Richter} MJ, {Doppmann} GW.
  2021.
\textit{ApJ} 908:171

\bibitem[{{Najita} et~al.(2013){Najita}, {Carr}, {Pontoppidan}, {Salyk}, {van
  Dishoeck} \& {Blake}}]{Najita13}
{Najita} JR, {Carr} JS, {Pontoppidan} KM, {Salyk} C, {van Dishoeck} EF, {Blake}
  GA. 2013.
\textit{ApJ} 766:134

\bibitem[{{Najita} et~al.(2018){Najita}, {Carr}, {Salyk}, {Lacy}, {Richter} \&
  {DeWitt}}]{Najita18}
{Najita} JR, {Carr} JS, {Salyk} C, {Lacy} JH, {Richter} MJ, {DeWitt} C. 2018.
\textit{ApJ} 862:122

\bibitem[{{Najita} et~al.(2010){Najita}, {Carr}, {Strom}, {Watson}, {Pascucci}
  et~al.}]{Najita10}
{Najita} JR, {Carr} JS, {Strom} SE, {Watson} DM, {Pascucci} I, et~al. 2010.
\textit{ApJ} 712:274--286

\bibitem[{{Natta} et~al.(2007){Natta}, {Testi}, {Calvet}, {Henning}, {Waters}
  \& {Wilner}}]{Natta07}
{Natta} A, {Testi} L, {Calvet} N, {Henning} T, {Waters} R, {Wilner} D. 2007.
\textit{{Dust in Protoplanetary Disks: Properties and Evolution}}. In
  \textit{Protostars and Planets V}, eds. B~{Reipurth}, D~{Jewitt}, K~{Keil}

\bibitem[{{Nomura} et~al.(2009){Nomura}, {Aikawa}, {Nakagawa} \&
  {Millar}}]{Nomura09}
{Nomura} H, {Aikawa} Y, {Nakagawa} Y, {Millar} TJ. 2009.
\textit{A\&A} 495:183--188

\bibitem[{{Nomura} et~al.(2007){Nomura}, {Aikawa}, {Tsujimoto}, {Nakagawa} \&
  {Millar}}]{Nomura07}
{Nomura} H, {Aikawa} Y, {Tsujimoto} M, {Nakagawa} Y, {Millar} TJ. 2007.
\textit{ApJ} 661:334--353

\bibitem[{{Nomura} et~al.(2021){Nomura}, {Tsukagoshi}, {Kawabe}, {Muto},
  {Kanagawa} et~al.}]{Nomura21}
{Nomura} H, {Tsukagoshi} T, {Kawabe} R, {Muto} T, {Kanagawa} KD, et~al. 2021.
\textit{ApJ} 914:113

\bibitem[{{\"O}berg \& Bergin(2021)}]{Oberg21_Review}
{\"O}berg KI, Bergin EA. 2021.
\textit{Physics Reports} 893:1--48.
Astrochemistry and compositions of planetary systems

\bibitem[{{{\"O}berg} et~al.(2011{\natexlab{a}}){{\"O}berg}, {Boogert},
  {Pontoppidan}, {van den Broek}, {van Dishoeck} et~al.}]{Oberg11c}
{{\"O}berg} KI, {Boogert} ACA, {Pontoppidan} KM, {van den Broek} S, {van
  Dishoeck} EF, et~al. 2011{\natexlab{a}}.
\textit{ApJ} 740:109

\bibitem[{{{\"O}berg} et~al.(2021{\natexlab{a}}){{\"O}berg}, {Cleeves},
  {Bergner}, {Cavanaro}, {Teague} et~al.}]{Oberg21_TWHya}
{{\"O}berg} KI, {Cleeves} LI, {Bergner} JB, {Cavanaro} J, {Teague} R, et~al.
  2021{\natexlab{a}}.
\textit{AJ} 161:38

\bibitem[{{{\"O}berg} et~al.(2015{\natexlab{a}}){{\"O}berg}, {Furuya},
  {Loomis}, {Aikawa}, {Andrews} et~al.}]{Oberg15}
{{\"O}berg} KI, {Furuya} K, {Loomis} R, {Aikawa} Y, {Andrews} SM, et~al.
  2015{\natexlab{a}}.
\textit{ApJ} 810:112

\bibitem[{{{\"O}berg} et~al.(2015{\natexlab{b}}){{\"O}berg}, {Guzm{\'a}n},
  {Furuya}, {Qi}, {Aikawa} et~al.}]{Oberg15Natur}
{{\"O}berg} KI, {Guzm{\'a}n} VV, {Furuya} K, {Qi} C, {Aikawa} Y, et~al.
  2015{\natexlab{b}}.
\textit{Nature} 520:198--201

\bibitem[{{{\"O}berg} et~al.(2021{\natexlab{b}}){{\"O}berg}, {Guzm{\'a}n},
  {Walsh}, {Aikawa}, {Bergin} et~al.}]{Oberg21-maps}
{{\"O}berg} KI, {Guzm{\'a}n} VV, {Walsh} C, {Aikawa} Y, {Bergin} EA, et~al.
  2021{\natexlab{b}}.
\textit{ApJs} 257:1

\bibitem[{{{\"O}berg} et~al.(2011{\natexlab{b}}){{\"O}berg}, {Murray-Clay} \&
  {Bergin}}]{Oberg11e}
{{\"O}berg} KI, {Murray-Clay} R, {Bergin} EA. 2011{\natexlab{b}}.
\textit{ApJL} 743:L16

\bibitem[{{{\"O}berg} et~al.(2010){{\"O}berg}, {Qi}, {Fogel}, {Bergin},
  {Andrews} et~al.}]{Oberg10c}
{{\"O}berg} KI, {Qi} C, {Fogel} JKJ, {Bergin} EA, {Andrews} SM, et~al. 2010.
\textit{ApJ} 720:480--493

\bibitem[{{{\"O}berg} et~al.(2011{\natexlab{c}}){{\"O}berg}, {Qi}, {Fogel},
  {Bergin}, {Andrews} et~al.}]{Oberg11a}
{{\"O}berg} KI, {Qi} C, {Fogel} JKJ, {Bergin} EA, {Andrews} SM, et~al.
  2011{\natexlab{c}}.
\textit{ApJ} 734:98

\bibitem[{{{\"O}berg} \& {Wordsworth}(2019)}]{Oberg19}
{{\"O}berg} KI, {Wordsworth} R. 2019.
\textit{AJ} 158:194

\bibitem[{{Okuzumi} \& {Tazaki}(2019)}]{Okuzumi19}
{Okuzumi} S, {Tazaki} R. 2019.
\textit{ApJ} 878:132

\bibitem[{{Owen} et~al.(1999){Owen}, {Mahaffy}, {Niemann}, {Atreya}, {Donahue}
  et~al.}]{Owen99}
{Owen} T, {Mahaffy} P, {Niemann} HB, {Atreya} S, {Donahue} T, et~al. 1999.
\textit{Nature} 402:269--270

\bibitem[{{Paneque-Carre{\~n}o} et~al.(2022){Paneque-Carre{\~n}o}, {Miotello},
  {van Dishoeck}, {P{\'e}rez}, {Facchini} et~al.}]{Paneque22}
{Paneque-Carre{\~n}o} T, {Miotello} A, {van Dishoeck} EF, {P{\'e}rez} LM,
  {Facchini} S, et~al. 2022.
\textit{arXiv e-prints} :arXiv:2207.08827

\bibitem[{{Pascucci} et~al.(2009){Pascucci}, {Apai}, {Luhman}, {Henning},
  {Bouwman} et~al.}]{Pascucci09}
{Pascucci} I, {Apai} D, {Luhman} K, {Henning} T, {Bouwman} J, et~al. 2009.
\textit{ApJ} 696:143--159

\bibitem[{{Pascucci} et~al.(2022){Pascucci}, {Cabrit}, {Edwards}, {Gorti},
  {Gressel} \& {Suzuki}}]{Pascucci22}
{Pascucci} I, {Cabrit} S, {Edwards} S, {Gorti} U, {Gressel} O, {Suzuki} T.
  2022.
\textit{arXiv e-prints} :arXiv:2203.10068

\bibitem[{{Pascucci} et~al.(2013){Pascucci}, {Herczeg}, {Carr} \&
  {Bruderer}}]{Pascucci13}
{Pascucci} I, {Herczeg} G, {Carr} JS, {Bruderer} S. 2013.
\textit{ApJ} 779:178

\bibitem[{{Pegues} et~al.(2021){Pegues}, {{\"O}berg}, {Bergner}, {Huang},
  {Pascucci} et~al.}]{Pegues21}
{Pegues} J, {{\"O}berg} KI, {Bergner} JB, {Huang} J, {Pascucci} I, et~al. 2021.
\textit{ApJ} 911:150

\bibitem[{{Pegues} et~al.(2020){Pegues}, {{\"O}berg}, {Bergner}, {Loomis}, {Qi}
  et~al.}]{Pegues20}
{Pegues} J, {{\"O}berg} KI, {Bergner} JB, {Loomis} RA, {Qi} C, et~al. 2020.
\textit{ApJ} 890:142

\bibitem[{{Phuong} et~al.(2018){Phuong}, {Chapillon}, {Majumdar}, {Dutrey},
  {Guilloteau} et~al.}]{Phuong18}
{Phuong} NT, {Chapillon} E, {Majumdar} L, {Dutrey} A, {Guilloteau} S, et~al.
  2018.
\textit{A\&A} 616:L5

\bibitem[{{Phuong} et~al.(2021){Phuong}, {Dutrey}, {Chapillon}, {Guilloteau},
  {Bary} et~al.}]{Phuong21}
{Phuong} NT, {Dutrey} A, {Chapillon} E, {Guilloteau} S, {Bary} J, et~al. 2021.
\textit{A\&A} 653:L5

\bibitem[{{Pickett} et~al.(1998){Pickett}, {Poynter}, {Cohen}, {Delitsky},
  {Pearson} \& {M{\"u}ller}}]{Pickett98}
{Pickett} HM, {Poynter} RL, {Cohen} EA, {Delitsky} ML, {Pearson} JC,
  {M{\"u}ller} HSP. 1998.
\textit{JQSRT} 60:883--890

\bibitem[{{Pi{\'e}tu} et~al.(2007){Pi{\'e}tu}, {Dutrey} \&
  {Guilloteau}}]{Pietu07}
{Pi{\'e}tu} V, {Dutrey} A, {Guilloteau} S. 2007.
\textit{A\&A} 467:163--178

\bibitem[{{Pinte} et~al.(2006){Pinte}, {M{\'e}nard}, {Duch{\^e}ne} \&
  {Bastien}}]{Pinte06}
{Pinte} C, {M{\'e}nard} F, {Duch{\^e}ne} G, {Bastien} P. 2006.
\textit{A\&A} 459:797--804

\bibitem[{{Pinte} et~al.(2018{\natexlab{a}}){Pinte}, {M{\'e}nard},
  {Duch{\^e}ne}, {Hill}, {Dent} et~al.}]{Pinte18b}
{Pinte} C, {M{\'e}nard} F, {Duch{\^e}ne} G, {Hill} T, {Dent} WRF, et~al.
  2018{\natexlab{a}}.
\textit{A\&A} 609:A47

\bibitem[{{Pinte} et~al.(2018{\natexlab{b}}){Pinte}, {Price}, {M{\'e}nard},
  {Duch{\^e}ne}, {Dent} et~al.}]{Pinte18}
{Pinte} C, {Price} DJ, {M{\'e}nard} F, {Duch{\^e}ne} G, {Dent} WRF, et~al.
  2018{\natexlab{b}}.
\textit{ApJL} 860:L13

\bibitem[{{Pinte} et~al.(2022){Pinte}, {Teague}, {Flaherty}, {Hall}, {Facchini}
  \& {Casassus}}]{Pinte22}
{Pinte} C, {Teague} R, {Flaherty} K, {Hall} C, {Facchini} S, {Casassus} S.
  2022.
\textit{arXiv e-prints} :arXiv:2203.09528

\bibitem[{{Pirovano} et~al.(2022){Pirovano}, {Fedele}, {van Dishoeck},
  {Hogerheijde}, {Lodato} \& {Bruderer}}]{Pirovano22}
{Pirovano} LM, {Fedele} D, {van Dishoeck} EF, {Hogerheijde} MR, {Lodato} G,
  {Bruderer} S. 2022.
\textit{arXiv e-prints} :arXiv:2207.10744

\bibitem[{{Piso} et~al.(2015){Piso}, {{\"O}berg}, {Birnstiel} \&
  {Murray-Clay}}]{Piso15}
{Piso} AMA, {{\"O}berg} KI, {Birnstiel} T, {Murray-Clay} RA. 2015.
\textit{ApJ} 815:109

\bibitem[{{Piso} et~al.(2016){Piso}, {Pegues} \& {{\"O}berg}}]{Piso16}
{Piso} AMA, {Pegues} J, {{\"O}berg} KI. 2016.
\textit{ApJ} 833:203

\bibitem[{{Podio} et~al.(2020){Podio}, {Garufi}, {Codella}, {Fedele}, {Bianchi}
  et~al.}]{Podio20}
{Podio} L, {Garufi} A, {Codella} C, {Fedele} D, {Bianchi} E, et~al. 2020.
\textit{A\&A} 642:L7

\bibitem[{{Pontoppidan} et~al.(2011){Pontoppidan}, {Blake} \&
  {Smette}}]{Pontoppidan11}
{Pontoppidan} KM, {Blake} GA, {Smette} A. 2011.
\textit{ApJ} 733:84

\bibitem[{{Pontoppidan} et~al.(2008){Pontoppidan}, {Blake}, {van Dishoeck},
  {Smette}, {Ireland} \& {Brown}}]{Pontoppidan08-spec}
{Pontoppidan} KM, {Blake} GA, {van Dishoeck} EF, {Smette} A, {Ireland} MJ,
  {Brown} J. 2008.
\textit{ApJ} 684:1323--1329

\bibitem[{{Pontoppidan} et~al.(2005){Pontoppidan}, {Dullemond}, {van Dishoeck},
  {Blake}, {Boogert} et~al.}]{Pontoppidan05}
{Pontoppidan} KM, {Dullemond} CP, {van Dishoeck} EF, {Blake} GA, {Boogert} ACA,
  et~al. 2005.
\textit{ApJ} 622:463--481

\bibitem[{{Pontoppidan} et~al.(2009){Pontoppidan}, {Meijerink}, {Dullemond} \&
  {Blake}}]{Pontoppidan09}
{Pontoppidan} KM, {Meijerink} R, {Dullemond} CP, {Blake} GA. 2009.
\textit{ApJ} 704:1482--1494

\bibitem[{{Pontoppidan} et~al.(2019){Pontoppidan}, {Salyk}, {Banzatti},
  {Blake}, {Walsh} et~al.}]{Pontoppidan19}
{Pontoppidan} KM, {Salyk} C, {Banzatti} A, {Blake} GA, {Walsh} C, et~al. 2019.
\textit{ApJ} 874:92

\bibitem[{{Pontoppidan} et~al.(2014){Pontoppidan}, {Salyk}, {Bergin},
  {Brittain}, {Marty} et~al.}]{Pontoppidan2014}
{Pontoppidan} KM, {Salyk} C, {Bergin} EA, {Brittain} S, {Marty} B, et~al. 2014.
\textit{{Volatiles in Protoplanetary Disks}}. In \textit{Protostars and Planets
  VI}, eds. H~{Beuther}, RS~{Klessen}, CP~{Dullemond}, T~{Henning}

\bibitem[{{Pontoppidan} et~al.(2010){Pontoppidan}, {Salyk}, {Blake},
  {Meijerink}, {Carr} \& {Najita}}]{Pontoppidan10}
{Pontoppidan} KM, {Salyk} C, {Blake} GA, {Meijerink} R, {Carr} JS, {Najita} J.
  2010.
\textit{ApJ} 720:887--903

\bibitem[{{Potapov} et~al.(2018{\natexlab{a}}){Potapov}, {J{\"a}ger} \&
  {Henning}}]{Potapov18}
{Potapov} A, {J{\"a}ger} C, {Henning} T. 2018{\natexlab{a}}.
\textit{ApJ} 865:58

\bibitem[{{Potapov} et~al.(2018{\natexlab{b}}){Potapov}, {J{\"a}ger} \&
  {Henning}}]{Poptapov18}
{Potapov} A, {J{\"a}ger} C, {Henning} T. 2018{\natexlab{b}}.
\textit{ApJ} 865:58

\bibitem[{{Powell} et~al.(2022){Powell}, {Gao}, {Murray-Clay} \&
  {Zhang}}]{Powell22}
{Powell} D, {Gao} P, {Murray-Clay} R, {Zhang} X. 2022.
\textit{Nature Astronomy}

\bibitem[{{Powell} et~al.(2019){Powell}, {Murray-Clay}, {P{\'e}rez},
  {Schlichting} \& {Rosenthal}}]{Powell19}
{Powell} D, {Murray-Clay} R, {P{\'e}rez} LM, {Schlichting} HE, {Rosenthal} M.
  2019.
\textit{ApJ} 878:116

\bibitem[{{Price} et~al.(2021){Price}, {Cleeves}, {Bodewits} \&
  {{\"O}berg}}]{Price21}
{Price} EM, {Cleeves} LI, {Bodewits} D, {{\"O}berg} KI. 2021.
\textit{ApJ} 913:9

\bibitem[{{Price} et~al.(2020){Price}, {Cleeves} \& {{\"O}berg}}]{Price20}
{Price} EM, {Cleeves} LI, {{\"O}berg} KI. 2020.
\textit{ApJ} 890:154

\bibitem[{{Prinn}(1993)}]{Prinn93}
{Prinn} RG. 1993.
\textit{{Chemistry and Evolution of Gaseous Circumstellar Disks}}. In
  \textit{Protostars and Planets III}, eds. EH~{Levy}, JI~{Lunine}

\bibitem[{{Qi} et~al.(2011){Qi}, {D'Alessio}, {{\"O}berg}, {Wilner}, {Hughes}
  et~al.}]{Qi11}
{Qi} C, {D'Alessio} P, {{\"O}berg} KI, {Wilner} DJ, {Hughes} AM, et~al. 2011.
\textit{ApJ} 740:84

\bibitem[{{Qi} et~al.(2003){Qi}, {Kessler}, {Koerner}, {Sargent} \&
  {Blake}}]{Qi03}
{Qi} C, {Kessler} JE, {Koerner} DW, {Sargent} AI, {Blake} GA. 2003.
\textit{ApJ} 597:986--997

\bibitem[{{Qi} et~al.(2019){Qi}, {{\"O}berg}, {Espaillat}, {Robinson},
  {Andrews} et~al.}]{Qi19}
{Qi} C, {{\"O}berg} KI, {Espaillat} CC, {Robinson} CE, {Andrews} SM, et~al.
  2019.
\textit{ApJ} 882:160

\bibitem[{{Qi} et~al.(2013){Qi}, {{\"O}berg}, {Wilner}, {D'Alessio}, {Bergin}
  et~al.}]{Qi13c}
{Qi} C, {{\"O}berg} KI, {Wilner} DJ, {D'Alessio} P, {Bergin} E, et~al. 2013.
\textit{Science} 341:630--632

\bibitem[{{Qi} et~al.(2008){Qi}, {Wilner}, {Aikawa}, {Blake} \&
  {Hogerheijde}}]{Qi08}
{Qi} C, {Wilner} DJ, {Aikawa} Y, {Blake} GA, {Hogerheijde} MR. 2008.
\textit{ApJ} 681:1396--1407

\bibitem[{{Rab} et~al.(2017){Rab}, {Elbakyan}, {Vorobyov}, {G{\"u}del},
  {Dionatos} et~al.}]{Rab17}
{Rab} C, {Elbakyan} V, {Vorobyov} E, {G{\"u}del} M, {Dionatos} O, et~al. 2017.
\textit{A\&A} 604:A15

\bibitem[{{Rab} et~al.(2018){Rab}, {G{\"u}del}, {Woitke}, {Kamp}, {Thi}
  et~al.}]{Rab18}
{Rab} C, {G{\"u}del} M, {Woitke} P, {Kamp} I, {Thi} WF, et~al. 2018.
\textit{A\&A} 609:A91

\bibitem[{{Rab} et~al.(2020){Rab}, {Kamp}, {Dominik}, {Ginski}, {Muro-Arena}
  et~al.}]{Rab20}
{Rab} C, {Kamp} I, {Dominik} C, {Ginski} C, {Muro-Arena} GA, et~al. 2020.
\textit{A\&A} 642:A165

\bibitem[{{Riviere-Marichalar} et~al.(2015){Riviere-Marichalar}, {Elliott},
  {Rebollido}, {Bayo}, {Ribas} et~al.}]{Riviere15}
{Riviere-Marichalar} P, {Elliott} P, {Rebollido} I, {Bayo} A, {Ribas} A, et~al.
  2015.
\textit{A\&A} 584:A22

\bibitem[{{Rivi{\`e}re-Marichalar} et~al.(2022){Rivi{\`e}re-Marichalar},
  {Fuente}, {Esplugues}, {Wakelam}, {le Gal} et~al.}]{Riviere22}
{Rivi{\`e}re-Marichalar} P, {Fuente} A, {Esplugues} G, {Wakelam} V, {le Gal} R,
  et~al. 2022.
\textit{arXiv e-prints} :arXiv:2207.06716

\bibitem[{{Ros} \& {Johansen}(2013)}]{Ros13}
{Ros} K, {Johansen} A. 2013.
\textit{A\&A} 552:A137

\bibitem[{{Rosenfeld} et~al.(2012){Rosenfeld}, {Qi}, {Andrews}, {Wilner},
  {Corder} et~al.}]{Rosenfeld12-kinematics}
{Rosenfeld} KA, {Qi} C, {Andrews} SM, {Wilner} DJ, {Corder} SA, et~al. 2012.
\textit{ApJ} 757:129

\bibitem[{{Ruaud} \& {Gorti}(2019)}]{Ruaud19}
{Ruaud} M, {Gorti} U. 2019.
\textit{ApJ} 885:146

\bibitem[{{Salinas} et~al.(2016){Salinas}, {Hogerheijde}, {Bergin}, {Cleeves},
  {Brinch} et~al.}]{Salinas16}
{Salinas} VN, {Hogerheijde} MR, {Bergin} EA, {Cleeves} LI, {Brinch} C, et~al.
  2016.
\textit{A\&A} 591:A122

\bibitem[{{Salinas} et~al.(2017){Salinas}, {Hogerheijde}, {Mathews},
  {{\"O}berg}, {Qi} et~al.}]{Salinas17}
{Salinas} VN, {Hogerheijde} MR, {Mathews} GS, {{\"O}berg} KI, {Qi} C, et~al.
  2017.
\textit{A\&A} 606:A125

\bibitem[{{Salyk} et~al.(2011{\natexlab{a}}){Salyk}, {Blake}, {Boogert} \&
  {Brown}}]{Salyk11b}
{Salyk} C, {Blake} GA, {Boogert} ACA, {Brown} JM. 2011{\natexlab{a}}.
\textit{ApJ} 743:112

\bibitem[{{Salyk} et~al.(2019){Salyk}, {Lacy}, {Richter}, {Zhang},
  {Pontoppidan} et~al.}]{Salyk19}
{Salyk} C, {Lacy} J, {Richter} M, {Zhang} K, {Pontoppidan} K, et~al. 2019.
\textit{ApJ} 874:24

\bibitem[{{Salyk} et~al.(2011{\natexlab{b}}){Salyk}, {Pontoppidan}, {Blake},
  {Najita} \& {Carr}}]{Salyk11}
{Salyk} C, {Pontoppidan} KM, {Blake} GA, {Najita} JR, {Carr} JS.
  2011{\natexlab{b}}.
\textit{ApJ} 731:130

\bibitem[{{Sargent} \& {Beckwith}(1987)}]{Sargent87}
{Sargent} AI, {Beckwith} S. 1987.
\textit{ApJ} 323:294

\bibitem[{{Sch{\"o}ier} et~al.(2005){Sch{\"o}ier}, {van der Tak}, {van
  Dishoeck} \& {Black}}]{Schoier05}
{Sch{\"o}ier} FL, {van der Tak} FFS, {van Dishoeck} EF, {Black} JH. 2005.
\textit{A\&A} 432:369--379

\bibitem[{{Schwarz} et~al.(2016){Schwarz}, {Bergin}, {Cleeves}, {Blake},
  {Zhang} et~al.}]{Schwarz16}
{Schwarz} KR, {Bergin} EA, {Cleeves} LI, {Blake} GA, {Zhang} K, et~al. 2016.
\textit{ApJ} 823:91

\bibitem[{{Schwarz} et~al.(2018){Schwarz}, {Bergin}, {Cleeves}, {Zhang},
  {{\"O}berg} et~al.}]{Schwarz18}
{Schwarz} KR, {Bergin} EA, {Cleeves} LI, {Zhang} K, {{\"O}berg} KI, et~al.
  2018.
\textit{ApJ} 856:85

\bibitem[{{Seifert} et~al.(2021){Seifert}, {Cleeves}, {Adams} \&
  {Li}}]{Seifert21}
{Seifert} RA, {Cleeves} LI, {Adams} FC, {Li} ZY. 2021.
\textit{ApJ} 912:136

\bibitem[{{Semenov} \& {Wiebe}(2011)}]{Semenov11}
{Semenov} D, {Wiebe} D. 2011.
\textit{ApJs} 196:25

\bibitem[{{Semenov} et~al.(2004){Semenov}, {Wiebe} \& {Henning}}]{Semenov04}
{Semenov} D, {Wiebe} D, {Henning} T. 2004.
\textit{A\&A} 417:93--106

\bibitem[{{Shu} et~al.(1987){Shu}, {Adams} \& {Lizano}}]{Shu87}
{Shu} FH, {Adams} FC, {Lizano} S. 1987.
\textit{ARA\&A} 25:23--81

\bibitem[{{Siebenmorgen} \& {Heymann}(2012)}]{Siebenmorgen12}
{Siebenmorgen} R, {Heymann} F. 2012.
\textit{A\&A} 543:A25

\bibitem[{{Simon} et~al.(2019){Simon}, {Guilloteau}, {Beck}, {Chapillon}, {Di
  Folco} et~al.}]{Simon19}
{Simon} M, {Guilloteau} S, {Beck} TL, {Chapillon} E, {Di Folco} E, et~al. 2019.
\textit{ApJ} 884:42

\bibitem[{{Smirnov-Pinchukov} et~al.(2022){Smirnov-Pinchukov}, {Molyarova},
  {Semenov}, {Akimkin}, {van Terwisga} et~al.}]{Smirnov22}
{Smirnov-Pinchukov} GV, {Molyarova} T, {Semenov} DA, {Akimkin} VV, {van
  Terwisga} S, et~al. 2022.
\textit{A\&A} 666:L8

\bibitem[{{Smith} et~al.(2015){Smith}, {Pontoppidan}, {Young} \&
  {Morris}}]{Smith15}
{Smith} RL, {Pontoppidan} KM, {Young} ED, {Morris} MR. 2015.
\textit{ApJ} 813:120

\bibitem[{{Stevenson} \& {Lunine}(1988)}]{Stevenson88}
{Stevenson} DJ, {Lunine} JI. 1988.
\textit{Icarus} 75:146--155

\bibitem[{{Sturm} et~al.(2022){Sturm}, {McClure}, {Harsono}, {Facchini}, {Long}
  et~al.}]{Sturm22}
{Sturm} JA, {McClure} MK, {Harsono} D, {Facchini} S, {Long} F, et~al. 2022.
\textit{A\&A} 660:A126

\bibitem[{{Teague} et~al.(2019){Teague}, {Bae} \& {Bergin}}]{Teague19}
{Teague} R, {Bae} J, {Bergin} EA. 2019.
\textit{Nature} 574:378--381

\bibitem[{{Teague} et~al.(2020){Teague}, {Jankovic}, {Haworth}, {Qi} \&
  {Ilee}}]{Teague20c}
{Teague} R, {Jankovic} MR, {Haworth} TJ, {Qi} C, {Ilee} JD. 2020.
\textit{MNRAS} 495:451--459

\bibitem[{{Teague} \& {Loomis}(2020)}]{Teague20}
{Teague} R, {Loomis} R. 2020.
\textit{ApJ} 899:157

\bibitem[{{Teague} et~al.(2017){Teague}, {Semenov}, {Gorti}, {Guilloteau},
  {Henning} et~al.}]{Teague17}
{Teague} R, {Semenov} D, {Gorti} U, {Guilloteau} S, {Henning} T, et~al. 2017.
\textit{ApJ} 835:228

\bibitem[{{Teague} et~al.(2015){Teague}, {Semenov}, {Guilloteau}, {Henning},
  {Dutrey} et~al.}]{Teague15}
{Teague} R, {Semenov} D, {Guilloteau} S, {Henning} T, {Dutrey} A, et~al. 2015.
\textit{A\&A} 574:A137

\bibitem[{{Terada} \& {Tokunaga}(2017)}]{terada17}
{Terada} H, {Tokunaga} AT. 2017.
\textit{ApJ} 834:115

\bibitem[{{Terwisscha van Scheltinga} et~al.(2021){Terwisscha van Scheltinga},
  {Hogerheijde}, {Cleeves}, {Loomis}, {Walsh} et~al.}]{Terwisscha21}
{Terwisscha van Scheltinga} J, {Hogerheijde} MR, {Cleeves} LI, {Loomis} RA,
  {Walsh} C, et~al. 2021.
\textit{ApJ} 906:111

\bibitem[{{Thi} et~al.(2004){Thi}, {van Zadelhoff} \& {van Dishoeck}}]{Thi04}
{Thi} W, {van Zadelhoff} G, {van Dishoeck} EF. 2004.
\textit{A\&A} 425:955--972

\bibitem[{{Thi} et~al.(2020){Thi}, {Hocuk}, {Kamp}, {Woitke}, {Rab}
  et~al.}]{Thi20}
{Thi} WF, {Hocuk} S, {Kamp} I, {Woitke} P, {Rab} C, et~al. 2020.
\textit{A\&A} 635:A16

\bibitem[{{Thi} et~al.(2019){Thi}, {Lesur}, {Woitke}, {Kamp}, {Rab} \&
  {Carmona}}]{Thi19}
{Thi} WF, {Lesur} G, {Woitke} P, {Kamp} I, {Rab} C, {Carmona} A. 2019.
\textit{A\&A} 632:A44

\bibitem[{{Thi} et~al.(2010){Thi}, {Mathews}, {M{\'e}nard}, {Woitke}, {Meeus}
  et~al.}]{Thi10}
{Thi} WF, {Mathews} G, {M{\'e}nard} F, {Woitke} P, {Meeus} G, et~al. 2010.
\textit{A\&A} 518:L125

\bibitem[{{Thi} et~al.(2014){Thi}, {Pinte}, {Pantin}, {Augereau}, {Meeus}
  et~al.}]{Thi14}
{Thi} WF, {Pinte} C, {Pantin} E, {Augereau} JC, {Meeus} G, et~al. 2014.
\textit{A\&A} 561:A50

\bibitem[{{Thi} et~al.(2002){Thi}, {Pontoppidan}, {van Dishoeck}, {Dartois} \&
  {d'Hendecourt}}]{Thi02}
{Thi} WF, {Pontoppidan} KM, {van Dishoeck} EF, {Dartois} E, {d'Hendecourt} L.
  2002.
\textit{A\&A} 394:L27--L30

\bibitem[{{Tielens}(2008)}]{Tielens08}
{Tielens} AGGM. 2008.
\textit{ARA\&A} 46:289--337

\bibitem[{{Tielens} \& {Hagen}(1982)}]{Tielens82}
{Tielens} AGGM, {Hagen} W. 1982.
\textit{A\&A} 114:245--260

\bibitem[{{Trapman} et~al.(2022){Trapman}, {Zhang}, {van't Hoff}, {Hogerheijde}
  \& {Bergin}}]{Trapman22}
{Trapman} L, {Zhang} K, {van't Hoff} MLR, {Hogerheijde} MR, {Bergin} EA. 2022.
\textit{ApJL} 926:L2

\bibitem[{{van Boekel} et~al.(2004){van Boekel}, {Min}, {Leinert}, {Waters},
  {Richichi} et~al.}]{vanBoekel04}
{van Boekel} R, {Min} M, {Leinert} C, {Waters} LBFM, {Richichi} A, et~al. 2004.
\textit{Nature} 432:479--482

\bibitem[{{Van Clepper} et~al.(2022){Van Clepper}, {Bergner}, {Bosman},
  {Bergin} \& {Ciesla}}]{vanClepper22}
{Van Clepper} E, {Bergner} JB, {Bosman} AD, {Bergin} E, {Ciesla} FJ. 2022.
\textit{ApJ} 927:206

\bibitem[{{van den Ancker} et~al.(2000){van den Ancker}, {Bouwman},
  {Wesselius}, {Waters}, {Dougherty} \& {van Dishoeck}}]{vandenancker00}
{van den Ancker} ME, {Bouwman} J, {Wesselius} PR, {Waters} LBFM, {Dougherty}
  SM, {van Dishoeck} EF. 2000.
\textit{A\&A} 357:325--329

\bibitem[{{van der Marel} et~al.(2021){van der Marel}, {Booth}, {Leemker}, {van
  Dishoeck} \& {Ohashi}}]{vanderMarel21}
{van der Marel} N, {Booth} AS, {Leemker} M, {van Dishoeck} EF, {Ohashi} S.
  2021.
\textit{A\&A} 651:L5

\bibitem[{{van der Marel} et~al.(2016){van der Marel}, {van Dishoeck},
  {Bruderer}, {Andrews}, {Pontoppidan} et~al.}]{vanderMarel16}
{van der Marel} N, {van Dishoeck} EF, {Bruderer} S, {Andrews} SM, {Pontoppidan}
  KM, et~al. 2016.
\textit{A\&A} 585:A58

\bibitem[{{van der Marel} et~al.(2015){van der Marel}, {van Dishoeck},
  {Bruderer}, {P{\'e}rez} \& {Isella}}]{vanderMarel15}
{van der Marel} N, {van Dishoeck} EF, {Bruderer} S, {P{\'e}rez} L, {Isella} A.
  2015.
\textit{A\&A} 579:A106

\bibitem[{{van der Marel} et~al.(2018){van der Marel}, {Williams} \&
  {Bruderer}}]{vanderMarel18b}
{van der Marel} N, {Williams} JP, {Bruderer} S. 2018.
\textit{ApJL} 867:L14

\bibitem[{{van der Tak} et~al.(2007){van der Tak}, {Black}, {Sch{\"o}ier},
  {Jansen} \& {van Dishoeck}}]{vanderTak07}
{van der Tak} FFS, {Black} JH, {Sch{\"o}ier} FL, {Jansen} DJ, {van Dishoeck}
  EF. 2007.
\textit{A\&A} 468:627--635

\bibitem[{{van der Wiel} et~al.(2014{\natexlab{a}}){van der Wiel}, {Naylor},
  {Kamp}, {M{\'e}nard}, {Thi} et~al.}]{vanderWiel14}
{van der Wiel} MHD, {Naylor} DA, {Kamp} I, {M{\'e}nard} F, {Thi} WF, et~al.
  2014{\natexlab{a}}.
\textit{MNRAS} 444:3911--3925

\bibitem[{{van der Wiel} et~al.(2014{\natexlab{b}}){van der Wiel}, {Naylor},
  {Kamp}, {M{\'e}nard}, {Thi} et~al.}]{van_der_wiel14}
{van der Wiel} MHD, {Naylor} DA, {Kamp} I, {M{\'e}nard} F, {Thi} WF, et~al.
  2014{\natexlab{b}}.
\textit{MNRAS} 444:3911--3925

\bibitem[{{van Dishoeck} \& {Bergin}(2021)}]{vanDishoeck21b}
{van Dishoeck} EF, {Bergin} EA. 2021.
\textit{{Astrochemistry and Planet Formation}}. In \textit{ExoFrontiers; Big
  Questions in Exoplanetary Science}, ed. N~{Madhusudhan}.  14--1

\bibitem[{{van Dishoeck} et~al.(2021){van Dishoeck}, {Kristensen}, {Mottram},
  {Benz}, {Bergin} et~al.}]{vanDishoeck21}
{van Dishoeck} EF, {Kristensen} LE, {Mottram} JC, {Benz} AO, {Bergin} EA,
  et~al. 2021.
\textit{A\&A} 648:A24

\bibitem[{{van Dishoek} et~al.(1993){van Dishoek}, {Blake}, {Draine} \&
  {Lunine}}]{vanDishoeck93}
{van Dishoek} EF, {Blake} GA, {Draine} BT, {Lunine} JI. 1993.
\textit{{The chemical evolution of protostellar and protoplanetary matter}}. In
  \textit{Protostars and Planets III}, eds. EH~{Levy}, JI~{Lunine}

\bibitem[{{van 't Hoff} et~al.(2018{\natexlab{a}}){van 't Hoff}, {Persson},
  {Harsono}, {Taquet}, {J{\o}rgensen} et~al.}]{vantHoff18}
{van 't Hoff} MLR, {Persson} MV, {Harsono} D, {Taquet} V, {J{\o}rgensen} JK,
  et~al. 2018{\natexlab{a}}.
\textit{A\&A} 613:A29

\bibitem[{{van 't Hoff} et~al.(2018{\natexlab{b}}){van 't Hoff}, {Tobin},
  {Trapman}, {Harsono}, {Sheehan} et~al.}]{vantHoff18a}
{van 't Hoff} MLR, {Tobin} JJ, {Trapman} L, {Harsono} D, {Sheehan} PD, et~al.
  2018{\natexlab{b}}.
\textit{ApJL} 864:L23

\bibitem[{{van 't Hoff} et~al.(2017){van 't Hoff}, {Walsh}, {Kama}, {Facchini}
  \& {van Dishoeck}}]{vantHoff17}
{van 't Hoff} MLR, {Walsh} C, {Kama} M, {Facchini} S, {van Dishoeck} EF. 2017.
\textit{A\&A} 599:A101

\bibitem[{{van Terwisga} et~al.(2019){van Terwisga}, {van Dishoeck},
  {Cazzoletti}, {Facchini}, {Trapman} et~al.}]{vanTerwisga19}
{van Terwisga} SE, {van Dishoeck} EF, {Cazzoletti} P, {Facchini} S, {Trapman}
  L, et~al. 2019.
\textit{A\&A} 623:A150

\bibitem[{{van Zadelhoff} et~al.(2003){van Zadelhoff}, {Aikawa}, {Hogerheijde}
  \& {van Dishoeck}}]{vanZadelhoff03}
{van Zadelhoff} G, {Aikawa} Y, {Hogerheijde} MR, {van Dishoeck} EF. 2003.
\textit{A\&A} 397:789--802

\bibitem[{{Vasyunin} et~al.(2011){Vasyunin}, {Wiebe}, {Birnstiel}, {Zhukovska},
  {Henning} \& {Dullemond}}]{Vasyunin11}
{Vasyunin} AI, {Wiebe} DS, {Birnstiel} T, {Zhukovska} S, {Henning} T,
  {Dullemond} CP. 2011.
\textit{ApJ} 727:76

\bibitem[{{Visser} et~al.(2018){Visser}, {Bruderer}, {Cazzoletti}, {Facchini},
  {Heays} \& {van Dishoeck}}]{Visser18}
{Visser} R, {Bruderer} S, {Cazzoletti} P, {Facchini} S, {Heays} AN, {van
  Dishoeck} EF. 2018.
\textit{A\&A} 615:A75

\bibitem[{{Visser} et~al.(2009){Visser}, {van Dishoeck}, {Doty} \&
  {Dullemond}}]{Visser09}
{Visser} R, {van Dishoeck} EF, {Doty} SD, {Dullemond} CP. 2009.
\textit{A\&A} 495:881--897

\bibitem[{{Wakelam} et~al.(2012){Wakelam}, {Herbst}, {Loison}, {Smith},
  {Chandrasekaran} et~al.}]{Wakelam12}
{Wakelam} V, {Herbst} E, {Loison} JC, {Smith} IWM, {Chandrasekaran} V, et~al.
  2012.
\textit{ApJs} 199:21

\bibitem[{{Wakelam} et~al.(2016){Wakelam}, {Ruaud}, {Hersant}, {Dutrey},
  {Semenov} et~al.}]{Wakelam16}
{Wakelam} V, {Ruaud} M, {Hersant} F, {Dutrey} A, {Semenov} D, et~al. 2016.
\textit{A\&A} 594:A35

\bibitem[{{Walsh} et~al.(2010){Walsh}, {Millar} \& {Nomura}}]{Walsh10}
{Walsh} C, {Millar} TJ, {Nomura} H. 2010.
\textit{ApJ} 722:1607--1623

\bibitem[{{Walsh} et~al.(2013){Walsh}, {Millar} \& {Nomura}}]{Walsh13}
{Walsh} C, {Millar} TJ, {Nomura} H. 2013.
\textit{ApJL} 766:L23

\bibitem[{{Walsh} et~al.(2014){Walsh}, {Millar}, {Nomura}, {Herbst}, {Widicus
  Weaver} et~al.}]{Walsh14}
{Walsh} C, {Millar} TJ, {Nomura} H, {Herbst} E, {Widicus Weaver} S, et~al.
  2014.
\textit{A\&A} 563:A33

\bibitem[{{Walsh} et~al.(2015{\natexlab{a}}){Walsh}, {Nomura} \& {van
  Dishoeck}}]{Walsh15}
{Walsh} C, {Nomura} H, {van Dishoeck} E. 2015{\natexlab{a}}.
\textit{A\&A} 582:A88

\bibitem[{{Walsh} et~al.(2015{\natexlab{b}}){Walsh}, {Nomura} \& {van
  Dishoeck}}]{Walsh15-model}
{Walsh} C, {Nomura} H, {van Dishoeck} E. 2015{\natexlab{b}}.
\textit{A\&A} 582:A88

\bibitem[{{Wei} et~al.(2019){Wei}, {Nomura}, {Lee}, {Ip}, {Walsh} \&
  {Millar}}]{Wei19}
{Wei} CE, {Nomura} H, {Lee} JE, {Ip} WH, {Walsh} C, {Millar} TJ. 2019.
\textit{ApJ} 870:129

\bibitem[{{Willacy}(2007)}]{Willacy07}
{Willacy} K. 2007.
\textit{ApJ} 660:441--460

\bibitem[{{Willacy} \& {Millar}(1998)}]{Willacy98}
{Willacy} K, {Millar} TJ. 1998.
\textit{MNRAS} 298:562--568

\bibitem[{{Willacy} \& {Woods}(2009)}]{Willacy09}
{Willacy} K, {Woods} PM. 2009.
\textit{ApJ} 703:479--499

\bibitem[{{Williams} \& {Best}(2014)}]{Williams14}
{Williams} JP, {Best} WMJ. 2014.
\textit{ApJ} 788:59

\bibitem[{{Wilson} et~al.(2016){Wilson}, {G{\"a}nsicke}, {Farihi} \&
  {Koester}}]{Wilson16}
{Wilson} DJ, {G{\"a}nsicke} BT, {Farihi} J, {Koester} D. 2016.
\textit{MNRAS} 459:3282--3286

\bibitem[{{Woitke} et~al.(2019){Woitke}, {Kamp}, {Antonellini}, {Anthonioz},
  {Baldovin-Saveedra} et~al.}]{Woitke19}
{Woitke} P, {Kamp} I, {Antonellini} S, {Anthonioz} F, {Baldovin-Saveedra} C,
  et~al. 2019.
\textit{PASP} 131:064301

\bibitem[{{Woitke} et~al.(2009){Woitke}, {Kamp} \& {Thi}}]{Woitke09}
{Woitke} P, {Kamp} I, {Thi} WF. 2009.
\textit{A\&A} 501:383--406

\bibitem[{{Woitke} et~al.(2018){Woitke}, {Min}, {Thi}, {Roberts}, {Carmona}
  et~al.}]{Woitke18}
{Woitke} P, {Min} M, {Thi} WF, {Roberts} C, {Carmona} A, et~al. 2018.
\textit{A\&A} 618:A57

\bibitem[{{Xu} et~al.(2019){Xu}, {Bai}, {{\"O}berg} \& {Zhang}}]{Xu19}
{Xu} R, {Bai} XN, {{\"O}berg} K, {Zhang} H. 2019.
\textit{ApJ} 872:107

\bibitem[{{Young} et~al.(2021){Young}, {Alexander}, {Walsh}, {Nealon}, {Booth}
  \& {Pinte}}]{Young21}
{Young} AK, {Alexander} R, {Walsh} C, {Nealon} R, {Booth} A, {Pinte} C. 2021.
\textit{MNRAS} 505:4821--4837

\bibitem[{{Yu} et~al.(2021){Yu}, {Teague}, {Bae} \& {{\"O}berg}}]{Yu21}
{Yu} H, {Teague} R, {Bae} J, {{\"O}berg} K. 2021.
\textit{ApJL} 920:L33

\bibitem[{{Yu} et~al.(2017){Yu}, {Evans}, {Dodson-Robinson}, {Willacy} \&
  {Turner}}]{yu17}
{Yu} M, {Evans} Neal~J. I, {Dodson-Robinson} SE, {Willacy} K, {Turner} NJ.
  2017.
\textit{ApJ} 850:169

\bibitem[{{Zhang} et~al.(2017){Zhang}, {Bergin}, {Blake}, {Cleeves} \&
  {Schwarz}}]{Zhang17}
{Zhang} K, {Bergin} EA, {Blake} GA, {Cleeves} LI, {Schwarz} KR. 2017.
\textit{Nature Astronomy} 1:0130

\bibitem[{{Zhang} et~al.(2019){Zhang}, {Bergin}, {Schwarz}, {Krijt} \&
  {Ciesla}}]{Zhang19}
{Zhang} K, {Bergin} EA, {Schwarz} K, {Krijt} S, {Ciesla} F. 2019.
\textit{ApJ} 883:98

\bibitem[{{Zhang} et~al.(2015){Zhang}, {Blake} \& {Bergin}}]{Zhang15}
{Zhang} K, {Blake} GA, {Bergin} EA. 2015.
\textit{ApJL} 806:L7

\bibitem[{{Zhang} et~al.(2021{\natexlab{a}}){Zhang}, {Booth}, {Law}, {Bosman},
  {Schwarz} et~al.}]{Zhang21}
{Zhang} K, {Booth} AS, {Law} CJ, {Bosman} AD, {Schwarz} KR, et~al.
  2021{\natexlab{a}}.
\textit{ApJs} 257:5

\bibitem[{{Zhang} et~al.(2013){Zhang}, {Pontoppidan}, {Salyk} \&
  {Blake}}]{Zhang13}
{Zhang} K, {Pontoppidan} KM, {Salyk} C, {Blake} GA. 2013.
\textit{ApJ} 766:82

\bibitem[{Zhang et~al.(2018)Zhang, Zhu, Huang, Guzm{\'{a}}n, Andrews
  et~al.}]{Zhang18}
Zhang S, Zhu Z, Huang J, Guzm{\'{a}}n VV, Andrews SM, et~al. 2018.
\textit{The Astrophysical Journal} 869:L47

\bibitem[{{Zhang} et~al.(2021{\natexlab{b}}){Zhang}, {Snellen} \&
  {Molli{\`e}re}}]{Zhang21-exo}
{Zhang} Y, {Snellen} IAG, {Molli{\`e}re} P. 2021{\natexlab{b}}.
\textit{A\&A} 656:A76

\end{thebibliography}
\bibliographystyle{ar-style2}

\end{document}